\documentclass[twocolumn,twocolappendix,trackchanges]{aastex63}
\usepackage{amsmath}

\newcommand\kms{km s$^{-1}$}
\newcommand\masyr{mas yr$^{-1}$}
\newcommand\teff{$T_{eff}$}

\newcommand\av{$A_V$}

\defcitealias{kounkel2019a}{Paper I}
\defcitealias{kharchenko2013}{MWSC}
\defcitealias{cantat-gaudin2018a}{CG18}

\usepackage{listings}

\begin{document}
\title{Untangling the Galaxy. II. Structure within 3 kpc}

\author[0000-0002-5365-1267]{Marina Kounkel}
\affil{Department of Physics and Astronomy, Western Washington University, 516 High St, Bellingham, WA 98225}
\author[0000-0001-6914-7797]{Kevin Covey}
\affil{Department of Physics and Astronomy, Western Washington University, 516 High St, Bellingham, WA 98225}
\author[0000-0002-3481-9052]{Keivan G. Stassun}
\affil{Department of Physics and Astronomy, Vanderbilt University, VU Station 1807, Nashville, TN 37235, USA}
\email{marina.kounkel@wwu.edu}

\begin{abstract}
We present the results of the hierarchical clustering analysis of the \textit{Gaia} DR2 data to search for clusters, co-moving groups, and other stellar structures. The current paper builds on the sample from the previous work, extending it in distance from 1 kpc to 3 kpc, increasing the number of identified structures up to 8292. To aid in the analysis of the population properties, we developed a neural network called Auriga to robustly estimate the age, extinction, and distance of a stellar group based on the input photometry and parallaxes of the individual members. We apply Auriga to derive the properties of not only the structures found in this paper, but also previously identified open clusters. Through this work, we examine the temporal structure of the spiral arms. Specifically, we find that the Sagittarius arm has moved by $>$500 pc in the last 100 Myr, and the Perseus arm has been experiencing a relative lull in star formation activity over the last 25 Myr. We confirm the findings from the previous paper on the transient nature of the spiral arms, with the timescale of transition of a few 100 Myr. Finally, we find a peculiar $\sim$1 Gyr old stream of stars that appears to be heliocentric. It is unclear what is the origin of it.
\end{abstract}


\section{Introduction}

The unprecedented precision and sensitivity of \textit{Gaia} DR2 has \citep{gaia-collaboration2018} has resulted in a significantly improved understanding of the structure and the kinematics of the Galaxy. Among the topics that have benefited the most from Gaia is identification of clusters, associations, and co-moving groups, as it is now possible to not only robustly identify members of such populations based on their position in the phase space \citep[e.g.,][]{cantat-gaudin2018a,castro-ginard2019,cantat-gaudin2019a,sim2019,liu2019,castro-ginard2020}, but to also analyze their position on the HR diagram.

Previously, in \citet[hereafter Paper I]{kounkel2019a} we systematically clustered \textit{Gaia} DR2 data within $|b|<30^\circ$ and parallax $\pi>1$ mas using HDBSCAN \citep{hdbscan1,hdbscan} to identify 1640 populations containing in total 288,370 stars. Furthermore, we estimated ages of these structures via isochrone fitting.

Approximately half of the stars in these populations are found in coherent extended string-like populations that have a typical length of $\sim$200 pc, and width of $\sim$10 pc. They are ubiquitous among young populations: most of the sources younger than 300 Myr are found in such strings. Given their age, these structures are not likely to be a result of tidal stretching of previously compact clusters, as the tidal tails take several 100s Myr to stretch out to similar distances \citep{ernst2011,roser2019,roser2019a}. Thus, the number of long strings found at the ages significantly younger than 100 Myr suggests a different origin. Rather, the strings are likely to be primordial, preserving the shape of fillamentary giant molecular clouds from which the stars have formed. Some such strings have also been found by \citet{beccari2020,jerabkova2019a,meingast2019,sim2019}.

As the stars from these populations slowly dissolve into the field, after $\sim$300 Myr most of the stars in the low density regions of the strings would no longer be clusterable, typically leaving behind the densest, most cluster-like isolated and compact groups of stars.

The identified populations in \citetalias{kounkel2019a} exceed the spatial coverage of previously known open clusters \citep[e.g.,][]{dias2002,kharchenko2013,cantat-gaudin2018a}, making the catalog more effective at analyzing the distribution of a number of properties, from stellar to galactic, as a function of age. Examining the populations younger than 100 Myr, we found that there is a preferred orientation of the strings to be generally in parallel to one another, perpendicular to the Local Arm. It is possible that these identified strings are stellar analogs to various gaseous spurs or feathers that have been identified in observations of other galaxies \citep[e.g.,][]{schinnerer2017}. At the ages of 8--8.7 dex, the overall stream that can be traced by the superposition of the strings has shifted. While they were still oriented in parallel to one another, the stream was tilted by $\sim60^\circ$ relative to the Local Arm, with a very sharp transition in age between these two modes. Furthermore, yet another transition has been observed at the age of $\sim8.7$ dex, revealing two different streams with a very different morphology. The interpretation of this has been that the strings trace the spiral arms as they existed when the stars inside these populations has been forming, and that the the spiral arms are transient -- dissolving, twisting, and reforming in a similar region of the galaxy with a different underlying shape. The time scale for this transition would be on an order of a Galactic orbit, or a few 100 Myr. 

However, the catalog in \citetalias{kounkel2019a} extended only to 1 kpc, and included only the Local arm. Although astrometric precision does drop at larger distances, \textit{Gaia} DR2 does allow to extend the catalog further, increasing the census of clustered structures, as well as the census of stars with age estimates. In this work we build on the previous efforts to identify additional structures up to $\pi<0.2$ mas, towards the detection limit for clustering imposed by the extinction. In Section \ref{sec:methods} we discuss the clustering technique to identify structures and derive the limits of the clustering approach as a function of age due to extinction. In Appendix \ref{sec:auriga} we develop an algorithm to derive structure properties. In Section \ref{sec:3d} we show and discuss the 3-dimensional distribution of the structures in the context of the Galactic structure. Finally, in Section \ref{sec:concl} we discuss the results.

\section{Methods}\label{sec:methods}

In this section we describe the process of selecting the sample for the analysis. Section \ref{sec:cluster} discusses extending the clustering analysis originally presented in \citetalias{kounkel2019a} to the larger distances through a homogeneous search for statistically meaningful overdensities in 5-dimensional phase space, and deriving average properties of the stars in those identified groups. Section \ref{sec:contam} describes the validation of the identified populations through independent confirmations of derived ages, as well as identifying the highest confidence sample that can be separated from the randomly drawn field stars based on their collective photometry (the analysis in Section \ref{sec:3d} is limited to this highest confidence sample). In Section \ref{sec:complete} we examine various biases as well as estimate the completeness limits in distance along the various lines of sight as a function of age.

\subsection{Clustering}\label{sec:cluster}

\begin{figure*}
\begin{interactive}{js}{interactive.zip}
		\gridline{
             \fig{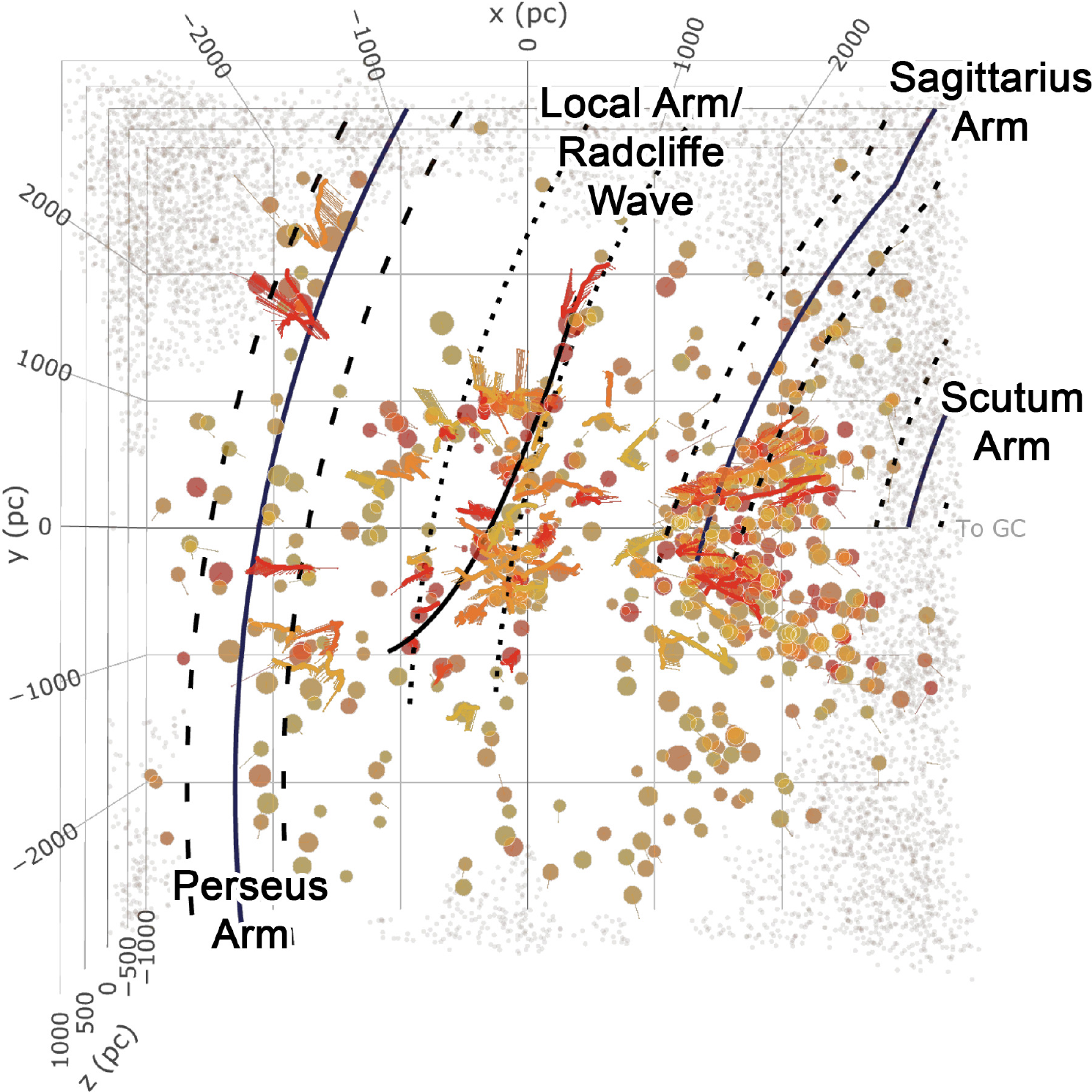}{0.35\textwidth}{}
             \fig{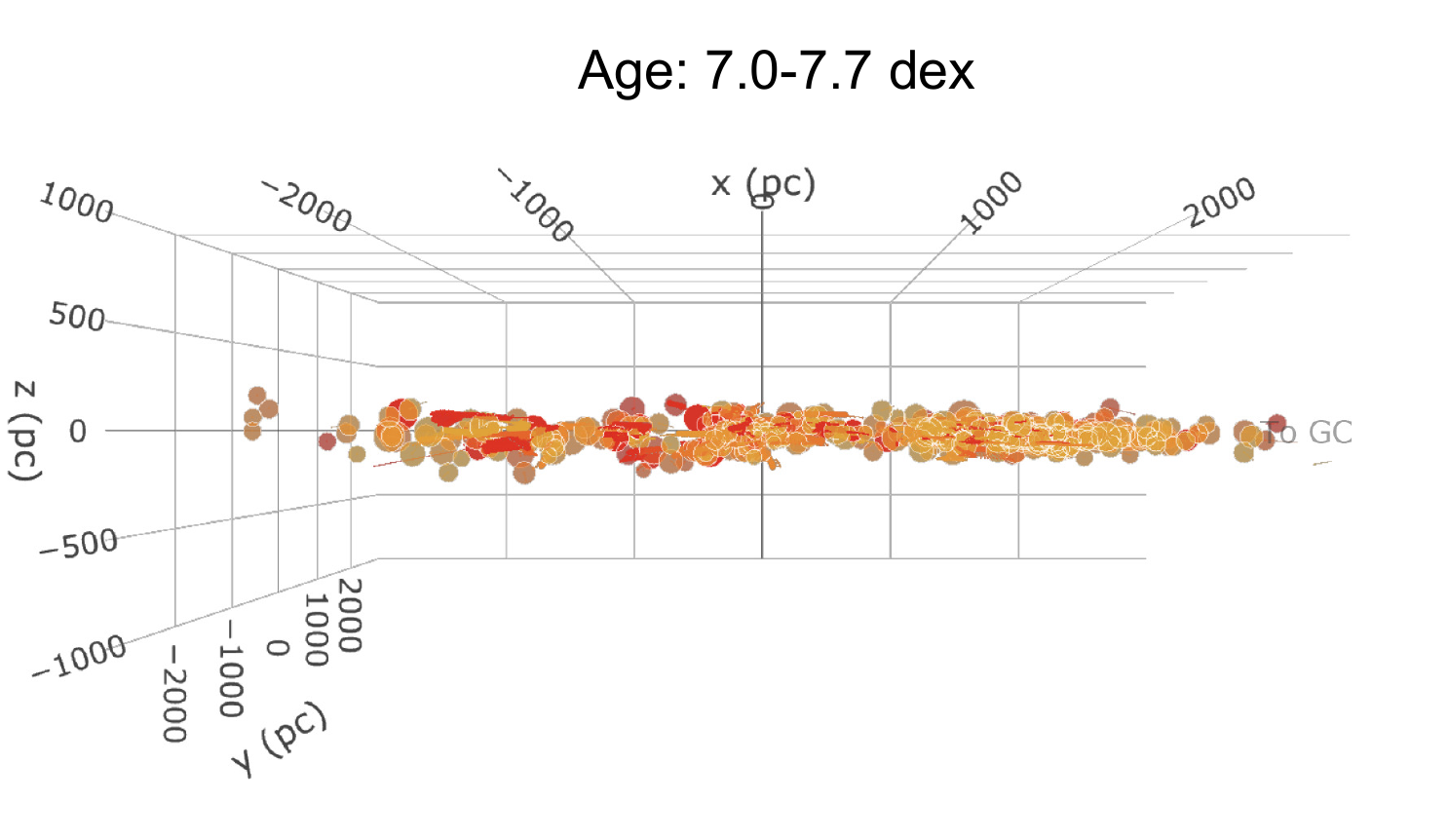}{0.6\textwidth}{}
        }\vspace{-1 cm}
        	\gridline{
             \fig{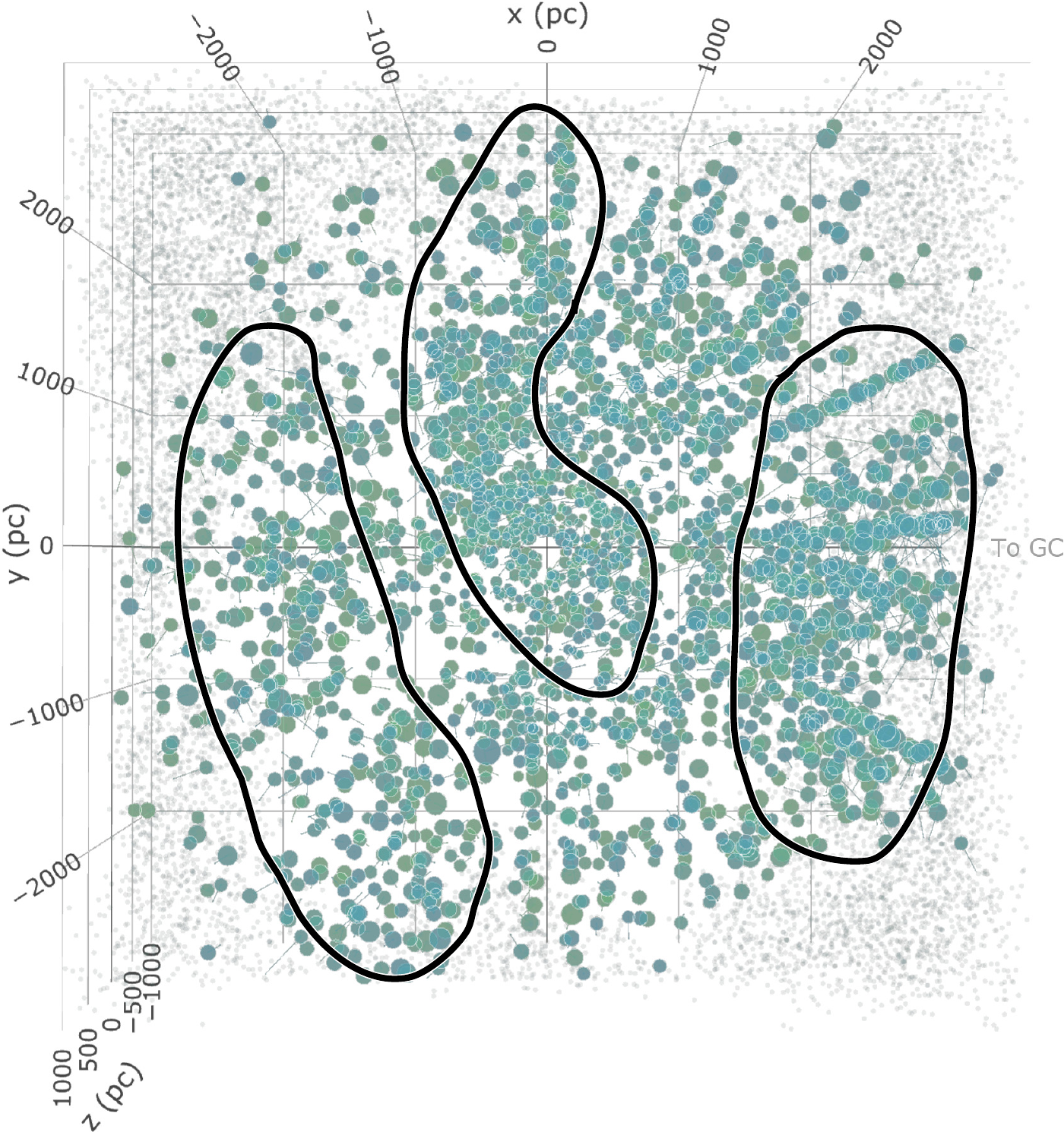}{0.35\textwidth}{}
             \fig{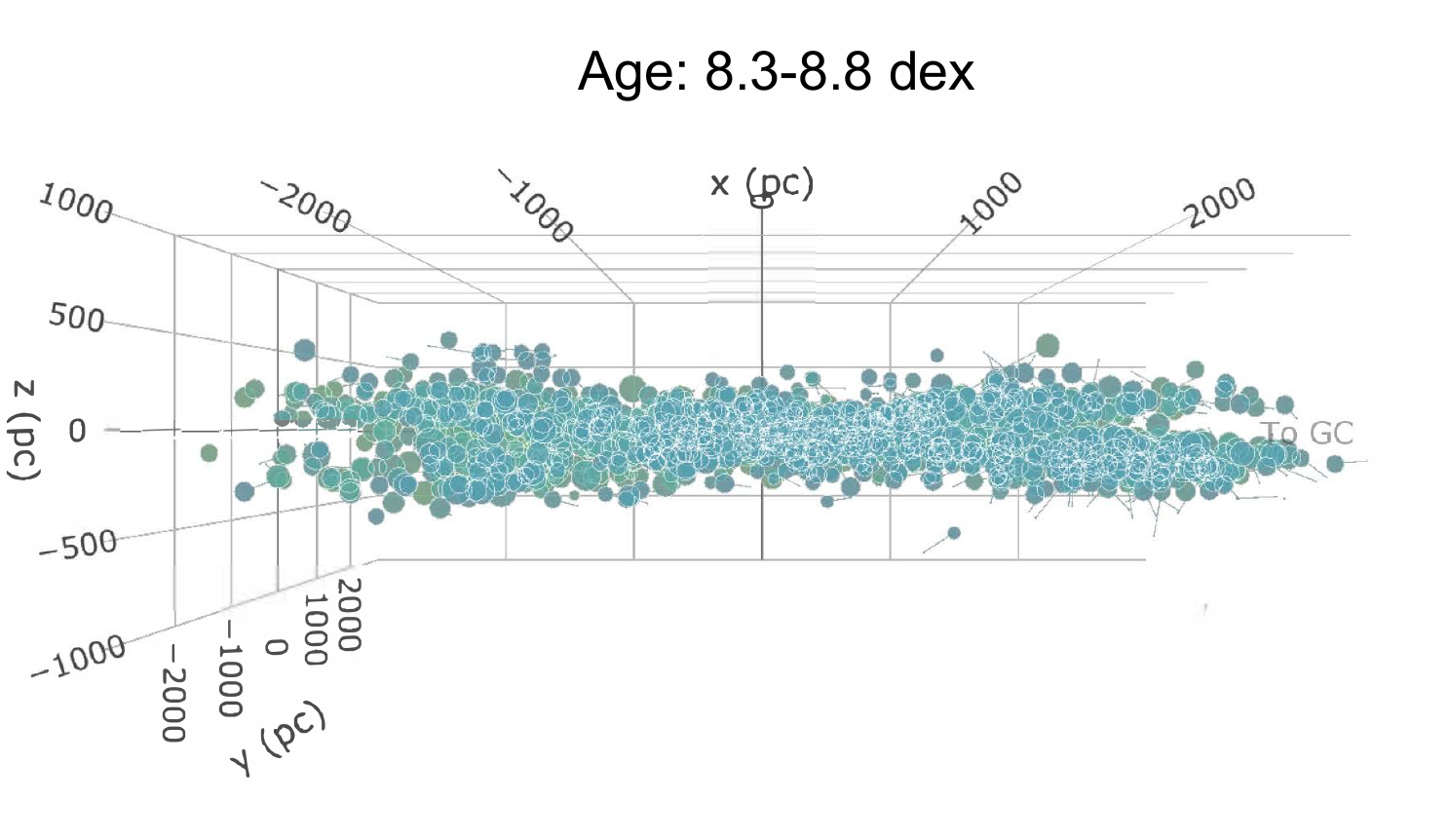}{0.6\textwidth}{}
        }\vspace{-1 cm}
		\gridline{
             \fig{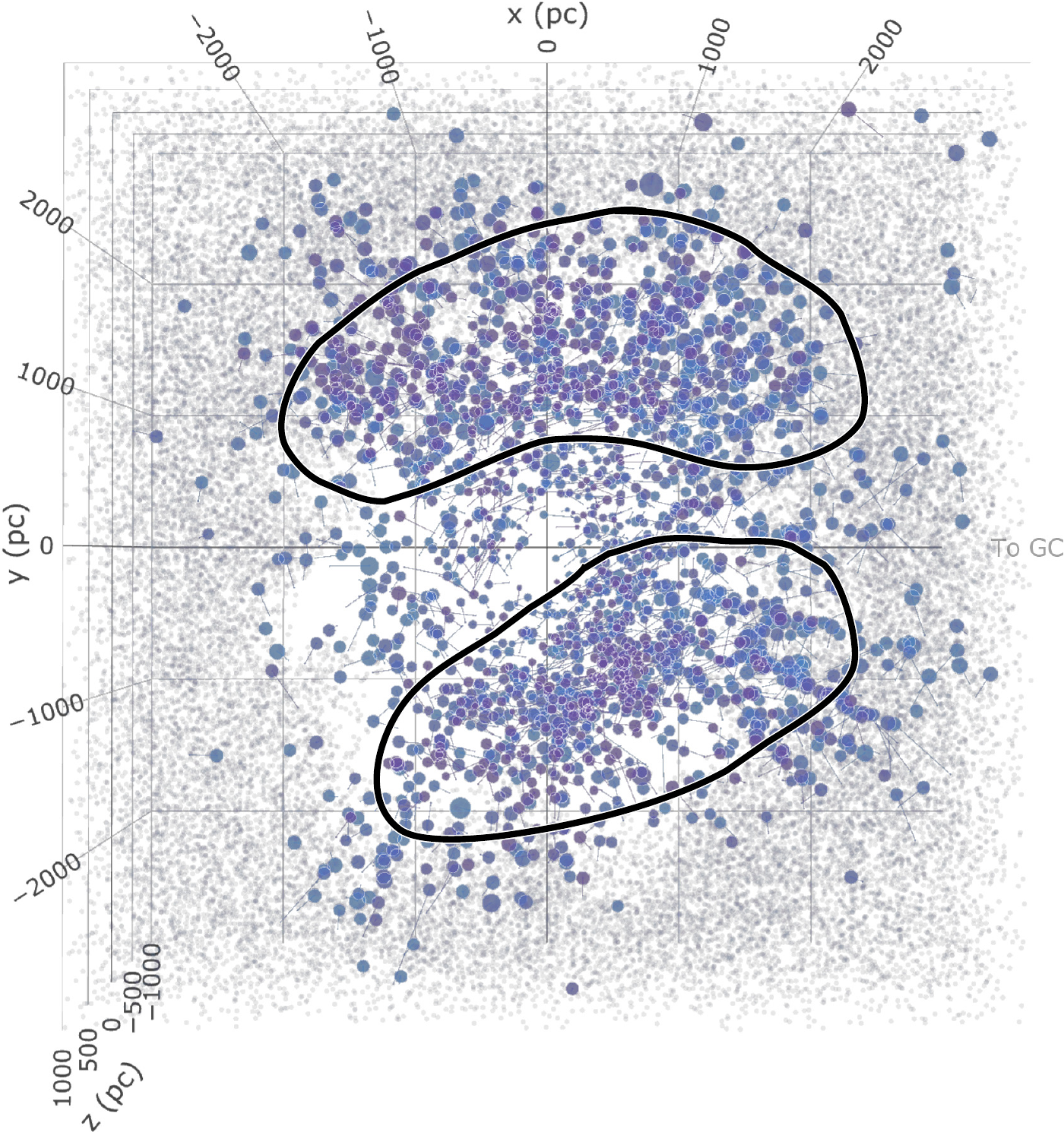}{0.35\textwidth}{}
             \fig{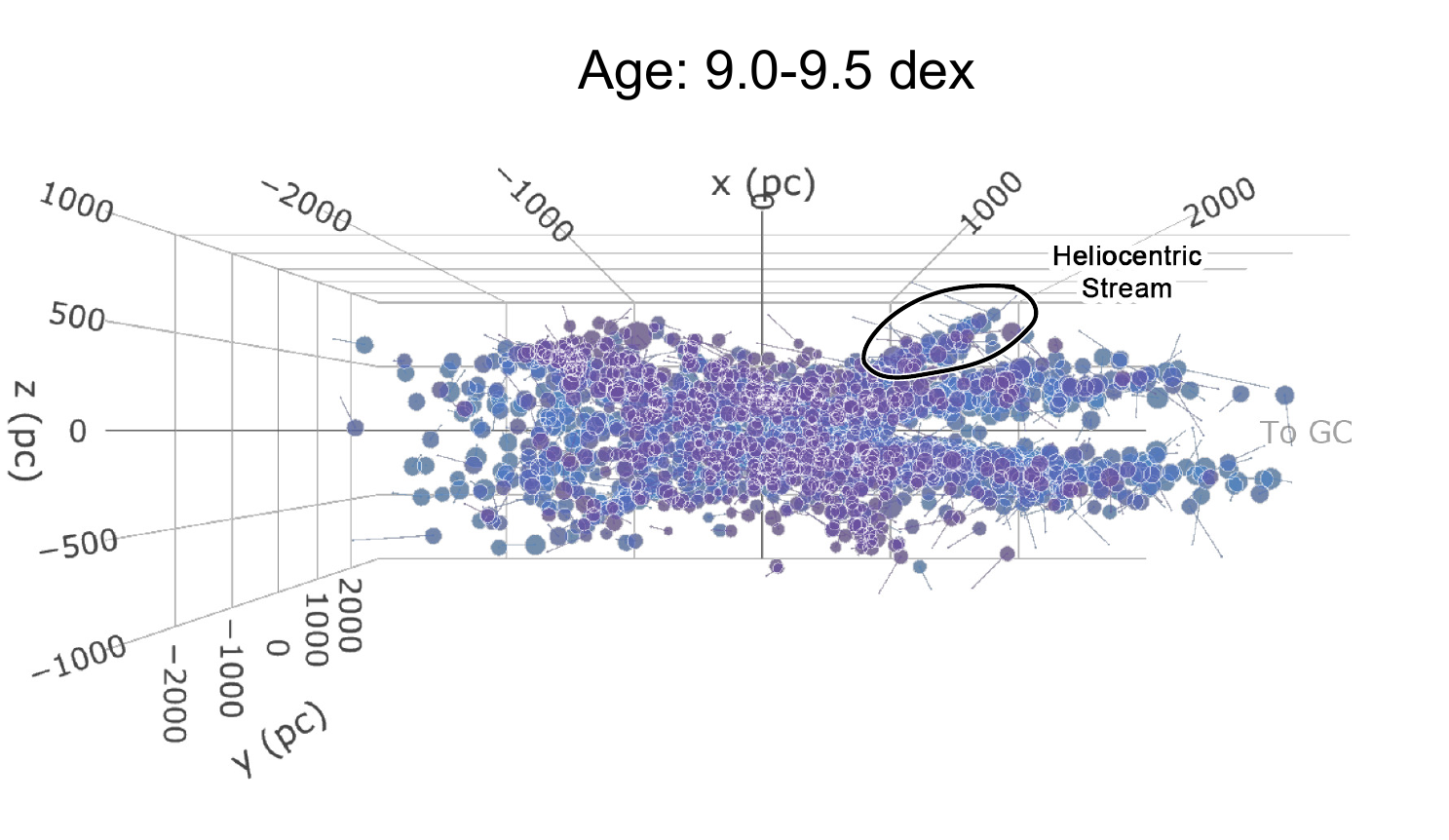}{0.6\textwidth}{}
        }\vspace{-1 cm}

\end{interactive}
\caption{3-dimensional distribution of the identified structures, color coded by their ages. Left panels show the face on view, right - edge on. Top row shows age ranges of 7--7.8 dex, middle - 8.3--8.8 dex, and bottom - 9--9.5 dex. Solid dots are the average position of the stars in the structures, their sizes correlates to a log of the number of sources with $M_G<4$. Thick lines on the top panel show the trace of the strings along their spine. Thinner lines show the typical kinematics of the groups over next 5 Myr, in the local standard of rest, corrected for the circular velocity of 220 \kms. Transparent dots on the face on view show the areas where the clustering is incomplete due to extinction. Black lines in the top panel show the location of the Perseus, Sagittarius, and Scutum arms from \citet{reid2019}, olive line, shows the Radcliffe Wave \citep{alves2020} which is similar to the position of the Local arm. In the bottom two panels, the outlines show the rough locations of the overdensities likely corresponding to the ancient spiral arms (left), as well as the Heliocentric Stream (right). Interactive version with full selection of the age ranges and layer control is available in the online version, temporarily at \url{http://mkounkel.com/mw3d/mw3dv2.html}. Rasterized movie of the interactive plot is available at \url{http://mkounkel.com/mw3d/mw3dv2.mp4}. \label{fig:3d}}
\end{figure*}

The original clustering of \textit{Gaia} DR2 data \citep{gaia-collaboration2018} in  \citetalias{kounkel2019a} was done in several onion-like layers, (processing the entire catalog up to a given distance, and merging it with a catalog that extends even further, joining overlapping structures) to preserve the same sensitivity in the nearby populations as in the more distant ones. The clustering was done in 5d space ($l$, $b$, $\mu_\alpha$, $\mu_\delta$, and $\pi$) in HDBSCAN \citep{hdbscan1,hdbscan}.

Extending the sample beyond 1 kpc, we used the same general approach. All the data quality cuts from \citetalias{kounkel2019a} (Section 2) were retained in this work, as were the cuts in the plane of the sky ($|b|<30^\circ$), and velocity ($|v_{lsr}^{\alpha,\delta}|<60$ \kms).

Three new layers were added - one extending up to $\pi>$0.6 mas, to $\pi>$0.3 mas, and the final one extending up to  $\pi>$0.2 mas. Note that all the sources within the solar neighborhood were preserved between different layers, i.e., $\pi>$0.6 mas slices retains all of the sources in the catalog from $0.6<\pi<1000$ mas, and, similarly, $\pi>$0.3 mas slice retains all the sources  $0.3<\pi<1000$ mas, to ensure overlap of the identified structures. As the data volume became too large due to an increasing number of sources, each one of these layers was further split in chunks of $l$ ranges: 0--65$^\circ$, 60--125$^\circ$ ... 300--5$^\circ$ for the $\pi>$0.6 mas sample, and 0--35$^\circ$, 30--65$^\circ$ ... 330--5$^\circ$ for the $\pi>$0.2 and 0.3 mas samples. These individual chunks were clustered separately using the `leaf' algorithm, with the minimum sample of 25, and a minimum cluster size of 40 stars. The `leaf' algorithm is more optimal for identifying more granular populations, selecting leaf nodes from the minimum spanning tree. On the other hand, the default `excess of mass' algorithm tends to perform poorly on \textit{Gaia} data, merging together most stars in the solar neighborhood in just a single population. The sample size is responsible for how conservative the clustering is in considering overdensities of stars and rejecting the noise, which helps set the characteristic scale of the identified populations. Finally, the cluster size further rejects the overdensities number that have fewer number of sources than the set threshold. See \citetalias{kounkel2019a} and HDBSCAN manual\footnote{\url{https://hdbscan.readthedocs.io/en/latest/parameter_selection.html}} for the more complete description of the process. 

The resulting outputs were stitched along the seams and joined into the merged catalog from \citetalias{kounkel2019a}. In some cases, some pairs of groups identified in \citetalias{kounkel2019a} were joined together if there were newly added stars that illustrated a common origin between them. The properties of the populations scale non-linearly with distance (e.g., more distant groups are smaller on the sky, have smaller and more uncertain parallaxes and proper motions, and there are more of them in that volume of space in comparison to those groups that are more nearby). Thus, the layers extending up to different distances have different `tuning' for characteristic scales of the identified structures, as such, those layers that include more distant stars have a sparser recovery of the stars in the more nearby populations. By in large, however, different slices trace the same underlying structure within the overlapping volume of space, even if the sensitivity to characteristic density may be somewhat different (See \citetalias{kounkel2019a} for full discussion).

In contrast, different slices in $l$ within the same distance limit have mostly negligible differences for the stars in the overlap area, as they are `tuned' to the same characteristic scale, and joining them together is mostly a trivial process. We further note that within each distance limit the performance by HDBSCAN identifies all overdensities in phase space that satisfy minimum sample and minimum cluster size. It is not intrinsically biased for or against populations in a given age range (as age, or anything regarding photometry of the stars is not provided to it) outside of the fact older populations tend to be more dispersed (both spatially and kinematically) and harder to identify. Nor is it intrinsically biased for or against any particular line of sight outside of astrophysically significant limitations such as extinction (Section \ref{sec:complete}). Although some inhomogeneity in sensitivity may occur right along the seams of different slices, overall the identified groups trace mostly the uniform census of the overdensities in the volume of space we examine.

A total of $\sim$82 million stars have been analyzed by HDBSCAN. The final catalog of clustered structures consists of 987,376 stars in 8292 groups, of which 6671 are new from \citetalias{kounkel2019a}. Their catalog is available in Table \ref{tab:data}. Their stellar properties, such as age, average extinction, and average distance are derived using deep learning with the Auriga neural network that is described in detail in the Appendix \ref{sec:auriga}, and presented in Table \ref{tab:theia}.

\begin{deluxetable}{cccc}
\tabletypesize{\scriptsize}
\tablewidth{0pt}
\tablecaption{Clustered sources\label{tab:data}}
\tablehead{
\colhead{Gaia DR2} &\colhead{$\alpha$} & \colhead{$\delta$} & \colhead{Theia}\\
\colhead{ID} &\colhead{(deg)} & \colhead{(deg)} & \colhead{Group ID}
}
\startdata
2172342682494385024 & 320.57220379 & 52.07733956 & 1\\
2170224782576556672  & 315.65202897 & 52.20035062 & 1\\
2168760576695182208  & 315.71659886 & 50.22889438 & 1\\
\enddata
\tablenotetext{}{Only a portion shown here. Full table with is available in an electronic form.}
\end{deluxetable}

Effort has been made to further identify strings in the increased sample. As in \citetalias{kounkel2019a}, strings were identified by searching for large axial elongation of the population in $l$ compared to $b$, $\mu_l$, $\mu_b$. Due to the lower precision in $\pi$ examination of that dimension was done only to ensure continuity. However due to the decreased size in the sky of the strings at larger distances coupled, they become significantly more difficult to disentangle from the more compact and more isolated groups. As such, while a few new extended string-like structures have been flagged in this work, their census, compared to 1 kpc sample, is by no means complete. Furthermore, due to larger uncertainties in distances, beyond 1 kpc conversion to the 3-dimensional Cartesian coordinates may be imprecise. In total, out of $\sim$8,000 identified structures presented in the paper, only $\sim$400 have been classified as strings. Of these, only $\sim$60 strings are located at distances $>$1 kpc. For this reason, as well, as well as degraded confidence in distance, we do not base the analysis presented in this paper on strings. Instead, we focus only on the average position and velocity of all of the identified structures.

The interactive 3d plot with selectable age range is presented in Figure \ref{fig:3d}. The static version shown in the pdf version of the paper shows only the examples of the typical patterns seen in the data, showing ranges of ages similar (albeit not exact) to those discussed at various points in the text.

\subsection{Contamination}\label{sec:contam}

\begin{figure*}
\epsscale{1.1}
\plottwo{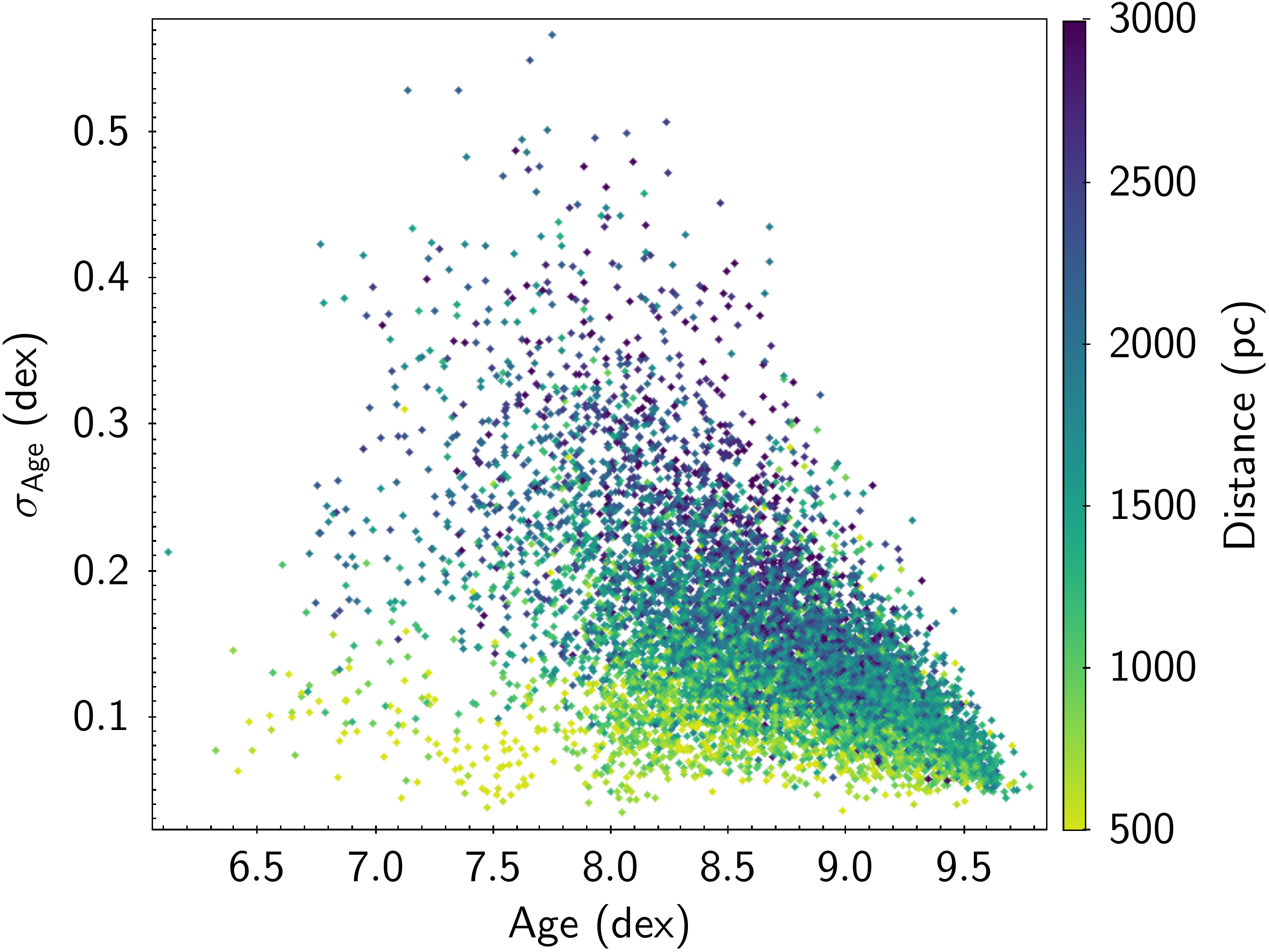}{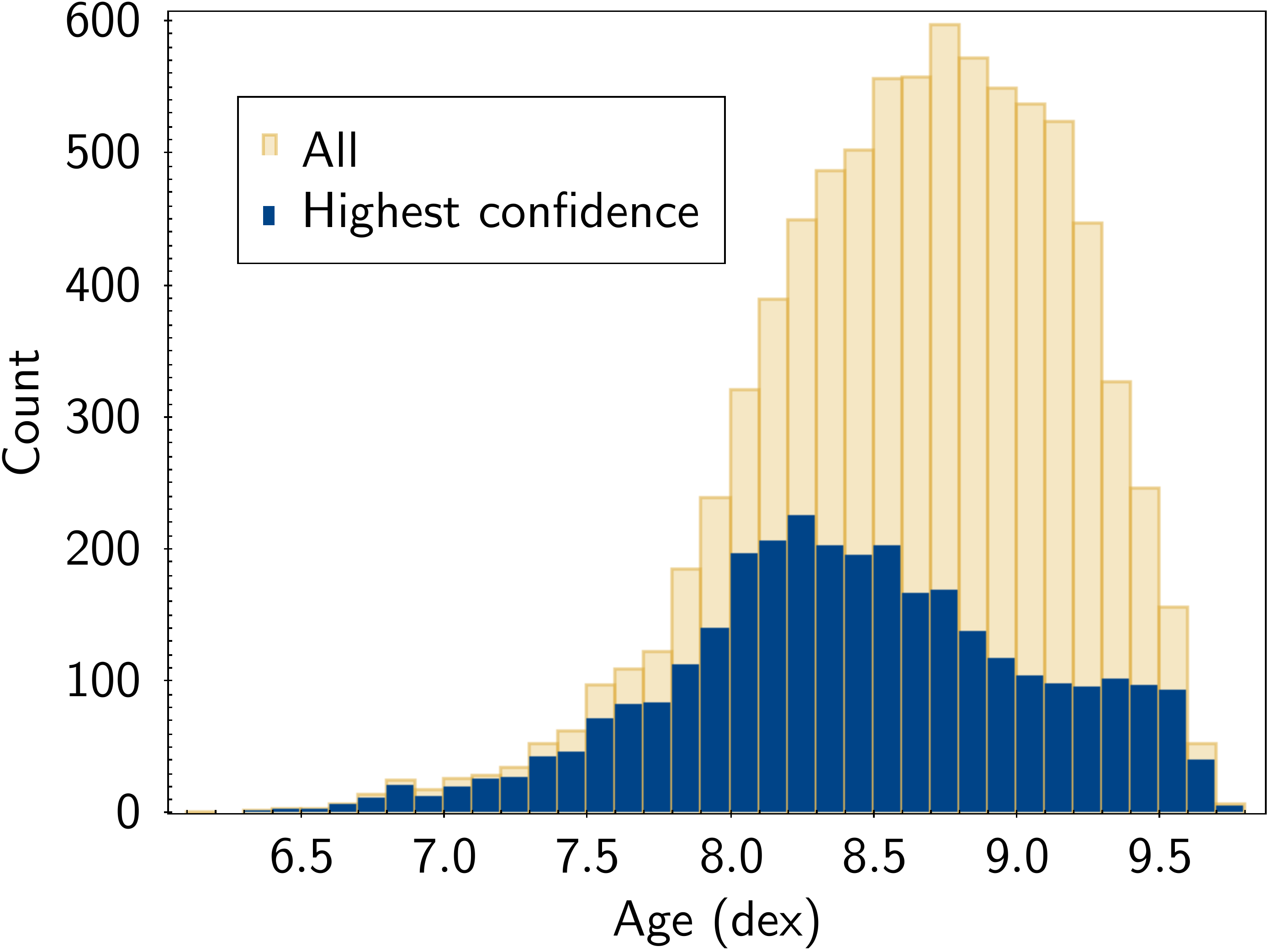}
\plottwo{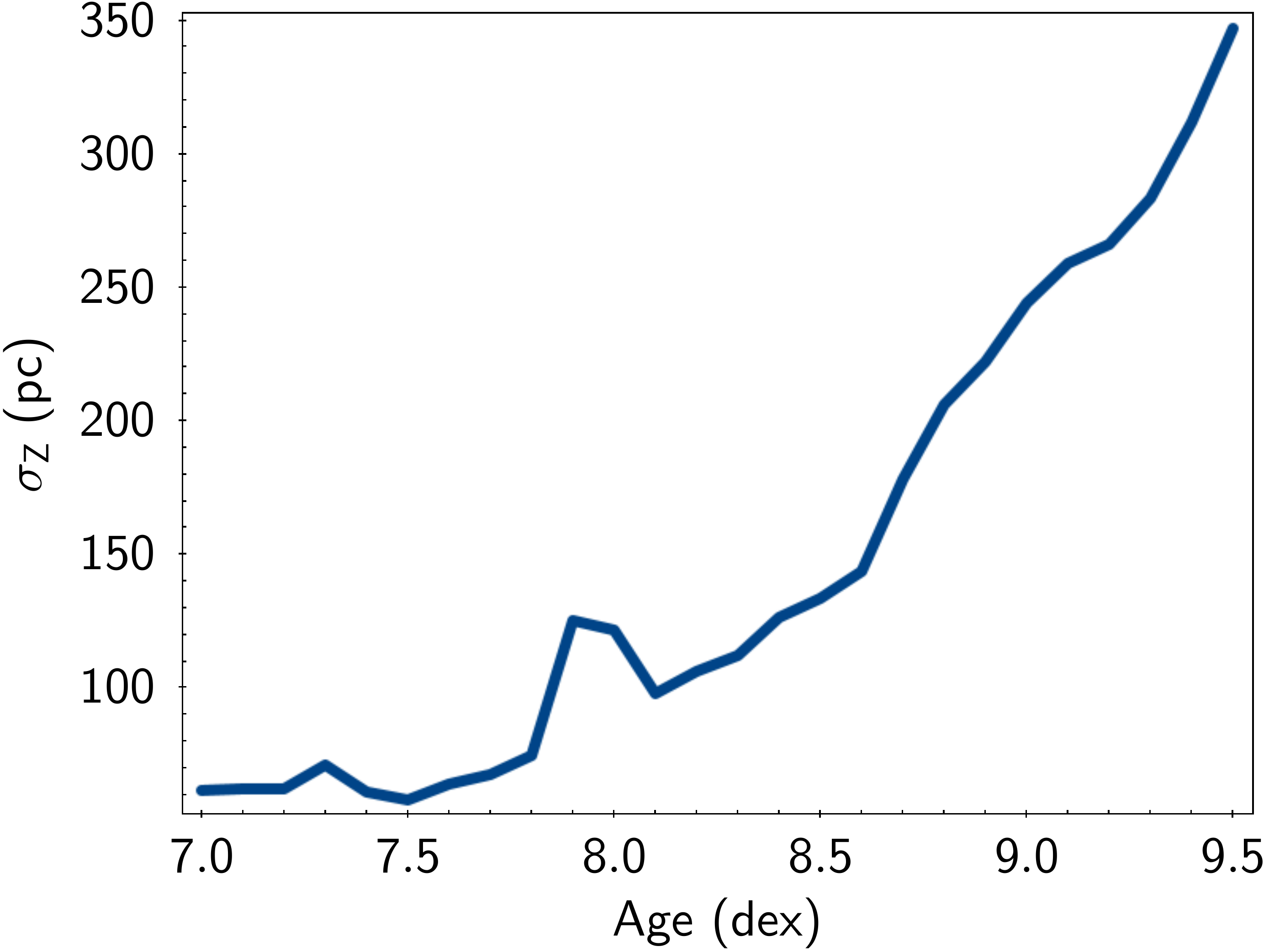}{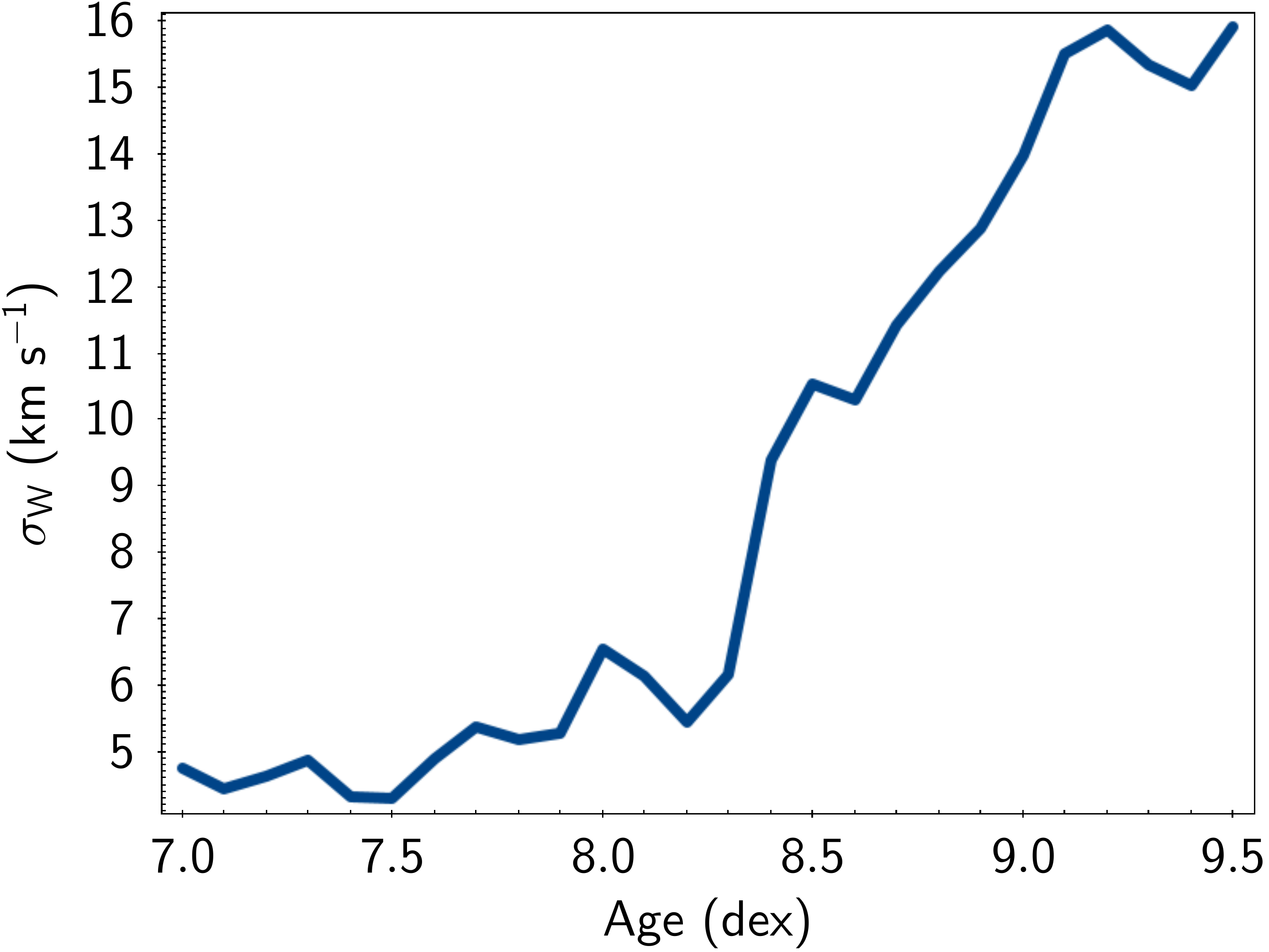}
\caption{Top Left: Ages and corresponding uncertainty of the identified populations as a function of distance. Top Right: the overall distribution of ages for the full and the highest confidence samples. Bottom Left: Dispersion in Z axis for all of the identified populations as a function of age. Bottom Right: Velocity dispersion in Z axis as a function of age. \label{fig:eage}}
\end{figure*}

Similarly as in the \citetalias{kounkel2019a}, some caution needs to be raised about both the contamination from the unrelated field stars that happen to coincide in the phase space with real comoving groups, as well as fake groupings that only appear to be comoving. At larger distances, with more uncertain parallaxes and proper motions, contamination is of even greater concern than in the solar neighborhood. 

For each star in the catalog we assign a randomly drawn field counterparts located at comparable $b$, $\pi$, and that have comparable spatial $A_V$. We then evaluate significance of each group relative to this random catalog based on any of the three of the following categories.

\begin{itemize}
\item First we compare the distribution of fluxes in each of the three \textit{Gaia} bands between the populations, to determine if all three have been drawn from different distribution with $>3\sigma$ significance based on the Kolmogorov-Smirnov test (other two-sample uni- or multivariate tests, such as Anderson Darling, or Cram\'{e}r's V tests produce qualitatively comparable outputs). This accounts for 1297 populations, of which 546 are unique to this category and do not overlap with the other two criteria. This tests whether the identified populations have comparable distribution of fluxes between high mass and low mass stars, such as if the mass function is consistent is consistent with the field or not, and how much scatter there is.
\item We also derive the `age' of the randomly drawn groups with Auriga, to test if the age of the groups in the catalog is different from the non-physical groups on $>3\sigma$ level, considering the reported uncertainties. This accounts for 2149 populations, of which 1522 are unique to this category. This examines the overall shape of the HR diagram in comparison to the field, whether there are prominent and rare pre-main sequence, of if there is a lack of early type stars in the environments where they may be expected to be found with some regularity. 
\item  Furthermore, we consider all groups that have counterparts to previously known populations \citep[e.g.,][]{cantat-gaudin2018a,cantat-gaudin2020} as real, independent of other metrics. This amounts to 783 populations (of which 226 are unique to this category). These groups are highlighted in Table 1 if they have an alternate name that has previously been used in the literature. The closest matching counterparts have been assessed through cross-matching the membership catalogs, identifying the population with the greatest number of matches and verifying that the overall properties of the identified group are consistent with the properties of the cross-matched population. Sometimes, it is impossible to identify a direct one-to-one match, such as, e.g., Orion Complex in this paper corresponds to a single structure, but contains multiple clusters that have been independently catalogued in other works. The inverse lookup of the clusters presented by \citet{cantat-gaudin2018a} to the identified structures is listed in Table \ref{tab:clusters}.
\end{itemize}

In total, 3165 out of 8293 populations belong to the higher confidence sample. Based on this analysis, the groups tend to be most robust if they are younger than 100 Myr (as such pre-main sequence stars are relatively rare in the field, and they occupy a distinct space on the HR diagram), and those groups that are massive, containing $>100$ stars (comparing to the minimum clustering size of 40 stars), although these criteria are not comprehensive. At larger distances, due to a lack of of detectable pre-main sequence stars, ages of younger populations may be more uncertain (Figure \ref{fig:eage}, top left). On the other hand, some of the older populations, due to their complete lack of early type stars, with the turnoff occurring at increasingly lower mass may stand out more clearly if the environment around them has early type stars in larger numbers. Nonetheless, most of the populations with the apparent age of $\sim$8.6 dex are suppressed in the highest confidence sample -- as this is a typical age at which most diffuse populations get dispersed into the field \citepalias{kounkel2019a}, replenishing it with the early type stars of this age, this is often returned as the typical ``age'' of the randomly drawn field groups (Figure \ref{fig:eage}, top right). We note that the precise typical ``age'' generally depends on the galactic latitude, with typically older ``ages'' at higher $|b|$. Thus, for example, a group identified by HDBSCAN with age of $\sim8.6$ dex would be distinct at higher $|b|$, while a group with the age of $>9$ dex would be more distinct at lower $|b|$.

\begin{figure*}
\epsscale{0.95}
\plottwo{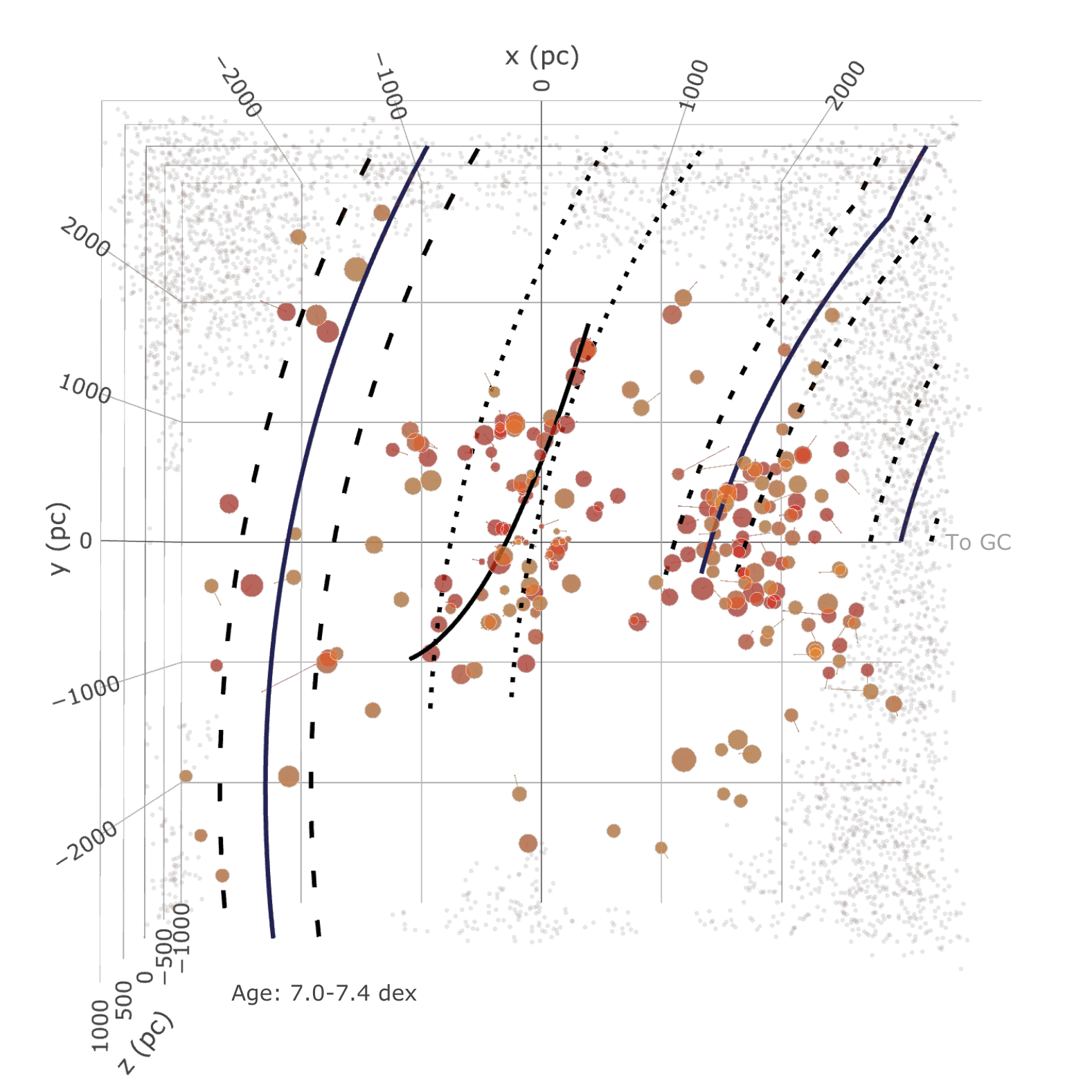}{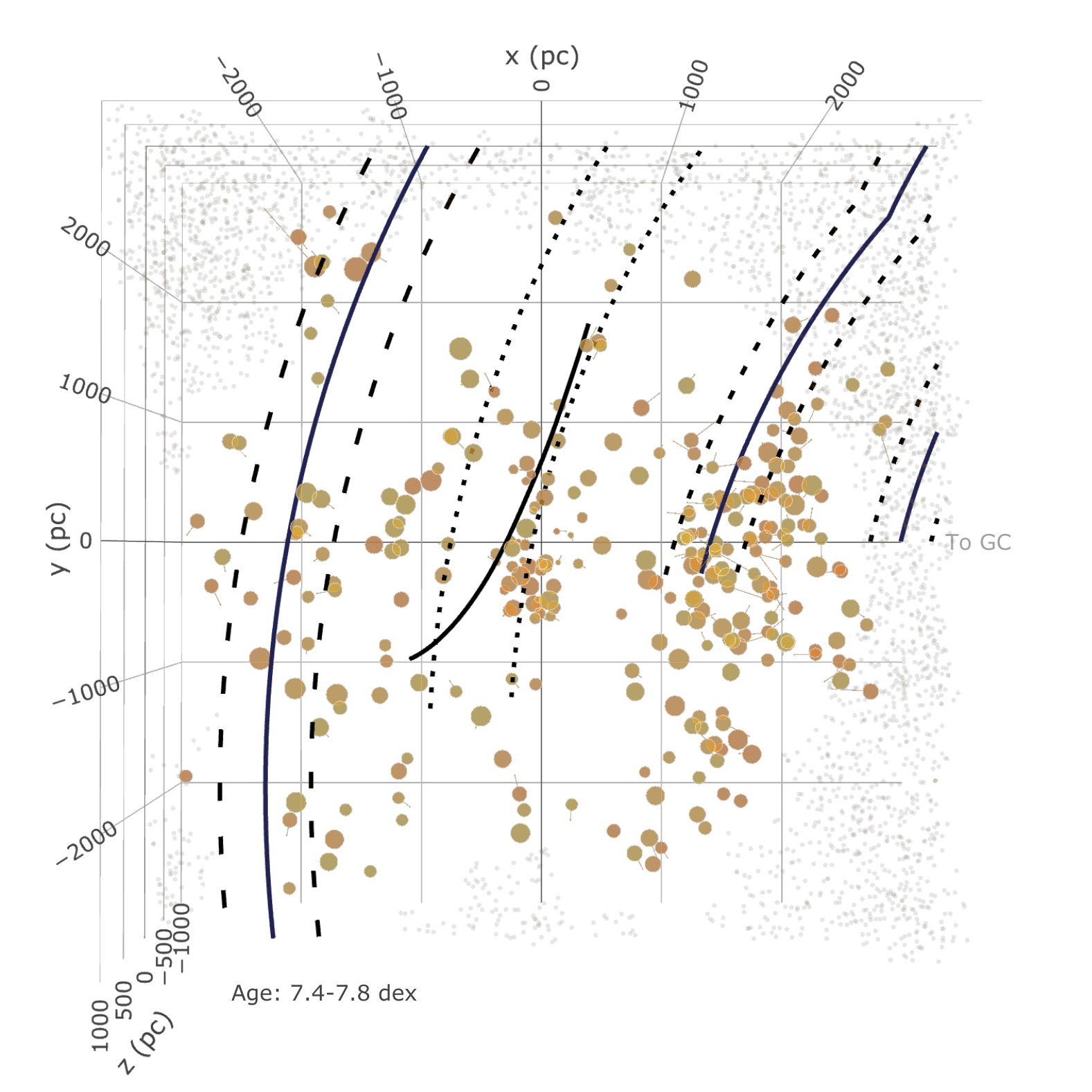}
\plottwo{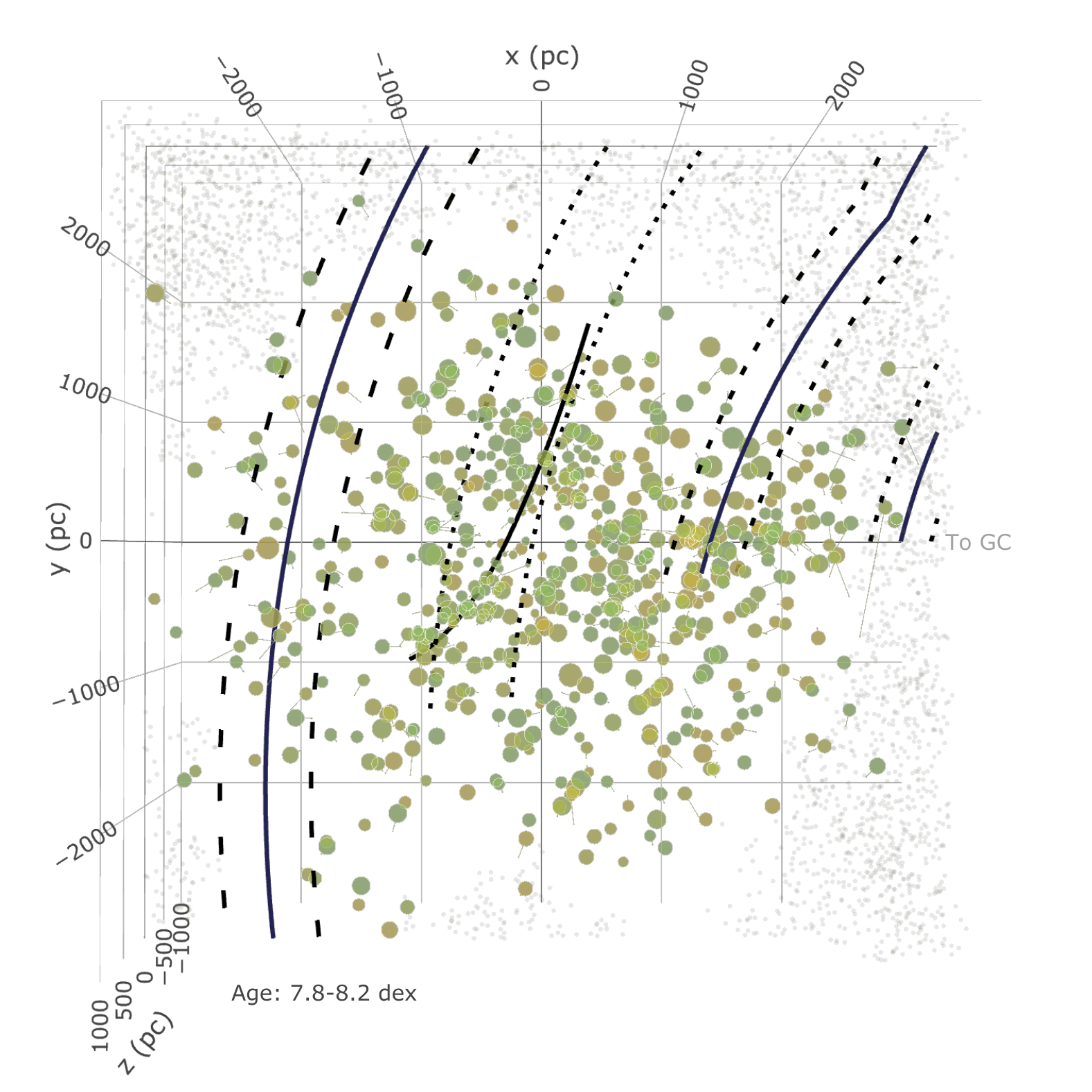}{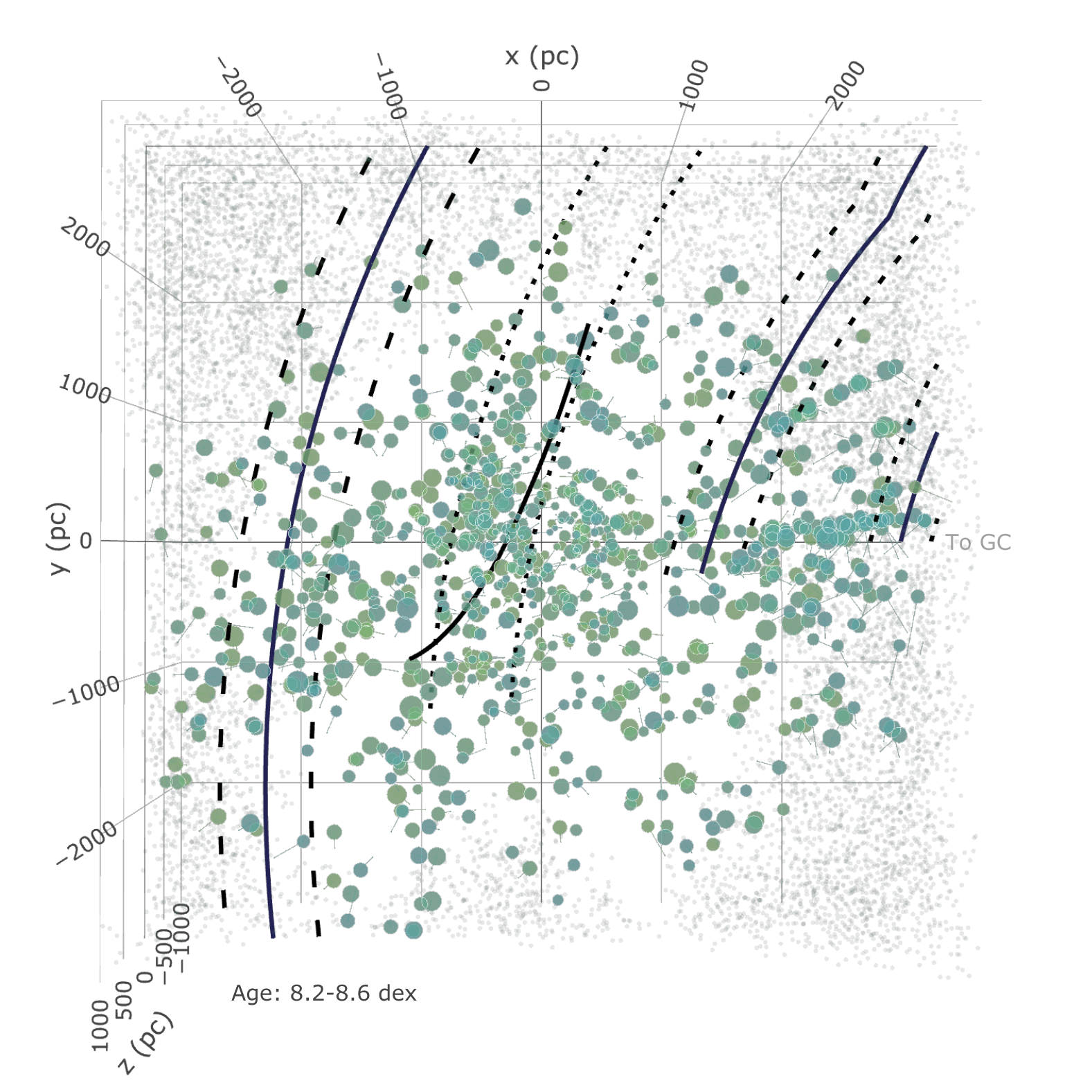}
\plottwo{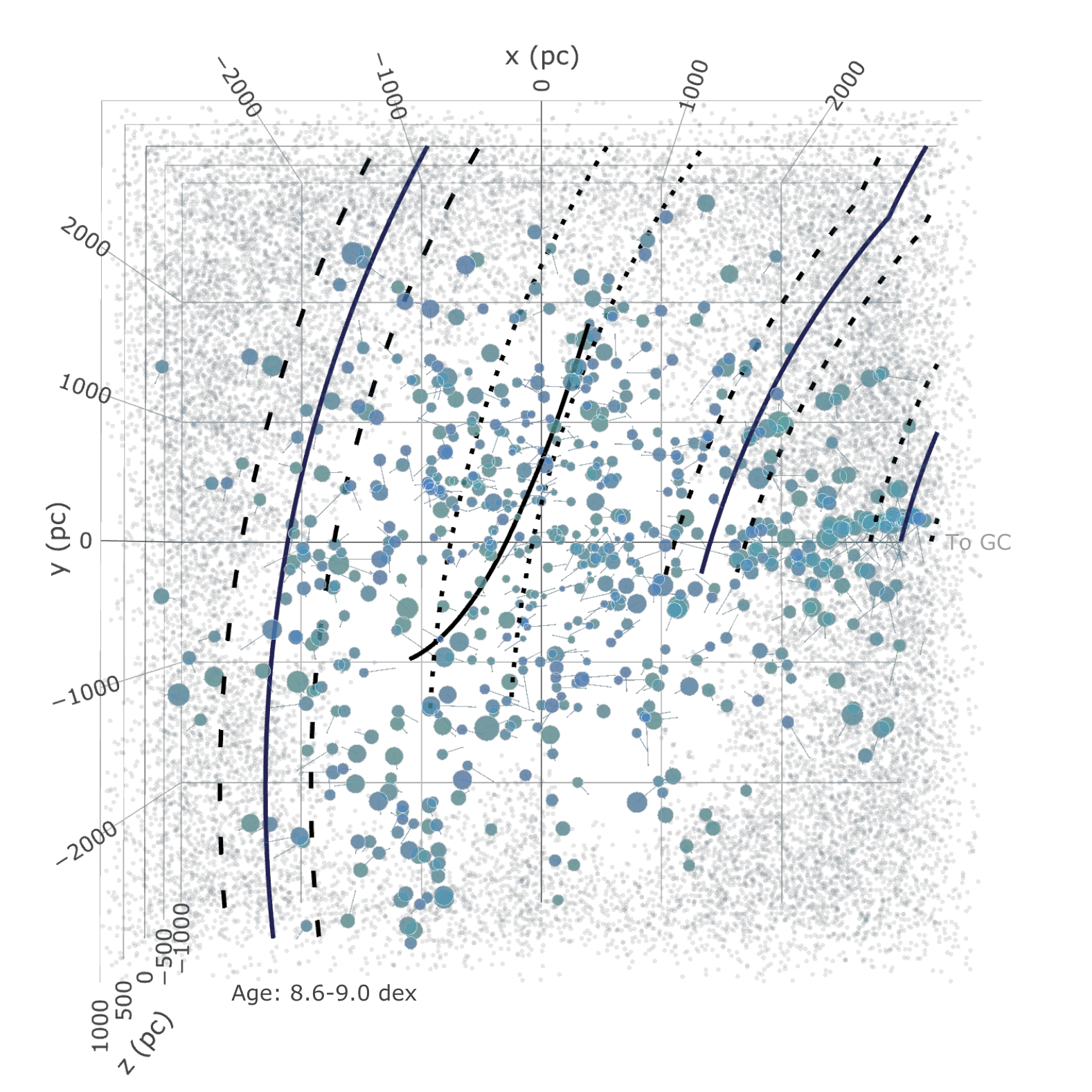}{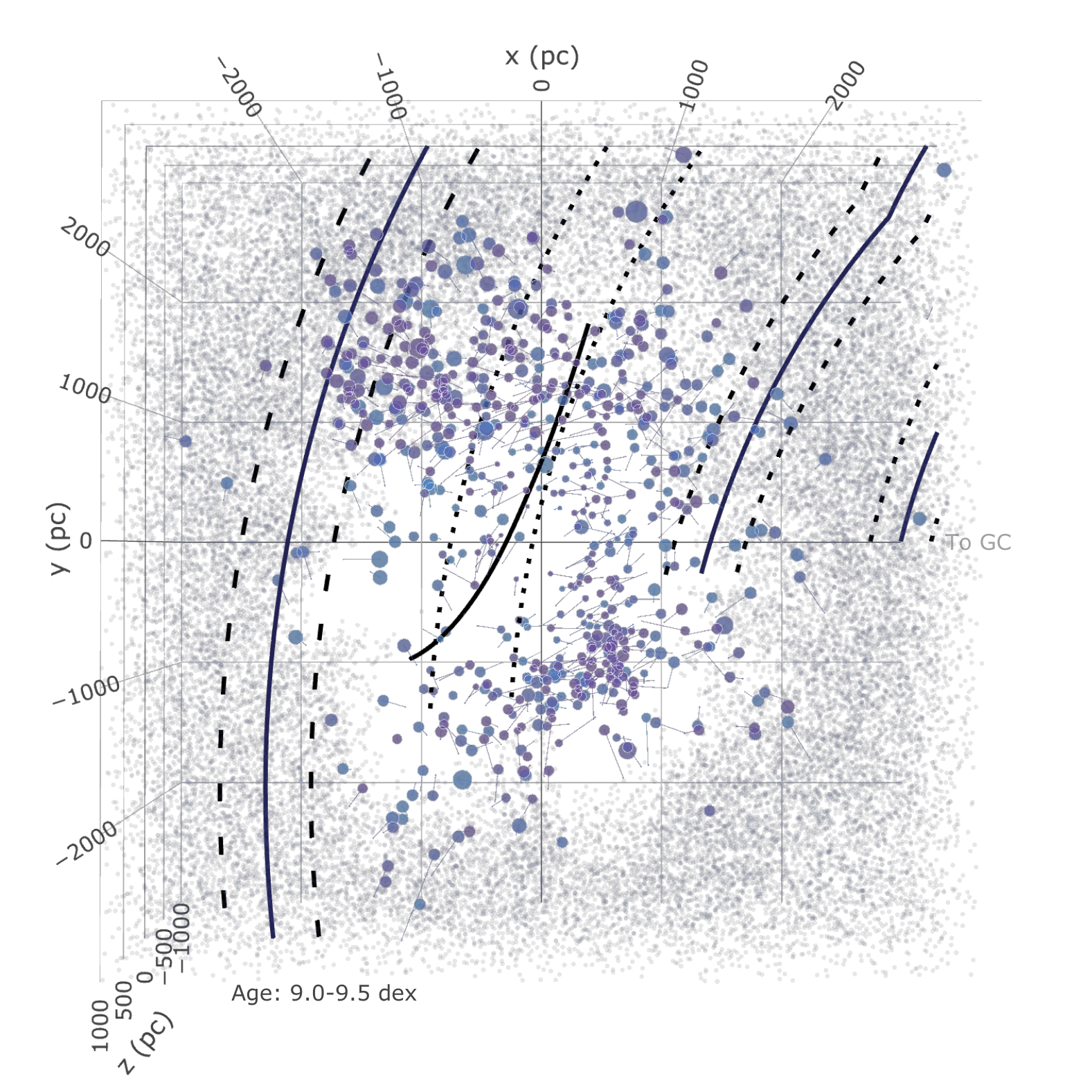}
\caption{Same as Figure \ref{fig:3d}, but restricted only to groups that have HR diagram different from randomly drawn field stars at $>3\sigma$ significance, (i.e, most distinct form the field) or that correspond to previously known populations. \label{fig:3dstat}}
\end{figure*}

We include all of the identified structures in tables and in Figure \ref{fig:3d}, as even fake groupings could be informative in regards to the kinematical structure of the Galaxy in bulk, even though some of the individual membership may not be. Nonetheless, we restrict the analysis in the subsequent sections to this highest confidence sample. Although the groups that satisfy the above criteria correspond to less than half of the total sample, we note that there is no systematic difference in the spatial distribution of the groups as a function of age between these groups vs the full sample, thus the analysis presented in Section \ref{sec:3d} is robust against contamination. (See interactive version of Figure \ref{fig:3d} and static version in Figure \ref{fig:3dstat}.)

If some of the groups in the sample are fake, consisting of physically unrelated stars that happen to be comoving, then the estimated ages from the isochrone would not necessarily be completely accurate for every single star in that population. Nonetheless, as clustering does not introduce any bias with mass, it would still be possible to distinguish between populations that are on average older or younger than one another, resulting in a representative age of the stars that is found in that phase space. Preliminary analysis confirms that the stars with \teff$>4000$ K for which rotational periods can be measured tend to have smooth and continuous evolution in their periods as a function of age estimated by Auriga, as is expected through gyrochronological relations \citep[e.g., compared to the works of][]{angus2019}. \citet{angus2020} use the velocity dispersion in Z as a proxy of age in comparison to the gyrochrones, showing that, as expected, dispersion increases at the older age bins. Although, in clustering the data together we do not pass on any information regarding the stellar age explicitly, HDBSCAN can discriminate in assigning stars to a given group based on their velocity. As is in \citetalias{kounkel2019a}, we also see velocity dispersion and the scale height of the disk increase as a function of age (Figure \ref{fig:eage} bottom). Thus, the agreement between the ages of the groups we measure compared to the trends we see relative to the gyrochrones is not necessarily surprising. We defer the full discussion of gyrochronological analysis to future works in the series.

\begin{figure*}
\epsscale{1.1}
\plottwo{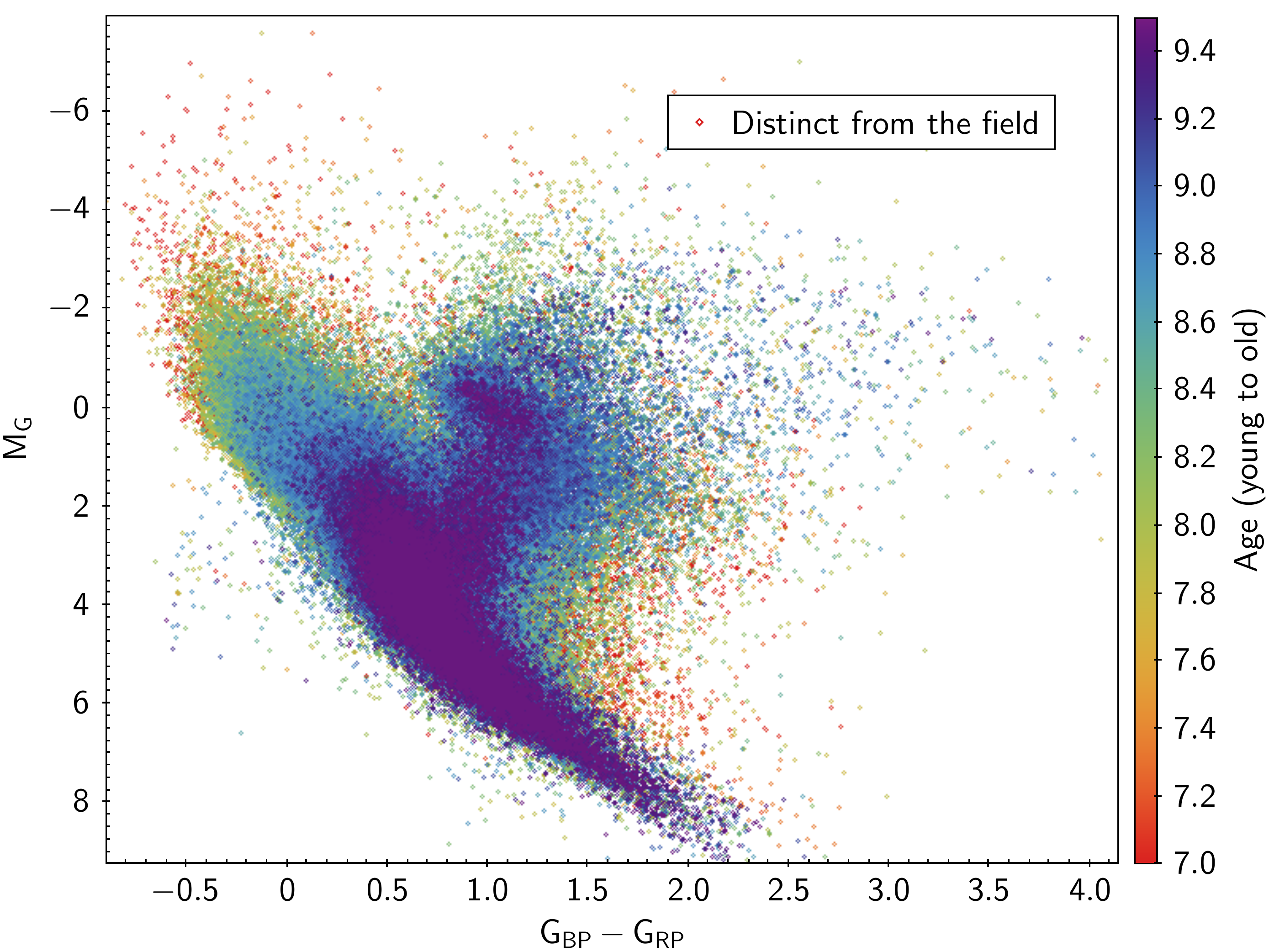}{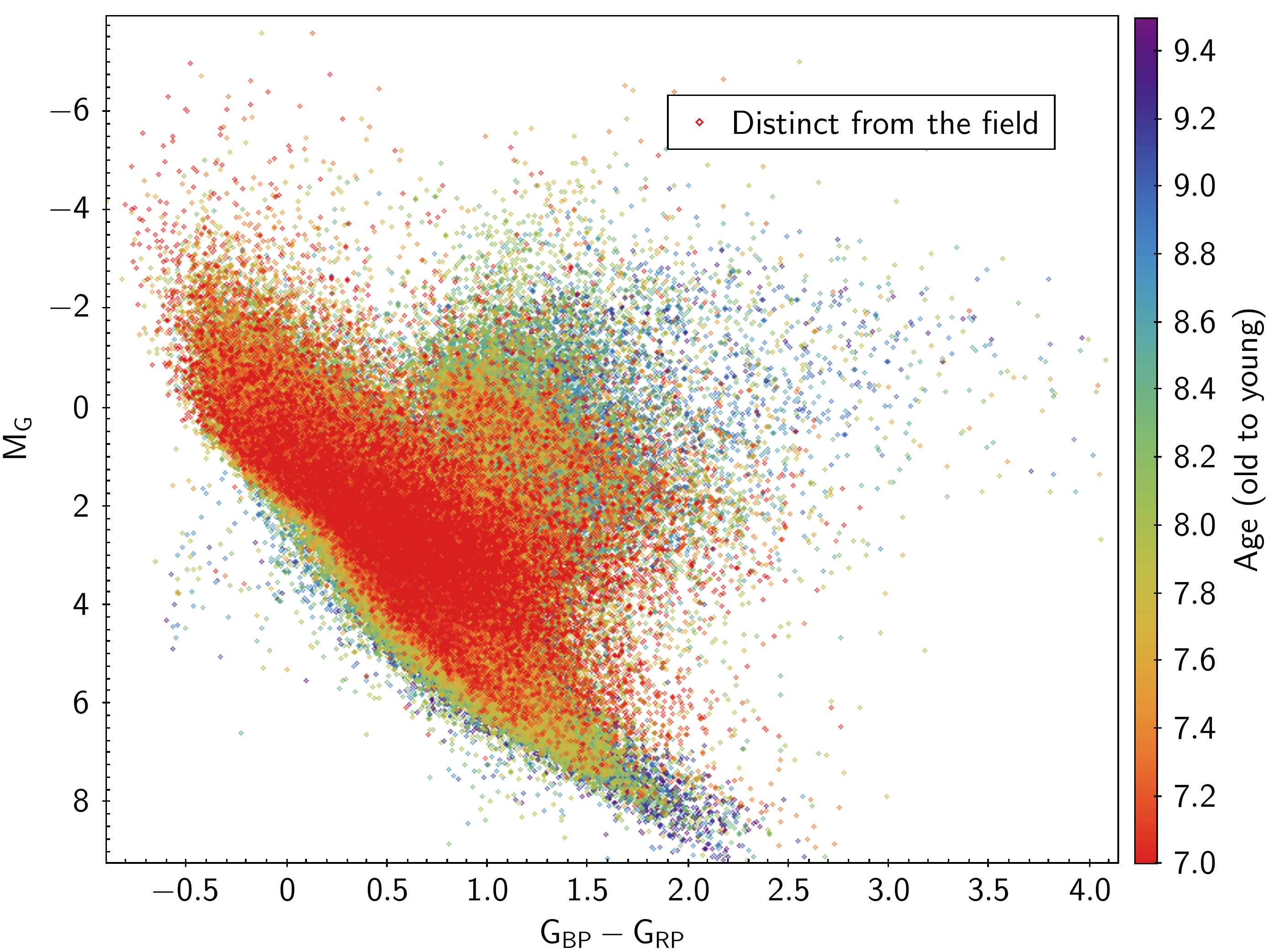}
\plottwo{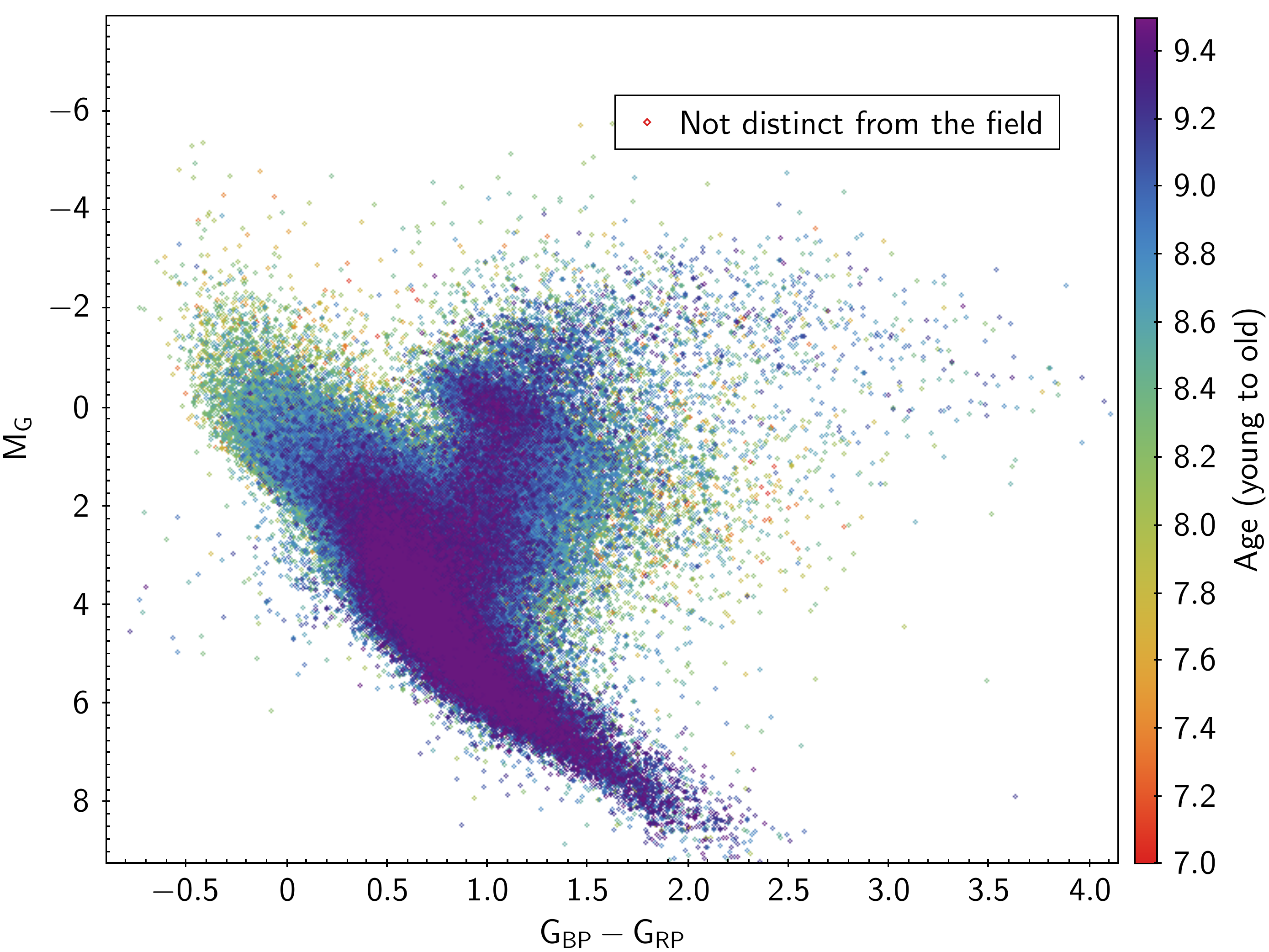}{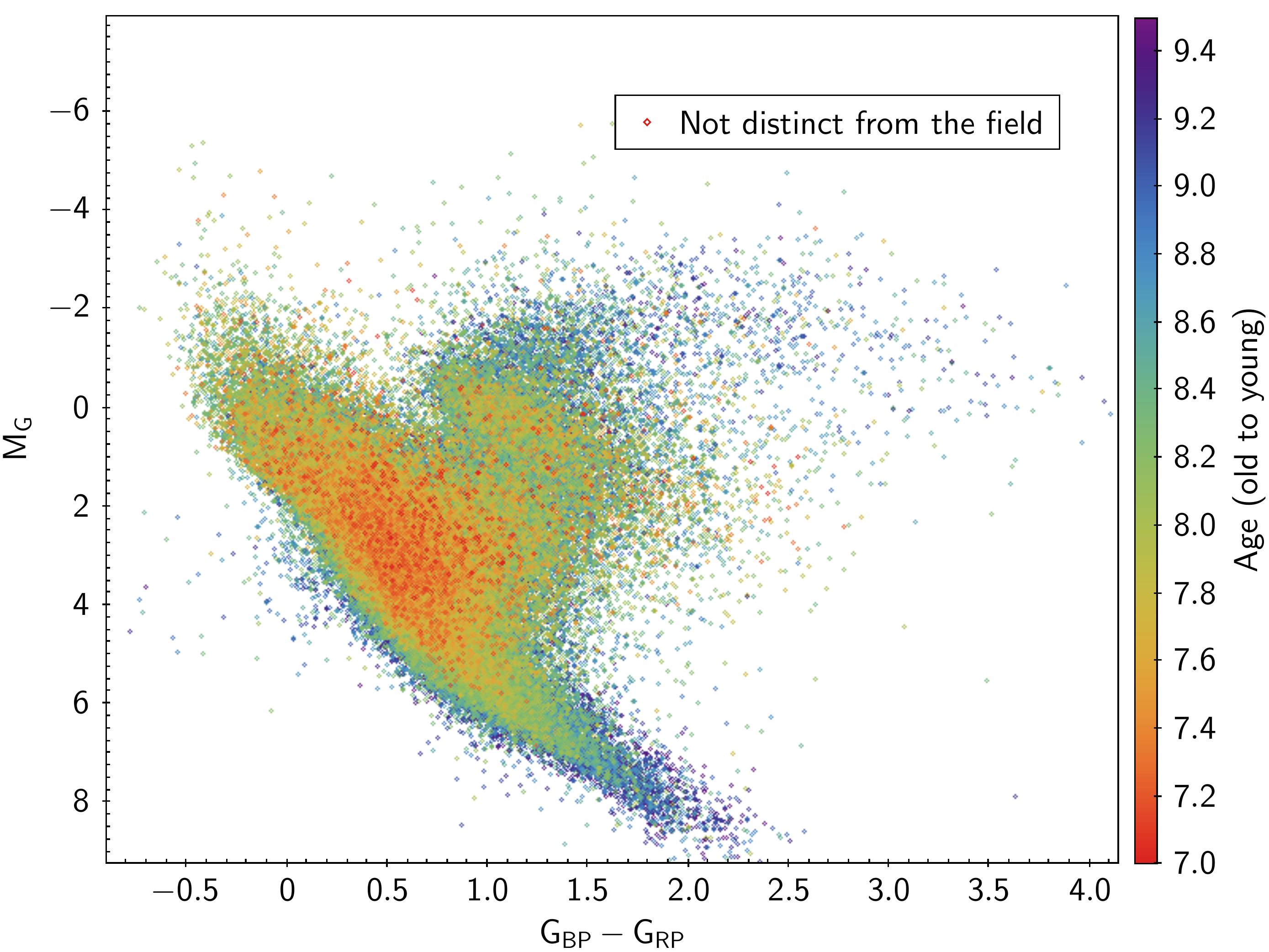}
\caption{Extinction corrected HR diagram of the catalog for populations with distance $>$1000 pc. They are color coded by the age of the structure, plotted from youngest to oldest (left), and oldest to youngest (right), for populations that can be distinguished from the field based on their HR diagram (top), and the sources that cannot (bottom). \label{fig:hr}}
\end{figure*}

The HR diagram of these two types of groups similarly shows a good agreement of the ages compared to the expected distribution (Figure \ref{fig:hr}). The vast majority of the populations do show good agreement with the isochrones. Validation of the performance of the is presented in Appendix \ref{sec:aurigavalidation}. We well recover ages of structures for which previous estimates were available.  Although some groups do have some contamination from the field that are clear outliers (e.g., a presence of an early type star in an otherwise old population, or a few sources on a red giant branch in an otherwise young population). However, that stellar contamination is regularly significant, either in the groups that can be distinguished from the field based on the HR diagram or those that cannot. In estimating the ages, Auriga focuses on the overall distribution of star (ratio of high mass to lower mass stars, ratio of red giants to main sequence stars, and a number of other correlations that are present in the data), thus it is able to disregard such stellar contamination in deriving the likeliest age, similar to the age that could be estimated through visual examination. And, although we cannot exclude the possibility that there are a number of older or younger stars among the lower mass stars hidden in the HR diagram, the estimate should be appropriate for the bulk of the stars in the group.

The identified groups would be found trace the most typical position-velocity components of the solar neighborhood, which allows for a comprehensive look at the Galactic structure. Of great interest in particular are the systematic gaps found in the spatial distribution of the populations as a function of age. Whether the populations consist of bona fide co-moving groups or fake groupings, presence of such gaps implies that a particular region of the phase space is unlikely to be inhabited by the stars of a given age range, even when allowed to pick and chose stars at random.

As the sample of sources with measured radial velocity increases, it should be possible to not only discriminate real groups more robustly, but to also clean up the membership of the real groupings from the contamination.

\subsection{Completeness}\label{sec:complete}

Selection based on the data quality of \textit{Gaia} data translate to a rough cut of $G<18$ mag. As older populations have a turn-off point that occurs at increasingly fainter magnitudes, it becomes more difficult to detect them at larger distances, both due to a larger distance modulus, and due to build up of extinction along the line of sight. As can be seen in Figure \ref{fig:3d}, there is a gap along the Galactic Plane. The interactive version of this figure demonstrates that the gap appears at ages $>$8 dex, and becomes increasingly more pronounced at the older ages. Extinction is largely responsible for this gap, as other old populations are found at similar distances at higher galactic latitudes.

Several studies have been conducted in the past to analyze the 3d distribution of the dust in the solar neighborhood based on \textit{Gaia} DR2, such as \citet{chen2019a} and \citet{green2019}, however, both of them have some limitations of the line of sights, the former completed the analysis within $|b|<10^\circ$, and the latter, due to the reliance on the Pan-STARRS data, to $\delta>-47.5^\circ$.

The \textit{Gaia} collaboration has released early estimates of extinction for a number of stars \citep{andrae2018}. Although individual measurements may be uncertain, it is possible to average them in bulk, as (outside of the young embedded objects), extinction depends only on three parameters: $l$, $b$, and $\pi$. We used a convolutional neural network with an architecture similar to Auriga that was trained on \textit{Gaia} DR2 $A_G$ values, with the training sample of top 3 million stars ordered by random\_index with $\pi>0.1$ mas. The advantage of a neural network in this case is that it is not necessary to either assume a functional form for the extinction along the line of sight nor force the computation to be based on a particular grid pattern. Using a transformation of 1$A_G$=0.85926$A_V$ \citep[as listed on the PARSEC isochrones web interface][]{marigo2017}, we predict $A_V$ at the given spatial coordinates. These $A_V$s are consistent to within 0.3 mag of the map of \citet{green2019} over the applicable volume.

\begin{figure}
\epsscale{1.1}
 \centering
\plotone{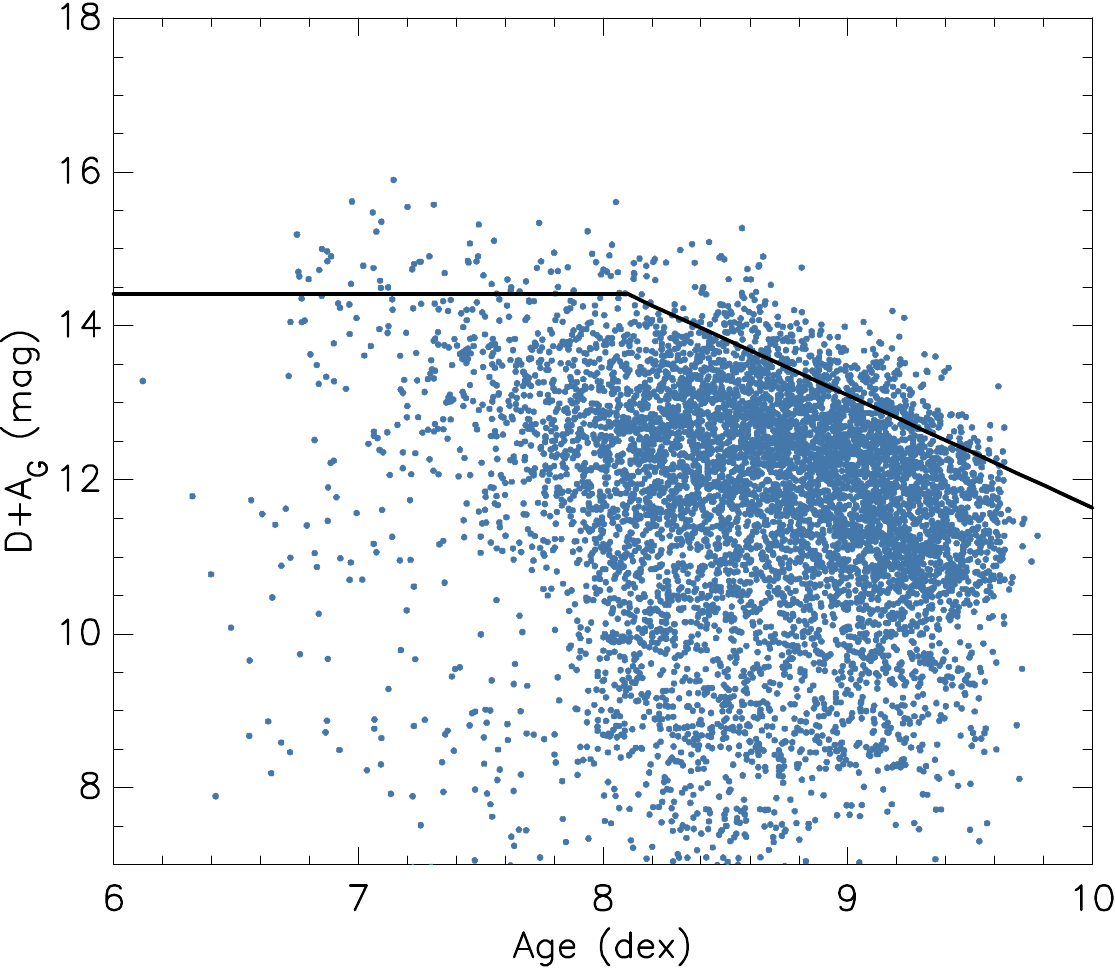}
\caption{Combination of extinction correction and the distance modulus of the detected structures as a function of age, with the threshold estimated by the Equation \ref{eqn:thr}.
\label{fig:threshold}}
\end{figure}

Through examining the absolute magnitude of 40th faintest star in each group as a function of age, we roughly parametrize the detection threshold of \begin{equation}\label{eqn:thr}D+A_G<26.2549-1.46189\times t \end{equation} where $D$ is the distance modulus, and $t$ is the age from 8.1 to 10 dex. As evolution of massive stars has only small effect in the shape of the HR diagram for populations younger than that, we adopt the same threshold for groups $<8.1$ dex as at 8.1 dex (Figure \ref{fig:threshold}). This threshold is only an approximation, and there are a few groups that can still be found at beyond it. Nonetheless, it can serve as an appropriate bound in Figure \ref{fig:3d} to highlight the regions in the 3d volume where the sample is expected to be largely complete at a given age range.

\begin{figure}
\epsscale{1.1}
 \centering
\plotone{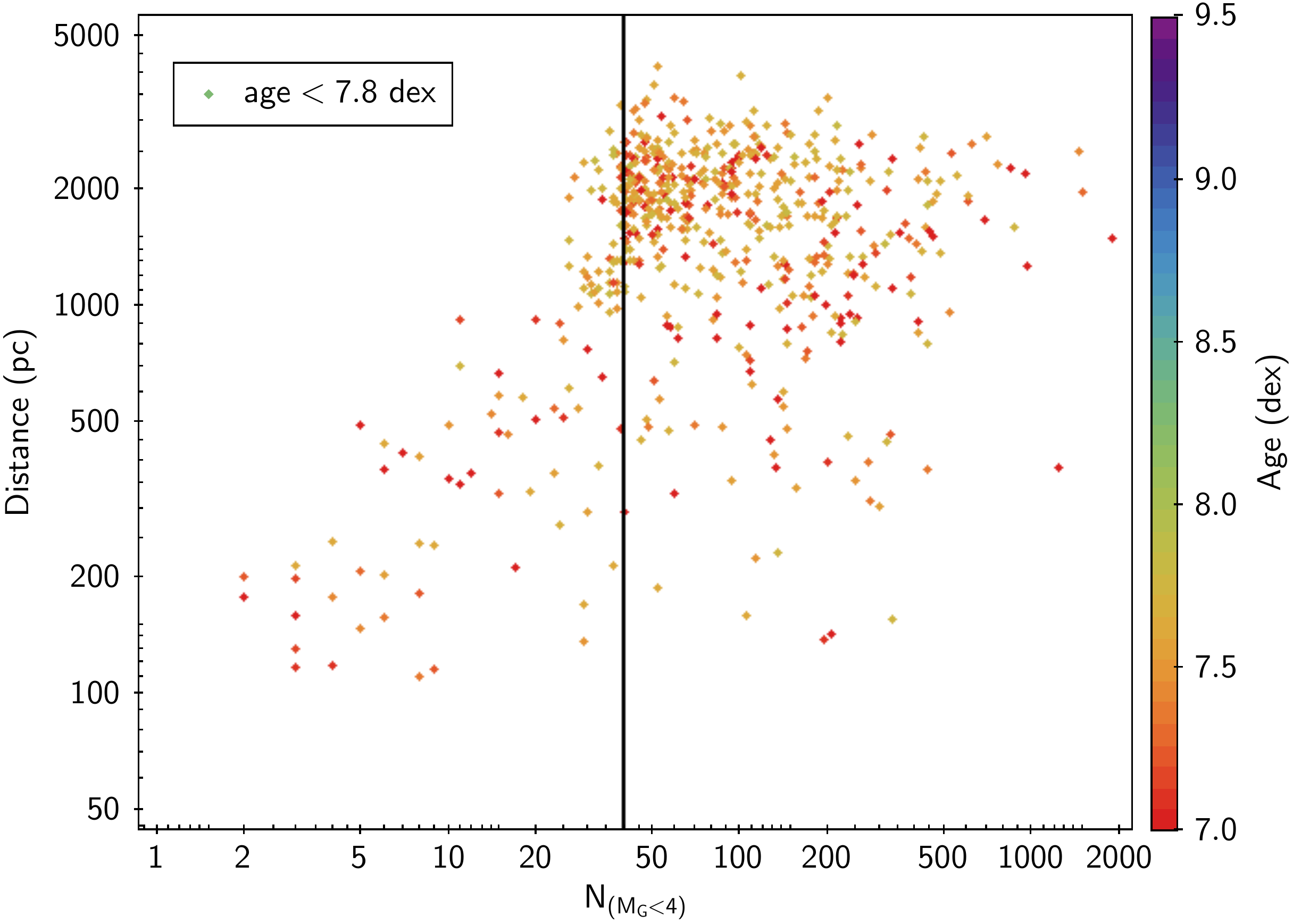}
\plotone{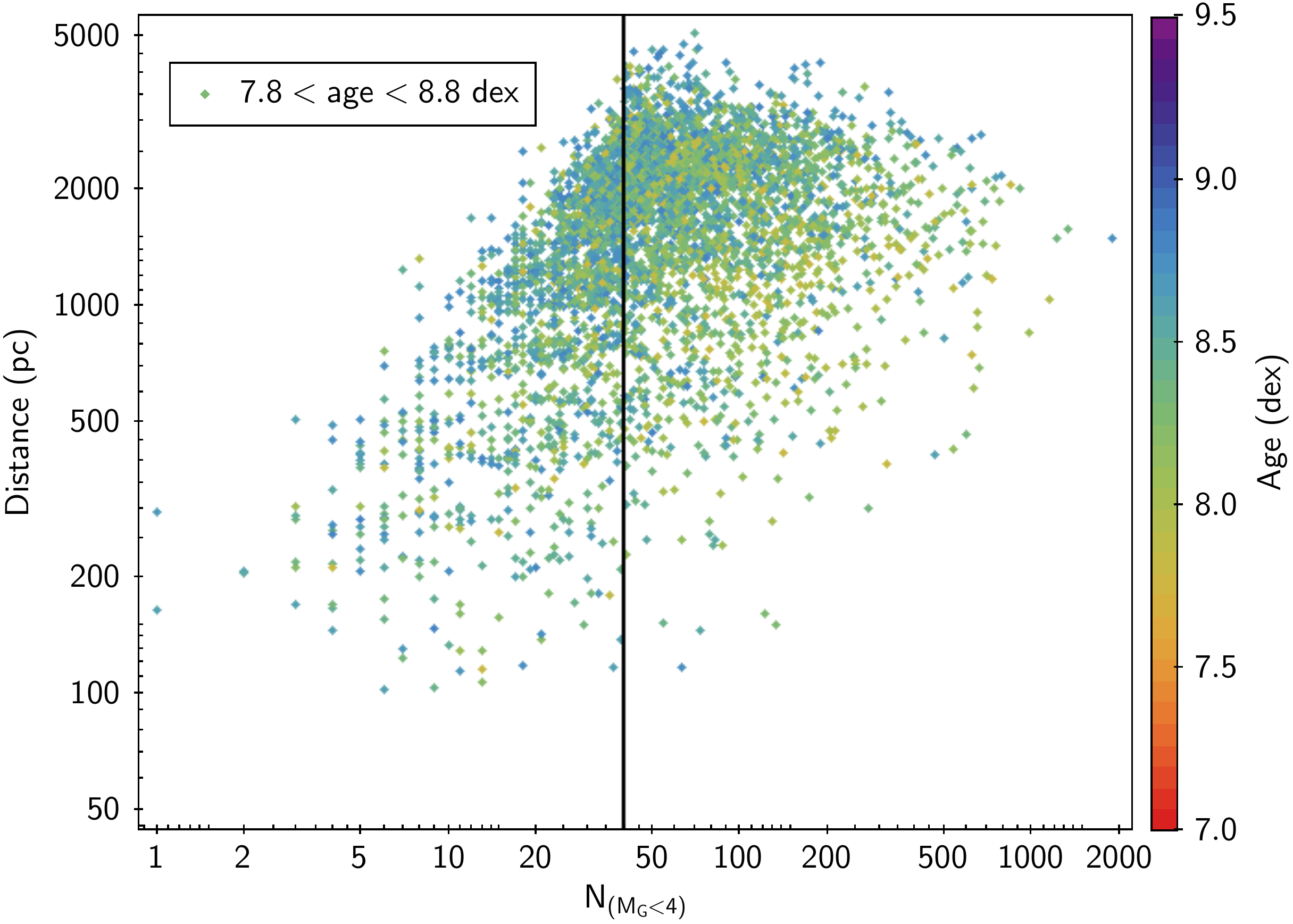}
\plotone{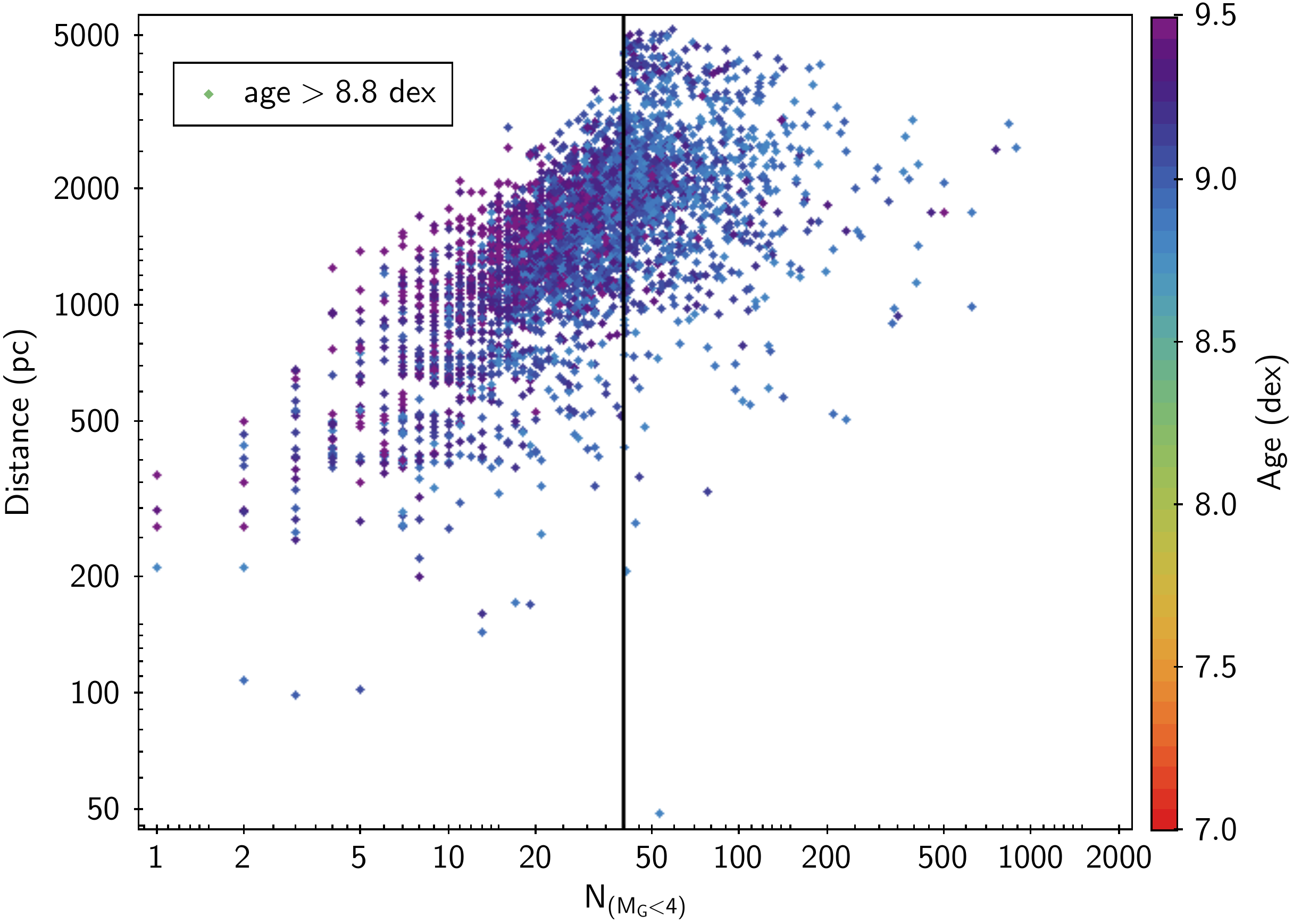}
\caption{Correlation in distance of the population versus the number of stars with $M_G<4$, as a function of age.
\label{fig:bright}}
\end{figure}

\begin{figure*}
\epsscale{0.95}
\plottwo{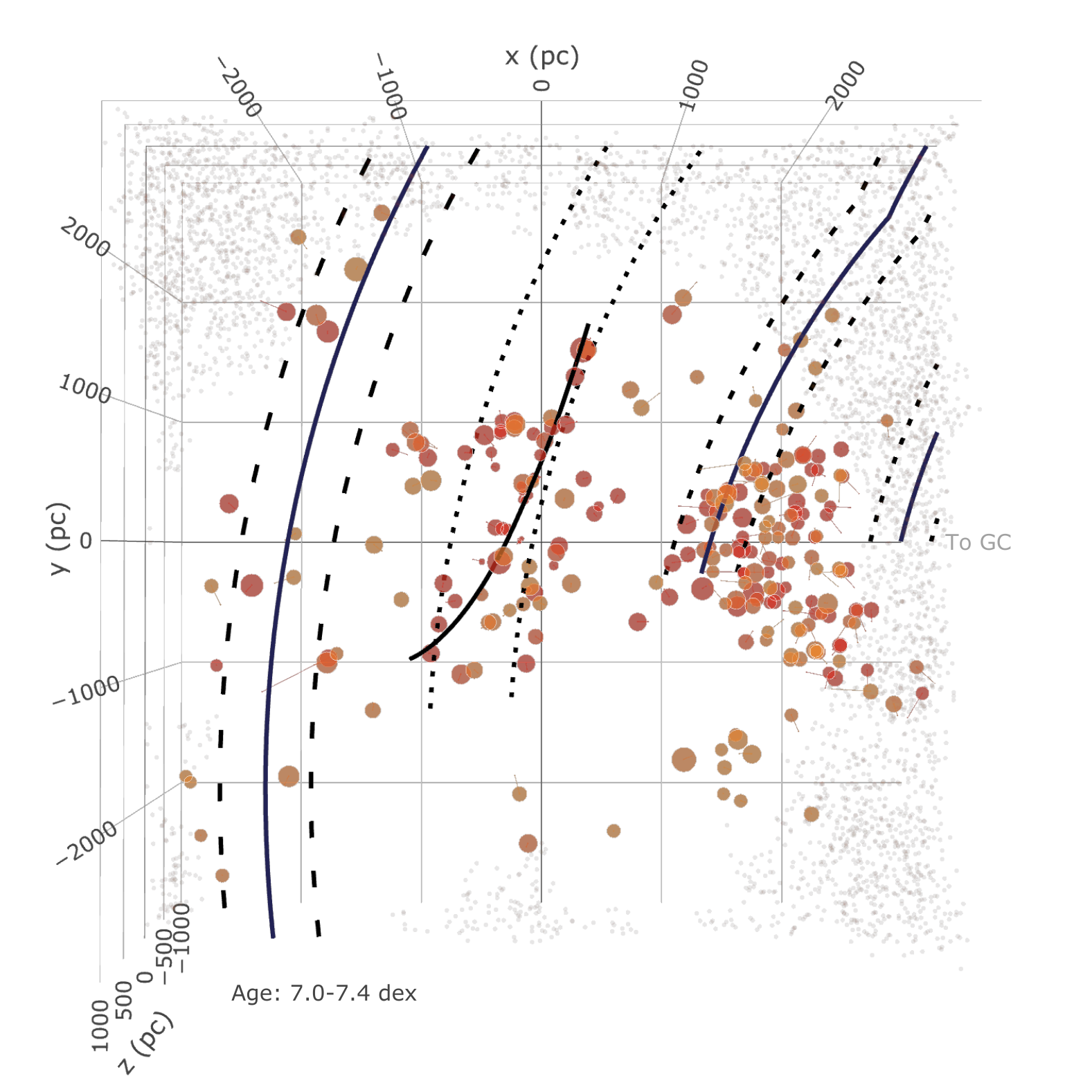}{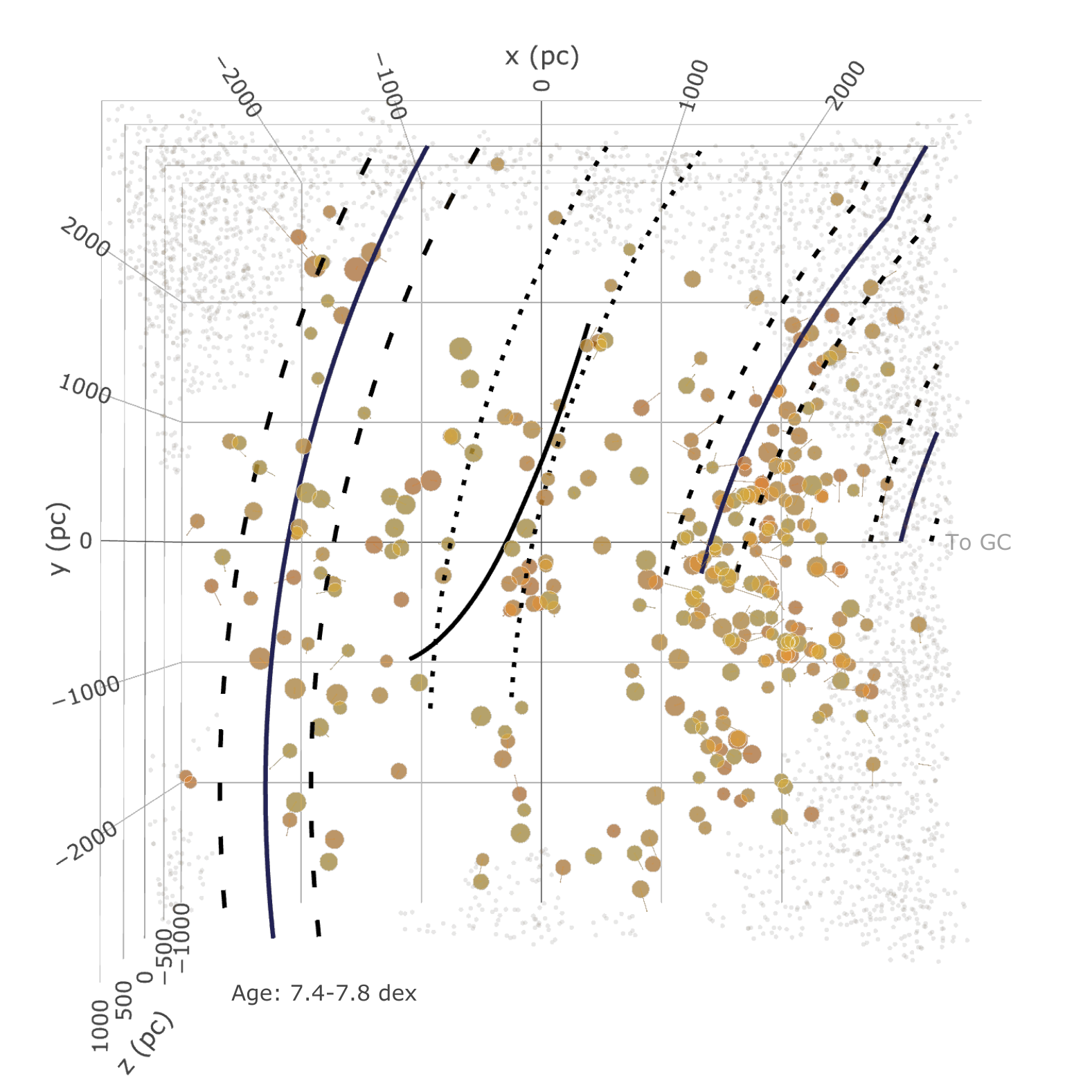}
\plottwo{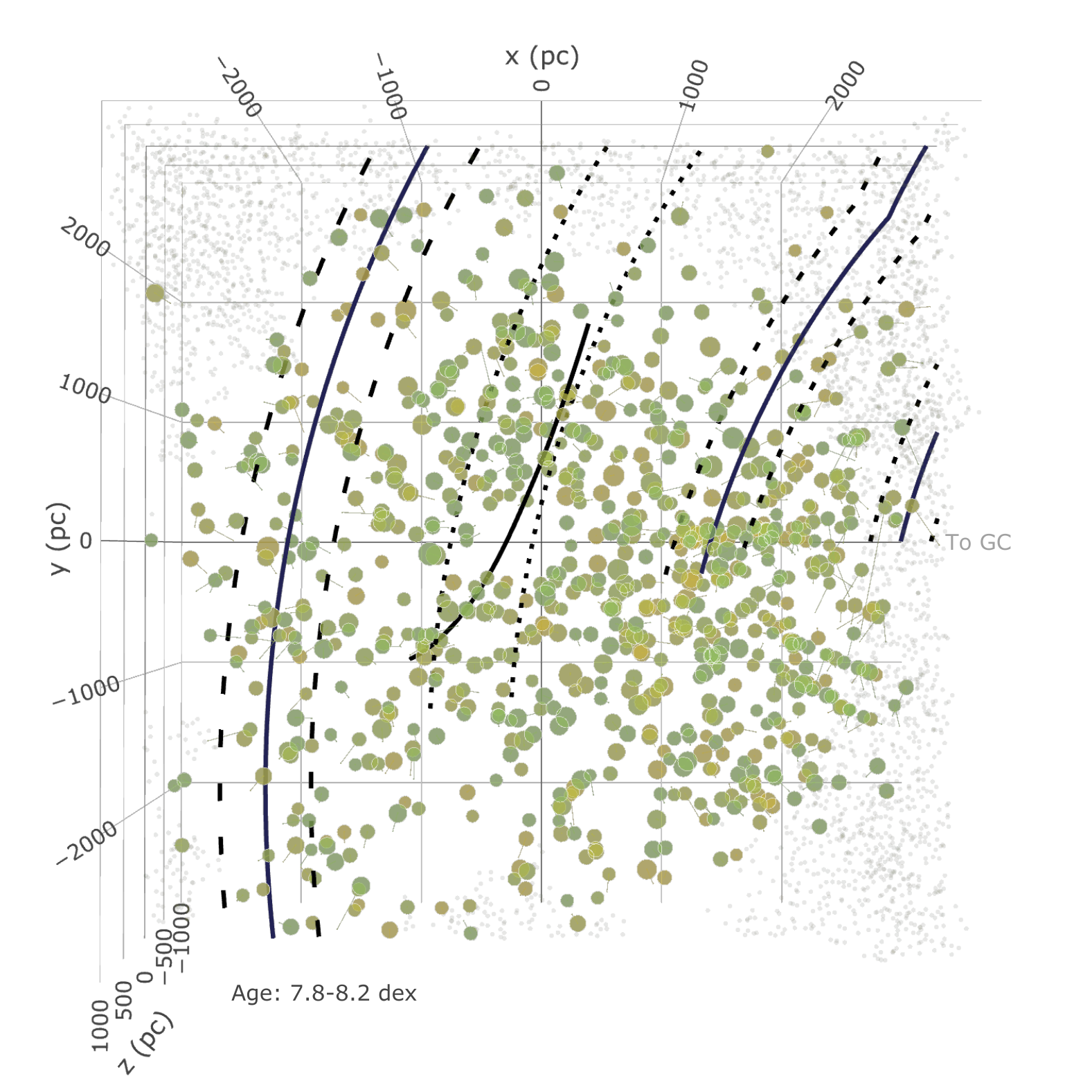}{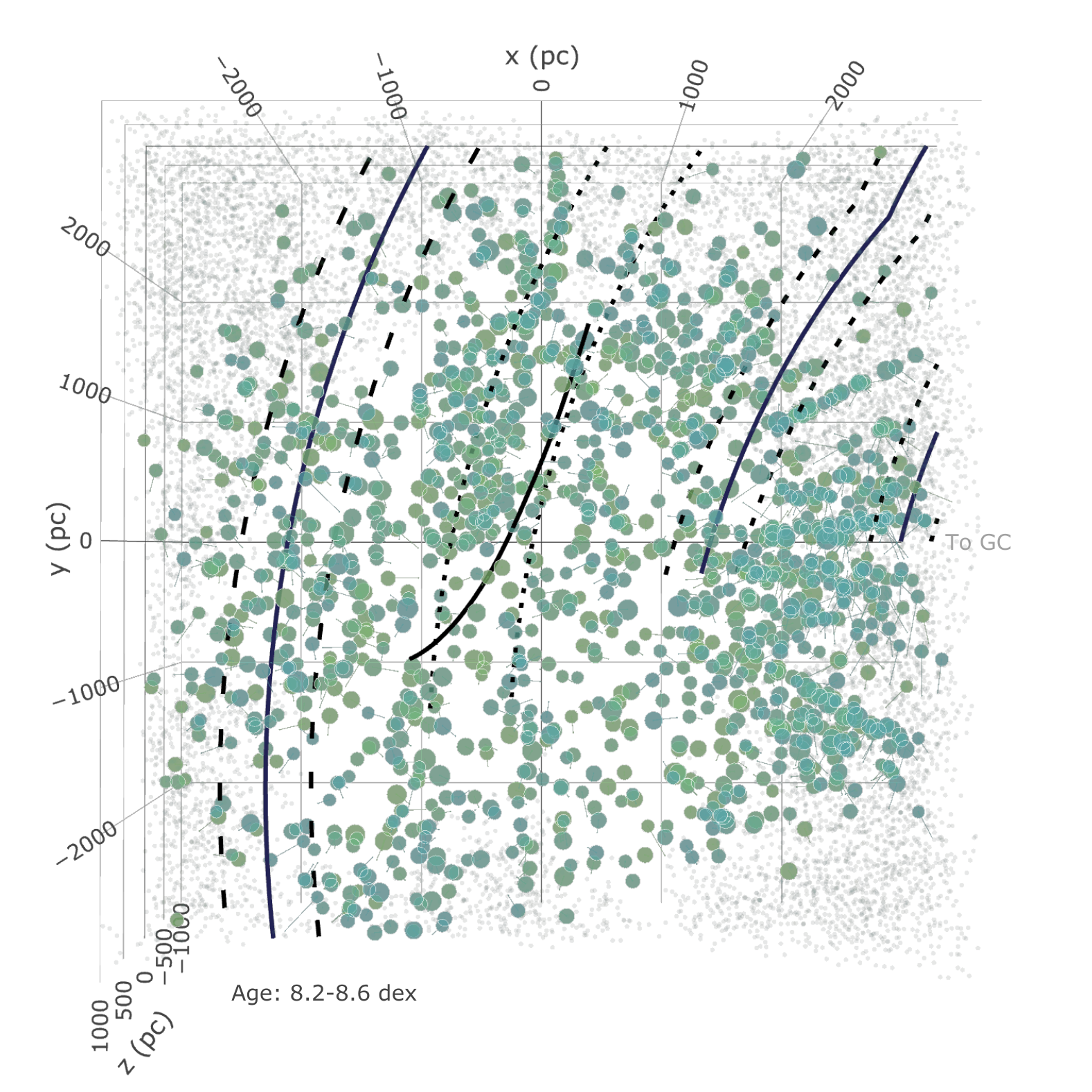}
\plottwo{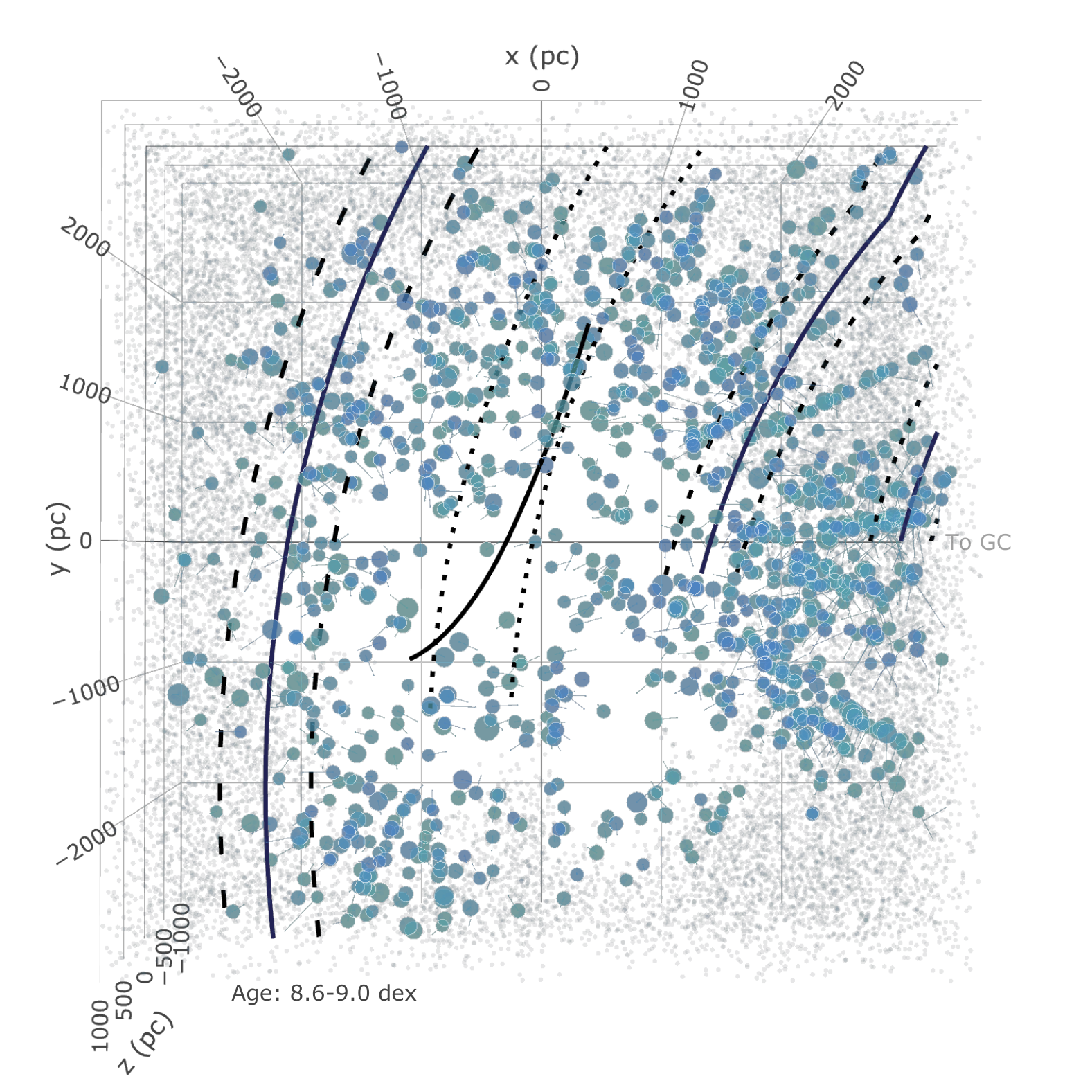}{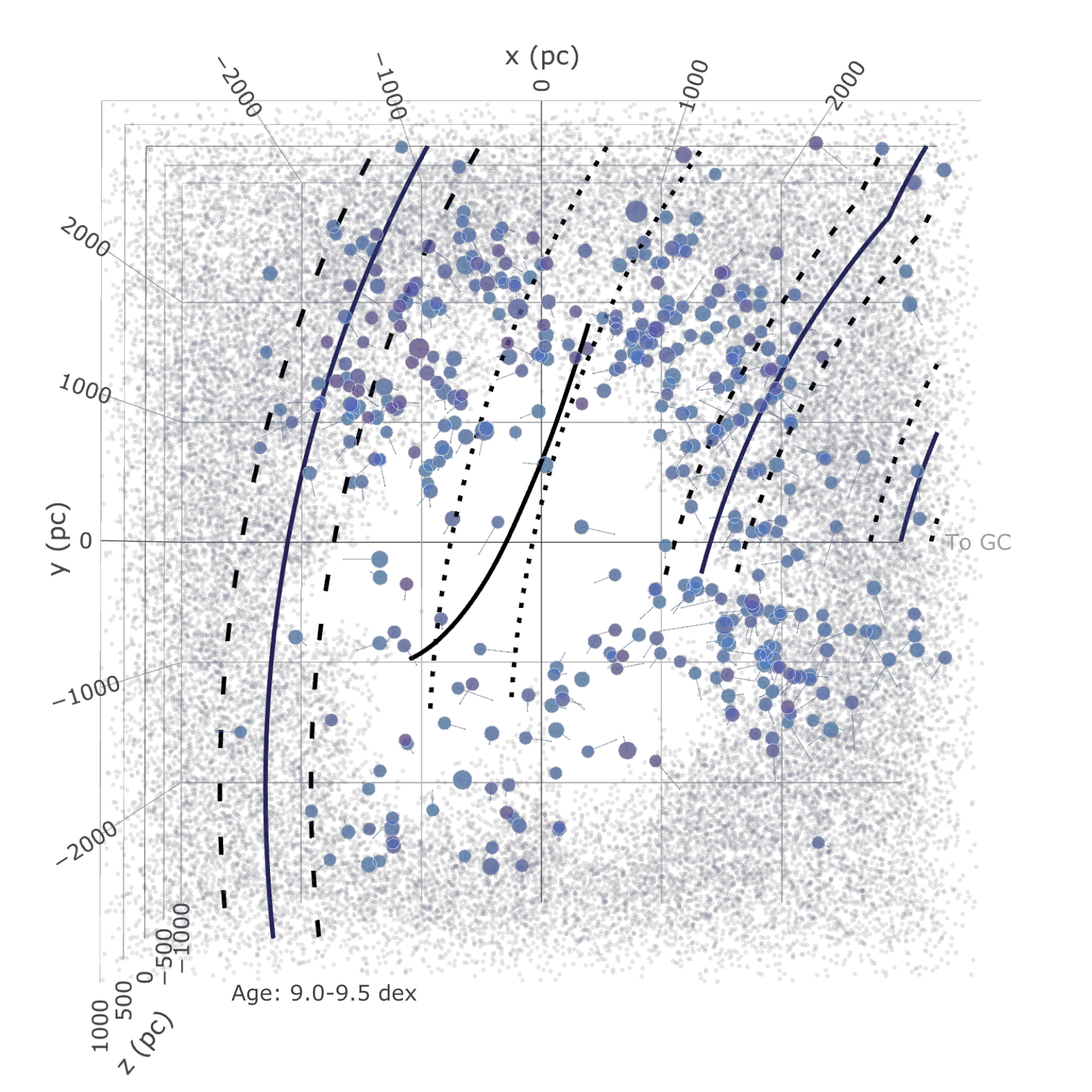}
\caption{Same as Figure \ref{fig:3d}, but restricted only to sources with $N_{(M_G<4)}>40$ to show a more uniform sensitivity in distance. \label{fig:3dbright}}
\end{figure*}

The effect of completeness from above does not take into the account the mass of the cluster. As at the larger distances, only the massive stars would be detected, using the same threshold of 40 stars as a minimum would result in a lack of detections of low mass populations further away (Figure \ref{fig:bright}).
%The sample is not entirely homogeneous due to being compiled from various slices, as discussed in Section \ref{sec:cluster}. Due to more distant stars being more numerous in the catalog, HDBSCAN is more atuned to the physical properties of the populations located at the boundary corresponding to the distance cutoff, resulting in a more optimal recovery of them with degraded performance in searching for more nearby populations. While the "onion layer" approach was chosen to mitigate these effects, some residual systematics of it remain.
All of the detected populations, regardless of the distance, do contain sources with extinction-corrected $M_G<4$. Thus, counting sources brighter than this threshold can serve as an age-dependent proxy for mass. The completeness volume estimate is appropriate $N_{(M_G<4)}\gtrapprox40$.  Figure \ref{fig:3dbright} shows the 3-dimensional distribution of only the populations that meet this cut, excluding lower mass nearby populations, for a more homogeneous view.

\begin{figure}
\epsscale{1.1}
 \centering
\plotone{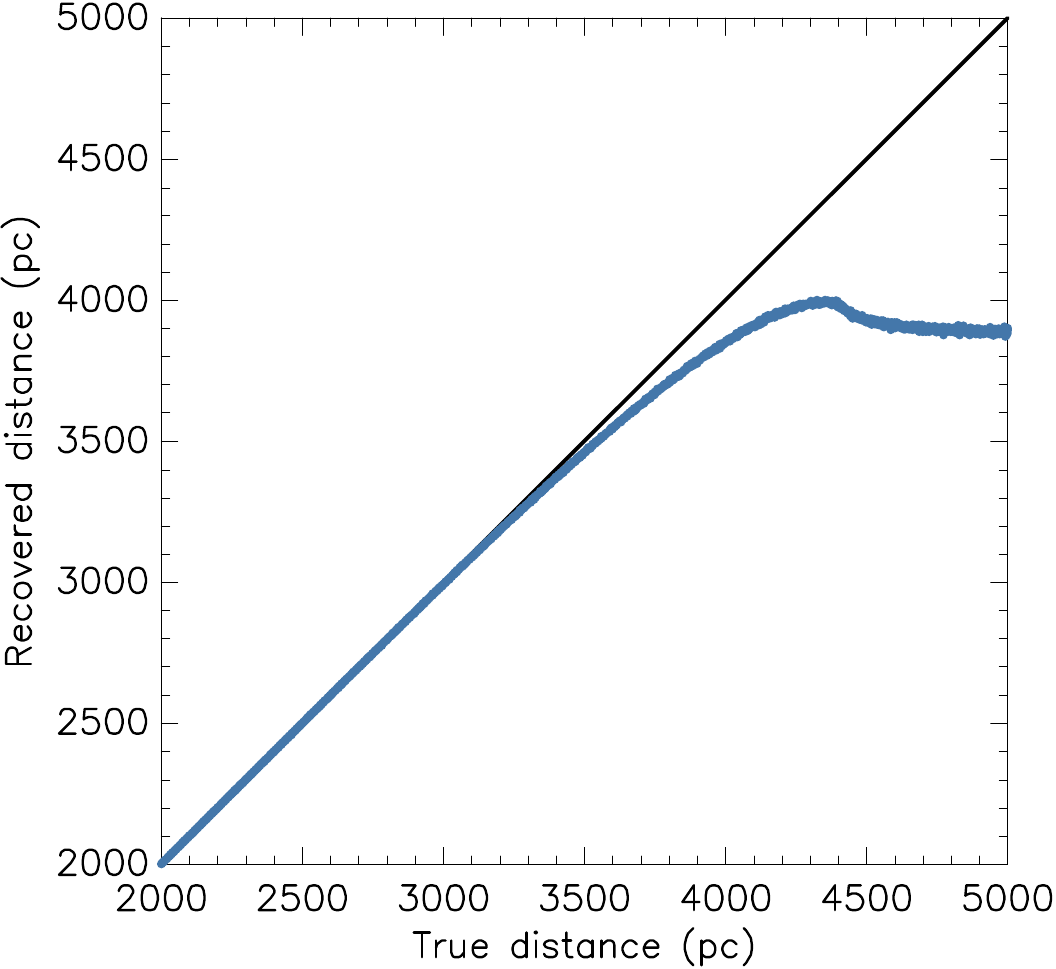}
\caption{The resulting average distance of the population with the parallax cut-off of 0.2 mas and the uncertainty in parallax of 0.1 mas.
\label{fig:distance}}
\end{figure}

Some caution should be exercised about the ``edge'' effect. The search of the structures is limited in the parallax space to 0.2 mas (ignoring the aforementioned biases), and the maximum uncertainty on the parallax is 0.1 mas. If a particular population has members with the resulting parallax measurement smaller than the limit, they would be excluded, and the resulting average distance for a population would appear to be closer than it really is, similarly to the Lutz-Kelker and Malmquist Biases \citep{lutz1973,malmquist1922,luri2018}. This effect is simulated in Figure \ref{fig:distance}.

Extinction moving the effective completeness limit closer in is not expected to have as stringent of an effect in this matter: if a population is not intrinsically embedded into a cloud, sources with smaller apparent parallaxes from the scatter due to uncertainties would not necessarily be excluded in clustering, even if this makes them appear projected behind the above detection threshold. However, extinction does have a role in shaping the recovered populations at large distances: as only a few lines of sight remain, these line of sights would create an overdensity of sources compared to their surroundings, and a clustering algorithm may group sources together more easily than if it would have been otherwise (See Section \ref{sec:old} for discussion). In the face on projection in Figure \ref{fig:3d}, this creates a number of groups ordered directly behind one another, protruding radially, like fingers. This can also be seen in the clusters in the sample from \citet{castro-ginard2020}. Although the fraction of real populations compared to random groupings along these lines of sight is likely comparable to what is found in the less extinct areas, we further caution for the need of independent vetting of these sources in the follow up studies. Furthermore, we note that their membership may be incomplete in such a way that wold affect their mean position.

\textit{Gaia} DR2 has a known systematic offset in parallaxes that could be as $\sim$80 $\mu$as \citep{stassun2018}. In estimating the distances with Auriga, this offset is partly taken into the account (Appendix \ref{sec:aurigavalidation}). Although there may be some not previously characterized second order effects that might influence the magnitude of the systematic offset on individual stars (such as, e.g., from color), this should not have a strong effect on the result presented in this work, as it only amounts to a systematic and mostly uniform scaling of the detected structures in distance with respect to the Sun, and doesn't strongly affect the location of the structures relative to one another.

\begin{splitdeluxetable*}{ccccccccccBccccccccc}
\tabletypesize{\scriptsize}
\tablewidth{0pt}
\tablecaption{Structure parameters\label{tab:theia}}
\tablehead{
\colhead{Theia} &\colhead{Common} & \colhead{String?} & \colhead{Distinct}  & \colhead{Age} & \colhead{$A_V$}& \colhead{Dist.}& \colhead{$N_*$}& \colhead{$l$}& \colhead{$b$}& \colhead{$\mu_l$\tablenotemark{a}}& \colhead{$\mu_b$\tablenotemark{a}}& \colhead{$v_r$}& \colhead{X}& \colhead{Y}& \colhead{Z}& \colhead{U\tablenotemark{ab}}& \colhead{V\tablenotemark{ab}}& \colhead{W\tablenotemark{ab}} \\
\colhead{ID} &\colhead{Name}  & \colhead{} & \colhead{from field?}& \colhead{(dex)} & \colhead{(mag)}& \colhead{(pc)}& \colhead{}& \colhead{(deg)}& \colhead{(deg)}& \colhead{(\masyr)}& \colhead{(\masyr)}& \colhead{(\kms)}& \colhead{(pc)}& \colhead{(pc)}& \colhead{(pc)}& \colhead{(\kms)}& \colhead{(\kms)}& \colhead{(\kms)}
}
\startdata
1 & LDN\_988e & F & Y & 6.48$\pm$0.08 & 1.16$\pm$0.20 & 655$\pm$30 & 194 & 90.9216 & 2.7943 & -5.88 & -1.13 & -9.25 & -10.52 & 654.4 & 31.9 & 0.83 & -8.07 & -3.96\\
2 & Chameleon\_I & F & Y & 6.48$\pm$0.06 & 1.48$\pm$0.31 & 210$\pm$7.8 & 193 & 297.0931 & -14.9898 & -2.88 & -2.01 & 2.40 & 92.5 & -180.9 & -54.4 & 3.20 & -2.86 & -2.56\\
3 &  & F & Y & 6.32$\pm$0.08 & 2.42$\pm$0.27 & 873$\pm$53 & 367 & 84.7852 & -0.1183 & -5.33 & 0.88 & -0.06 & 79.3 & 869.1 & -1.8 & -1.59 & -0.79 & 3.64\\
\enddata
\tablenotetext{}{Only a portion shown here. Full table is available in an electronic form.}
\tablenotetext{a}{In the local standard of rest.}
\tablenotetext{b}{Corrected for the bulk rotation, assuming solar velocity of 220 \kms, and the distance to the galactic center of 8.15 kpc \citep{reid2019}.}
\end{splitdeluxetable*}

\section{Results: 3D Structure and Evolution of the Milky Way}\label{sec:3d}

\subsection{Populations younger than 100 Myr}

\begin{figure*}
\epsscale{1}
		\gridline{
             \fig{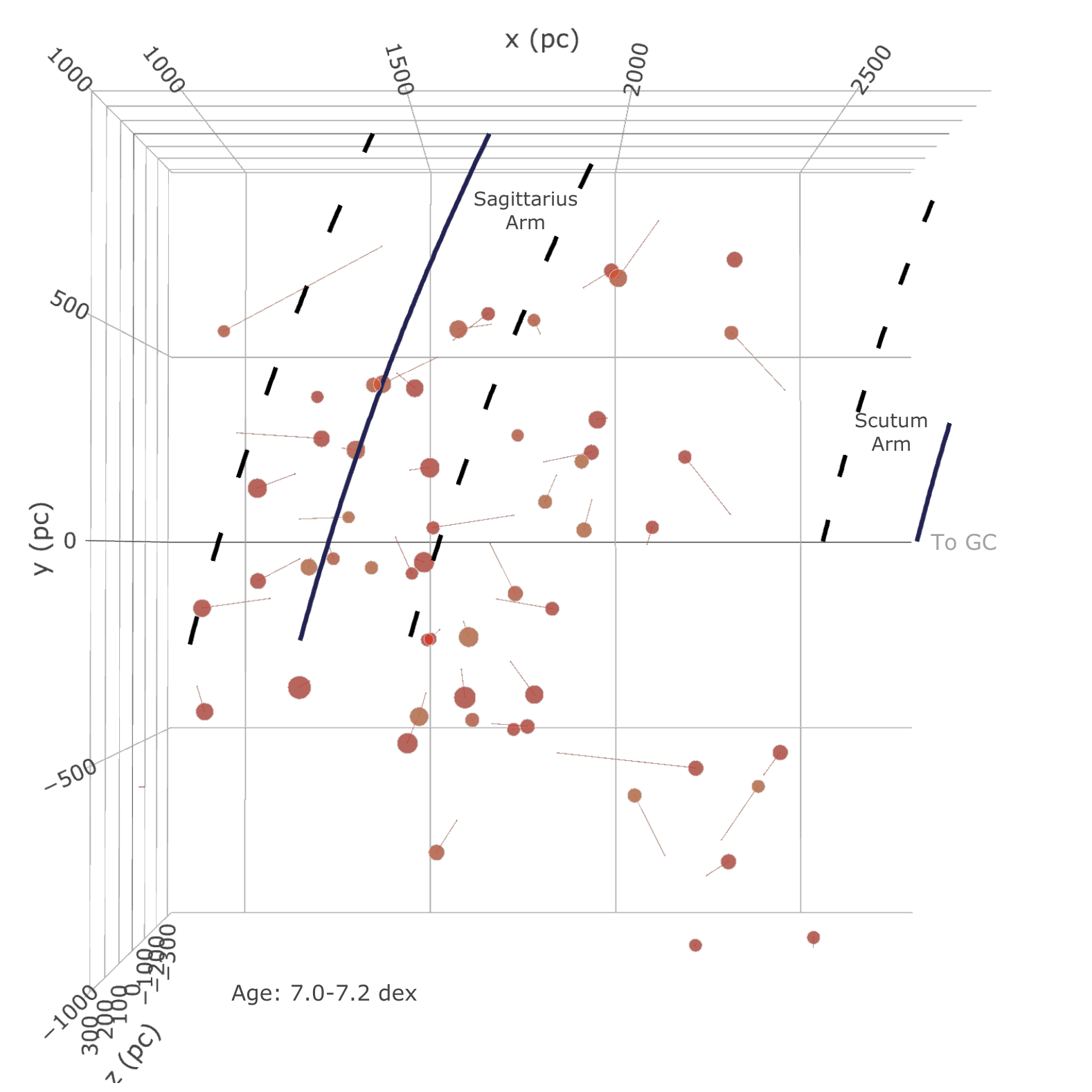}{0.4\textwidth}{}
             \fig{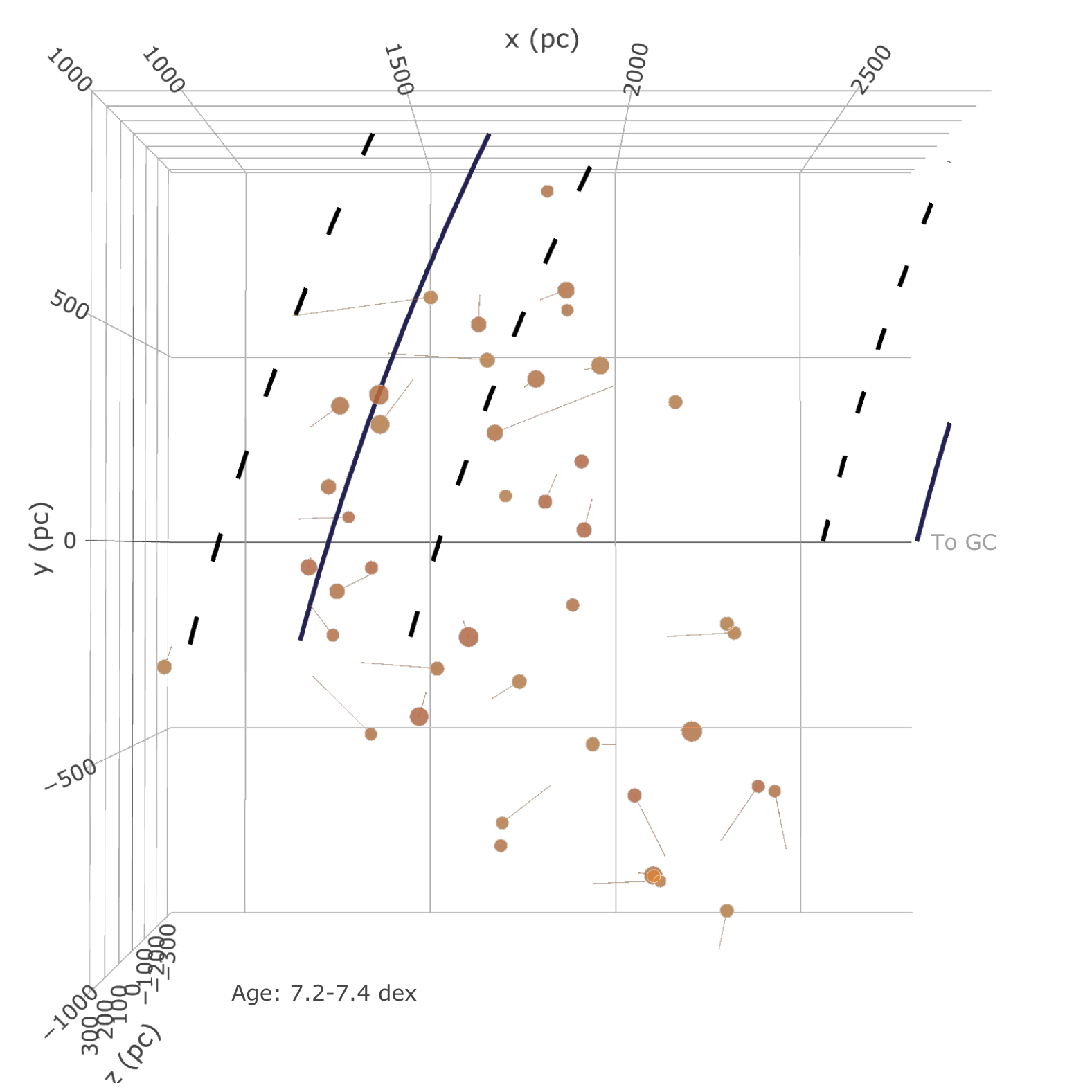}{0.4\textwidth}{}
        }\vspace{-1 cm}
		\gridline{
             \fig{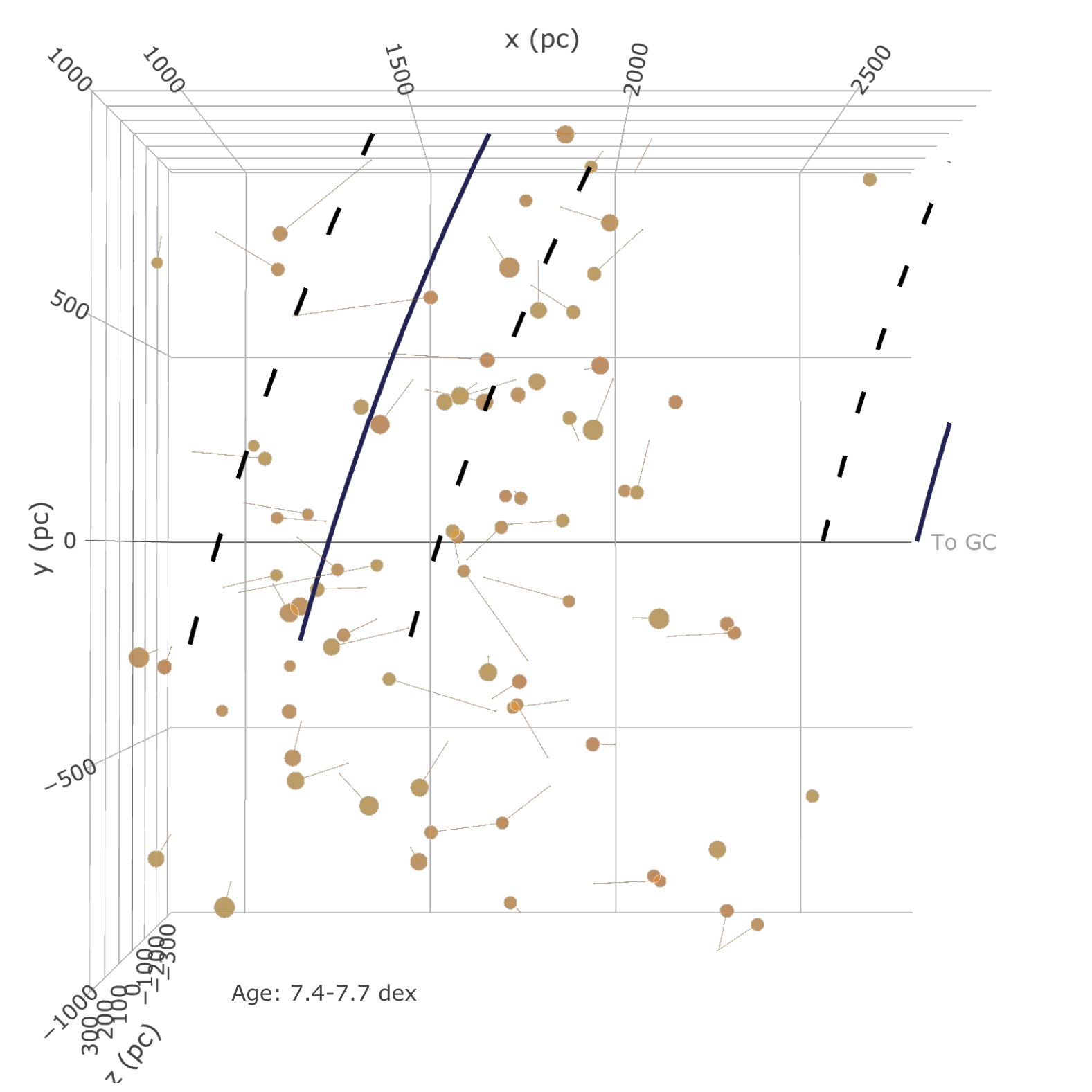}{0.4\textwidth}{}
             \fig{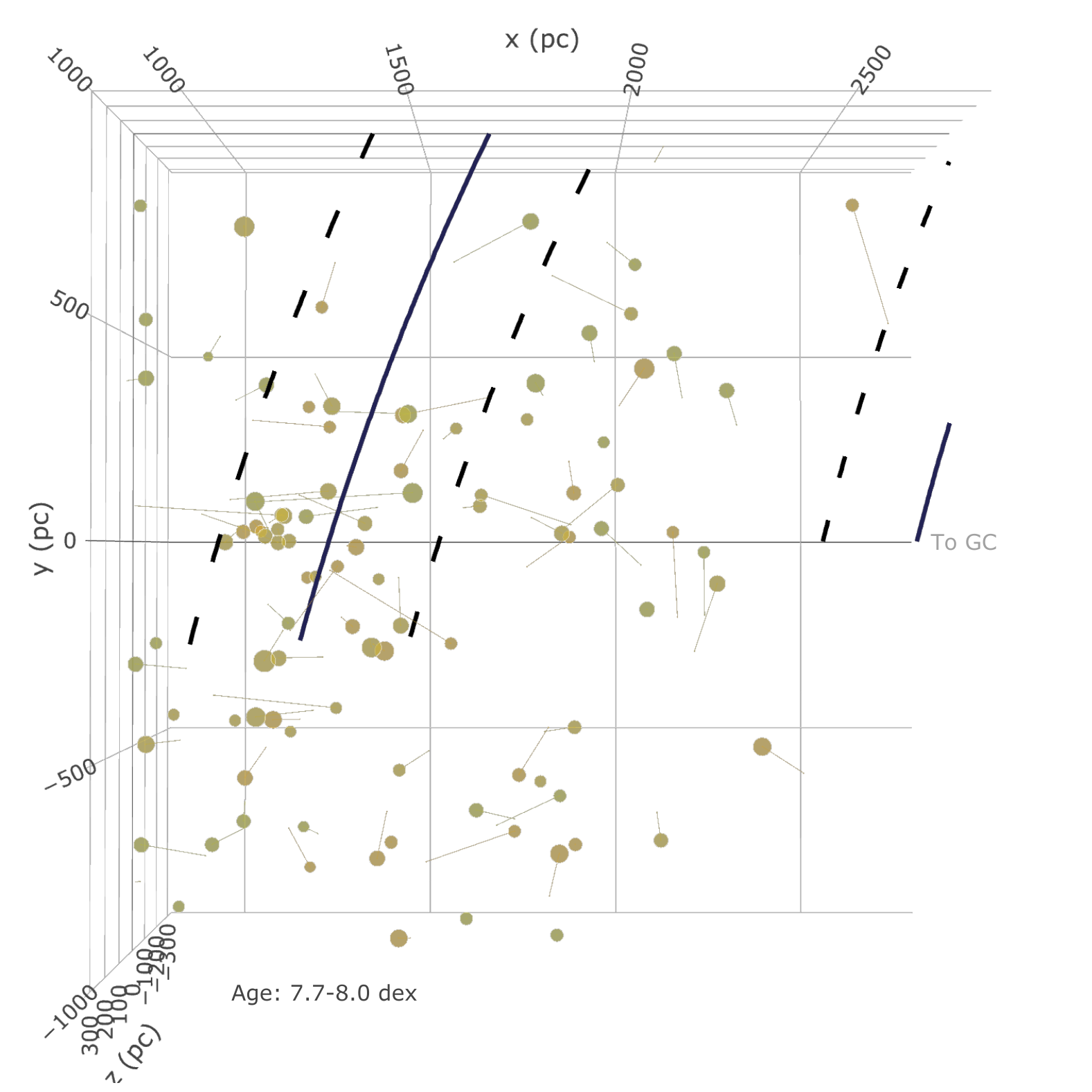}{0.4\textwidth}{}
        }\vspace{-1 cm}
        		\gridline{
             \fig{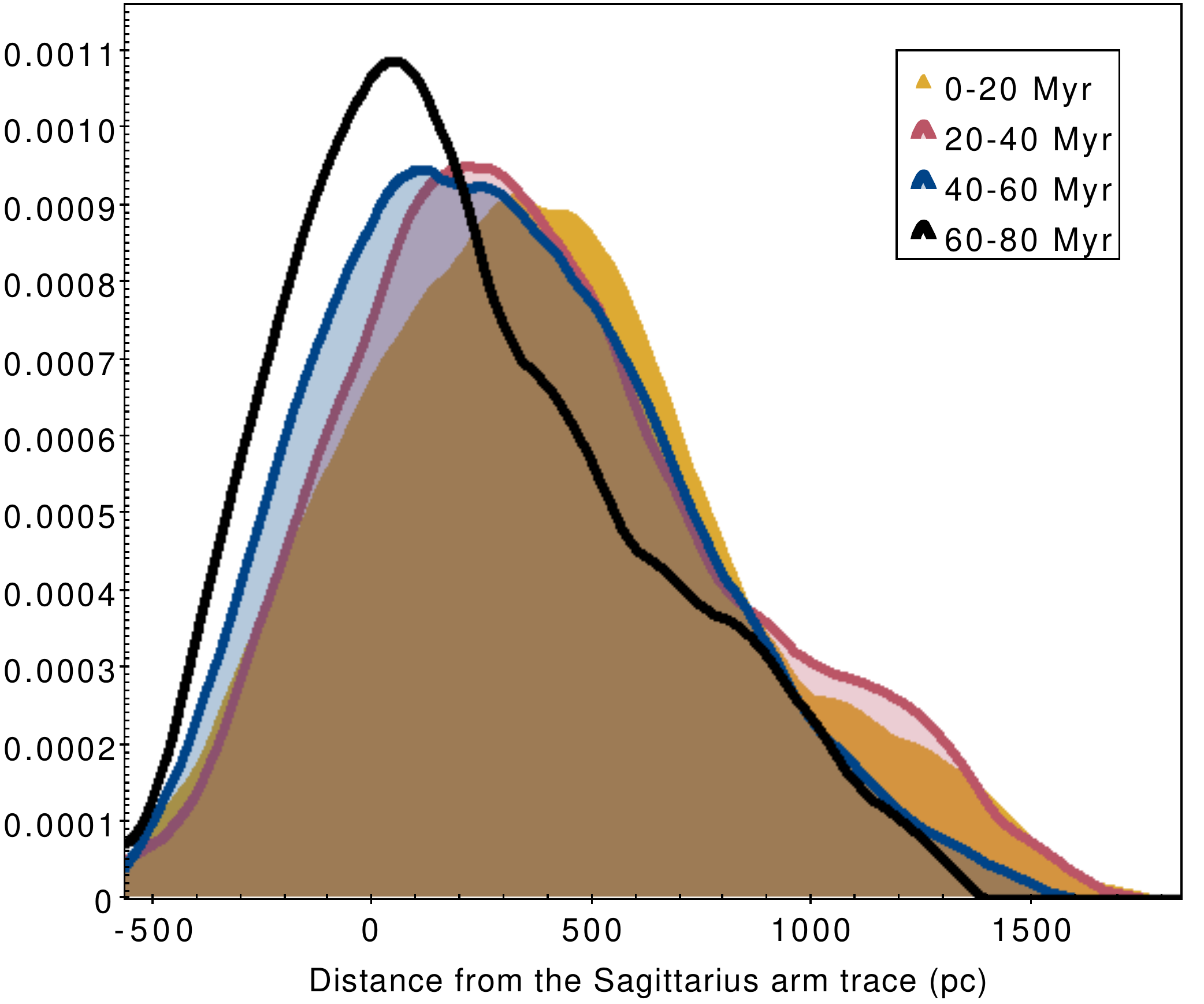}{0.3\textwidth}{}
             \fig{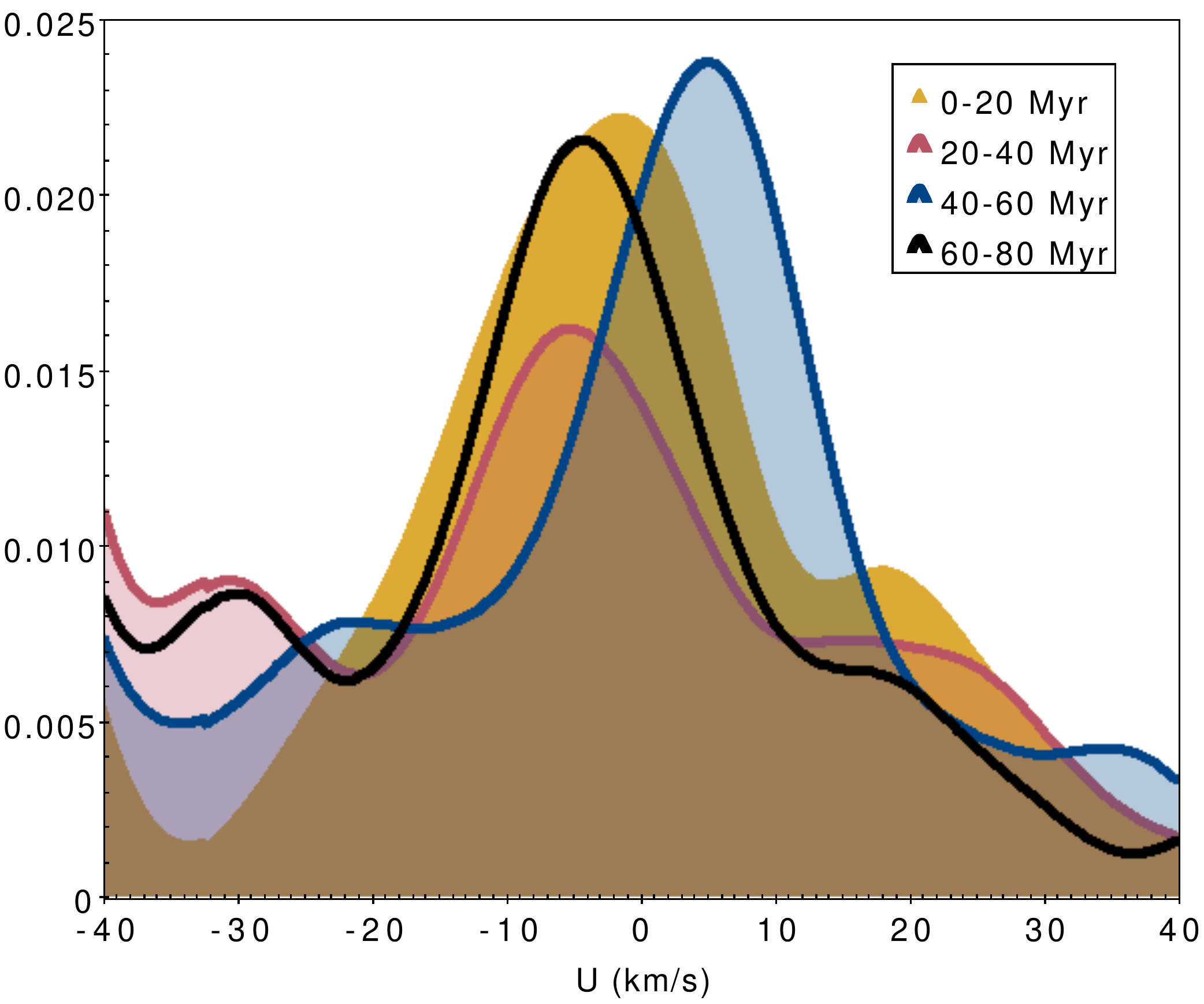}{0.3\textwidth}{}
             \fig{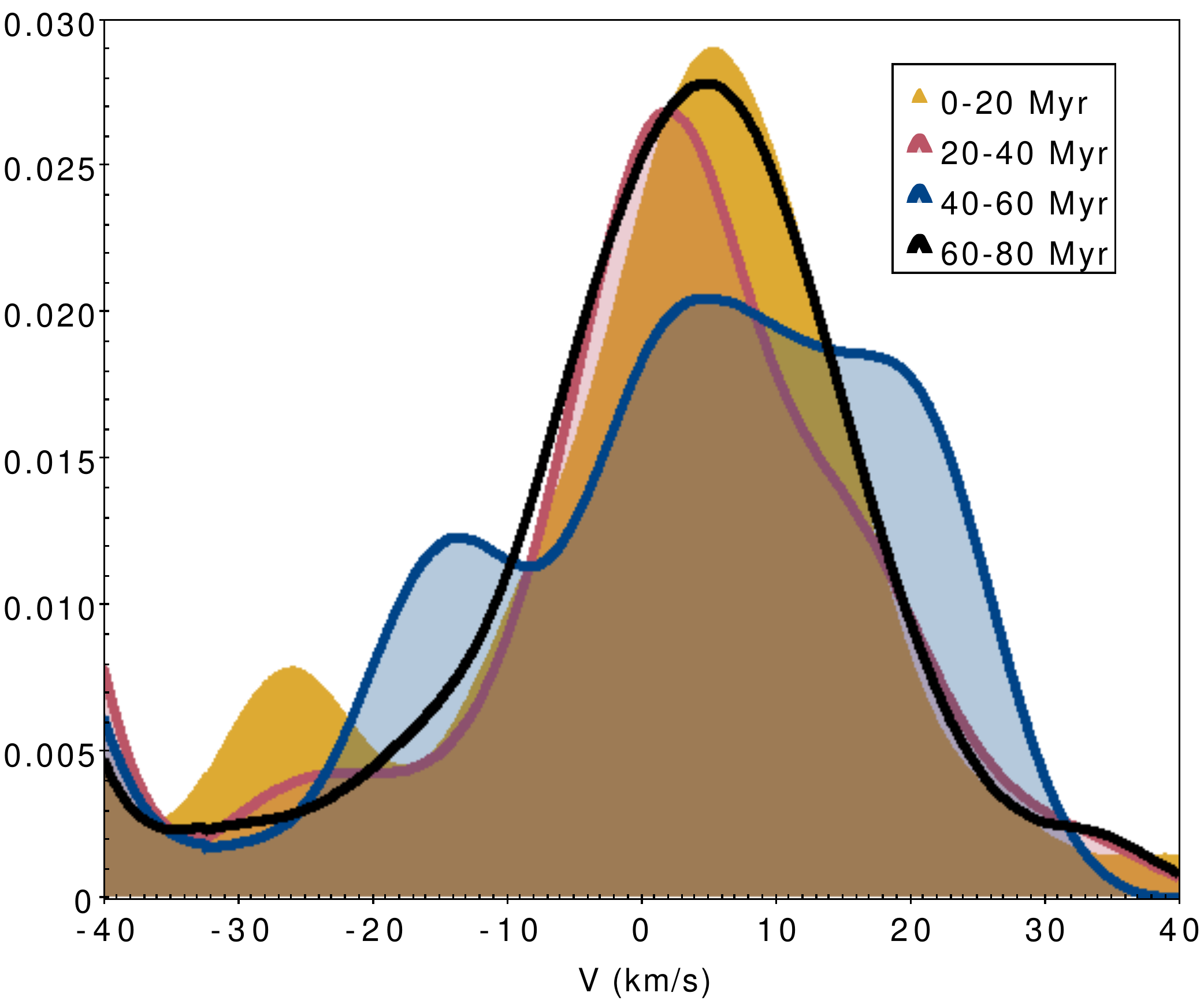}{0.3\textwidth}{}
        }\vspace{-1 cm}
\caption{Slices from Figure \ref{fig:3dstat} (highest confidence sample), but zooming in at the young populations towards the Sagittarius arm, at the age ranges of 7--7.2 dex, 7.2--7.4 dex, 7.4--7.7 dex, and 7.7--8 dex. Black lines show the location of the Sagittarius and Scutum arms from \citet{reid2019}. Bottom left plot shows the Kernel Density Estimate of the distributions of the populations that can be associated with the Sagittarius arm relative to the spiral arm trace from \citet{reid2019} as a function of age, limited to the highest confidence sample. Bottom middle and right plot shows the distribution of velocities as a function of age for this sample. \label{fig:sagittarius}}
\end{figure*}

\begin{figure}
\epsscale{1}
\plotone{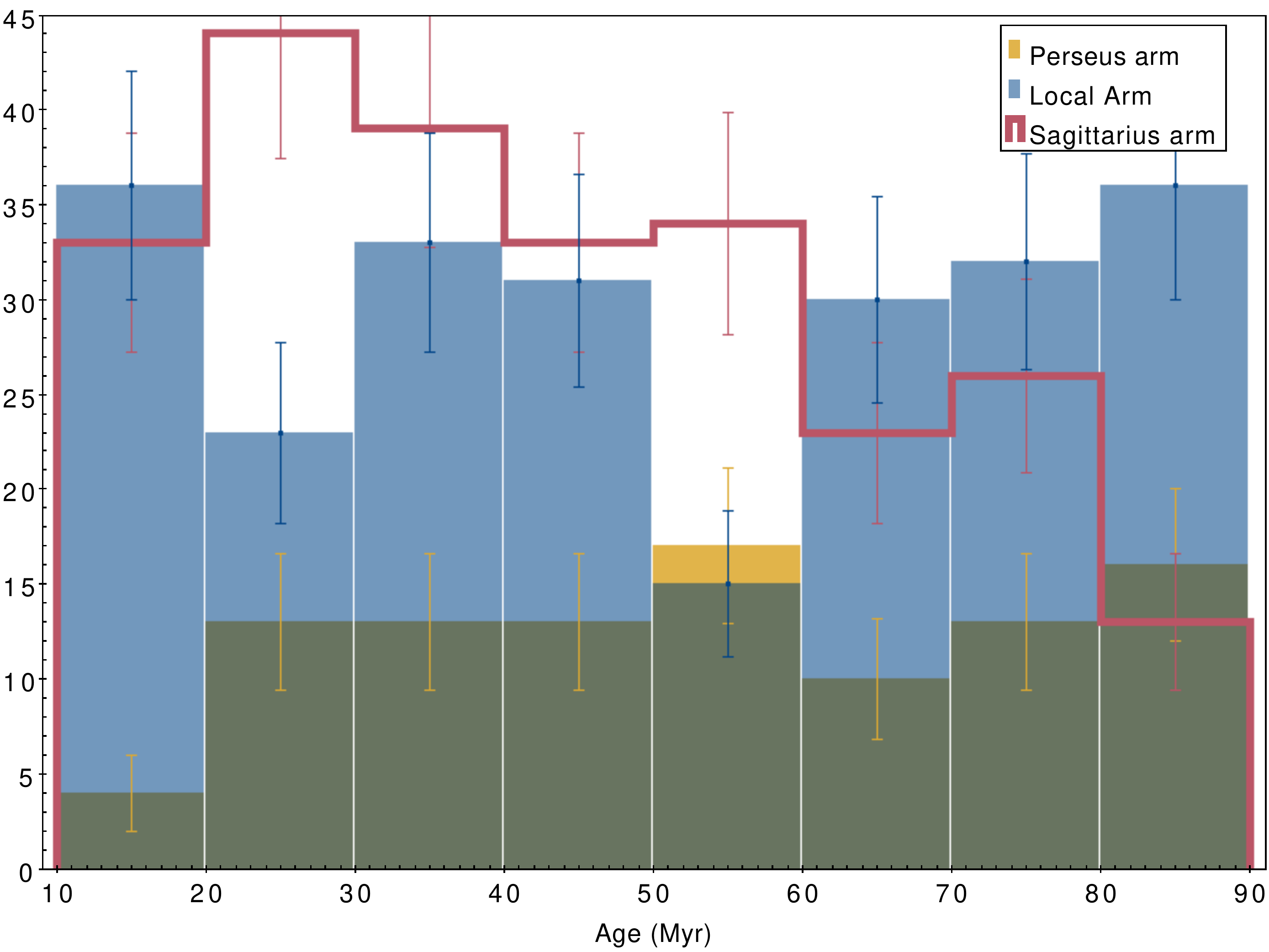}
\caption{Distribution of ages of the populations in the spiral arms, defining the Perseus arm at $-3000<X<-1000$ pc, Local arm at $-1000<X<1000$ pc, and the Sagittarius arm at $1000<X<3000$ pc, for the highest confidence sample. \label{fig:agebin}}
\end{figure}

\begin{figure}
\epsscale{1.1}
 \centering
\plotone{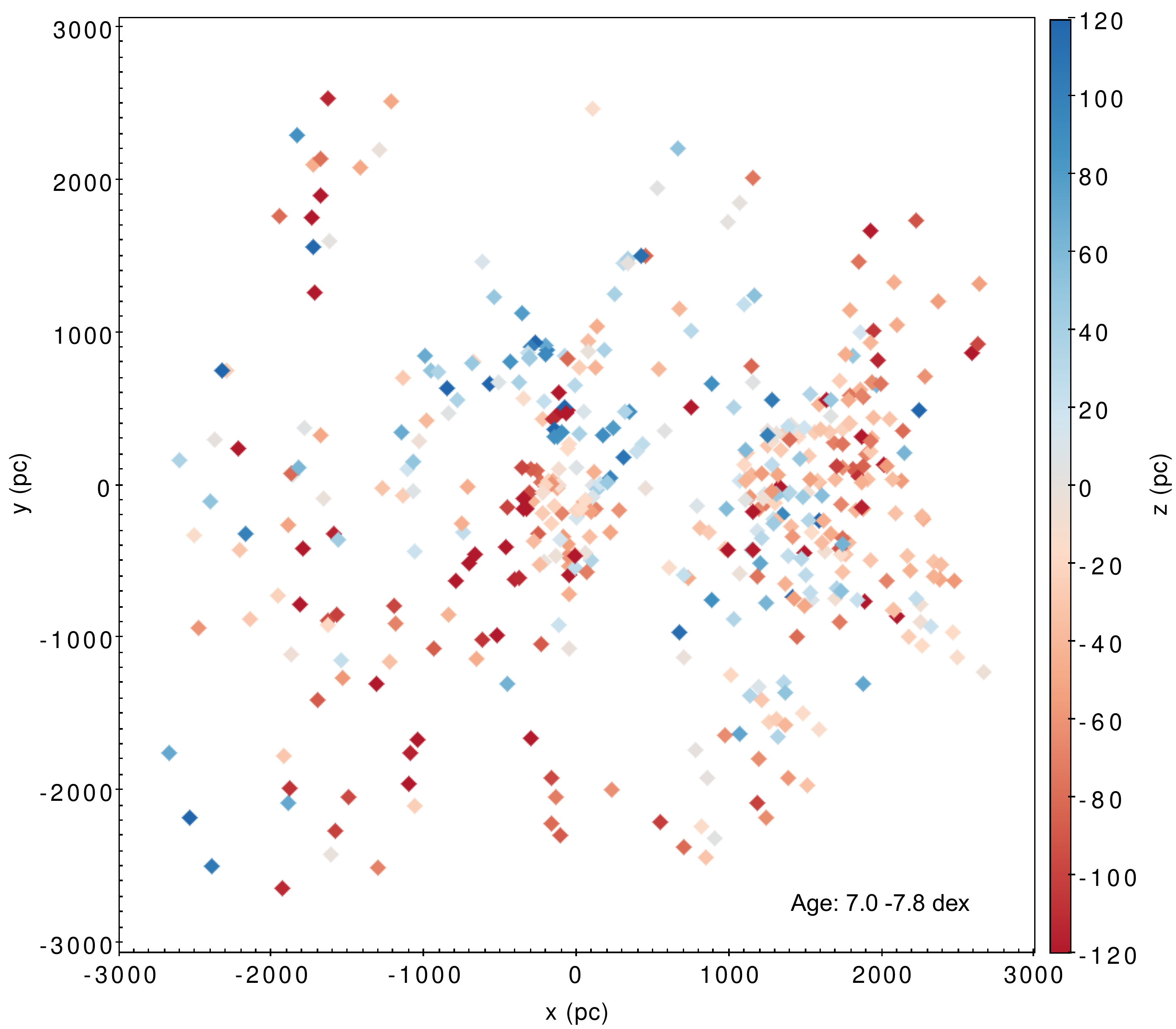}
\plotone{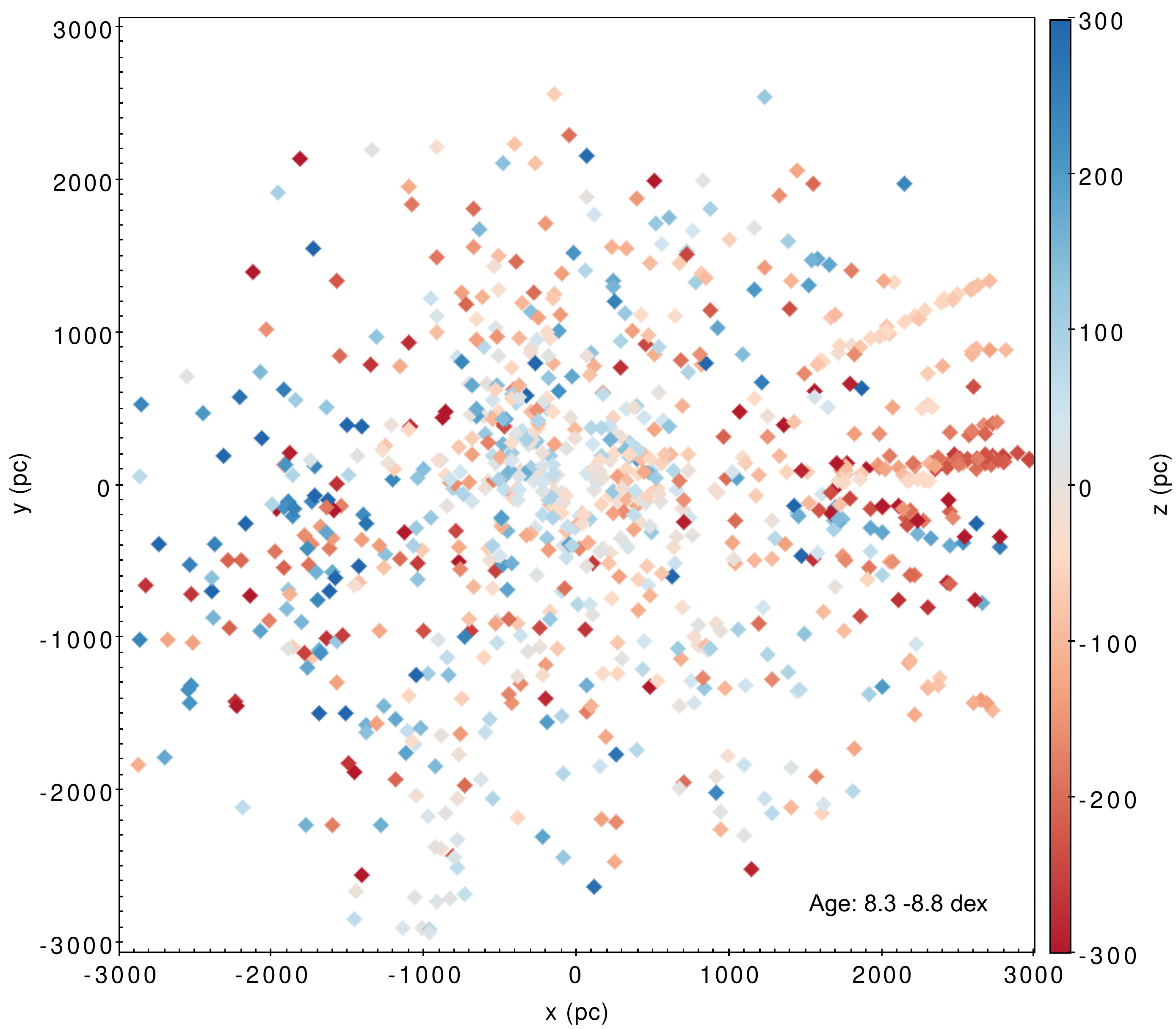}
\plotone{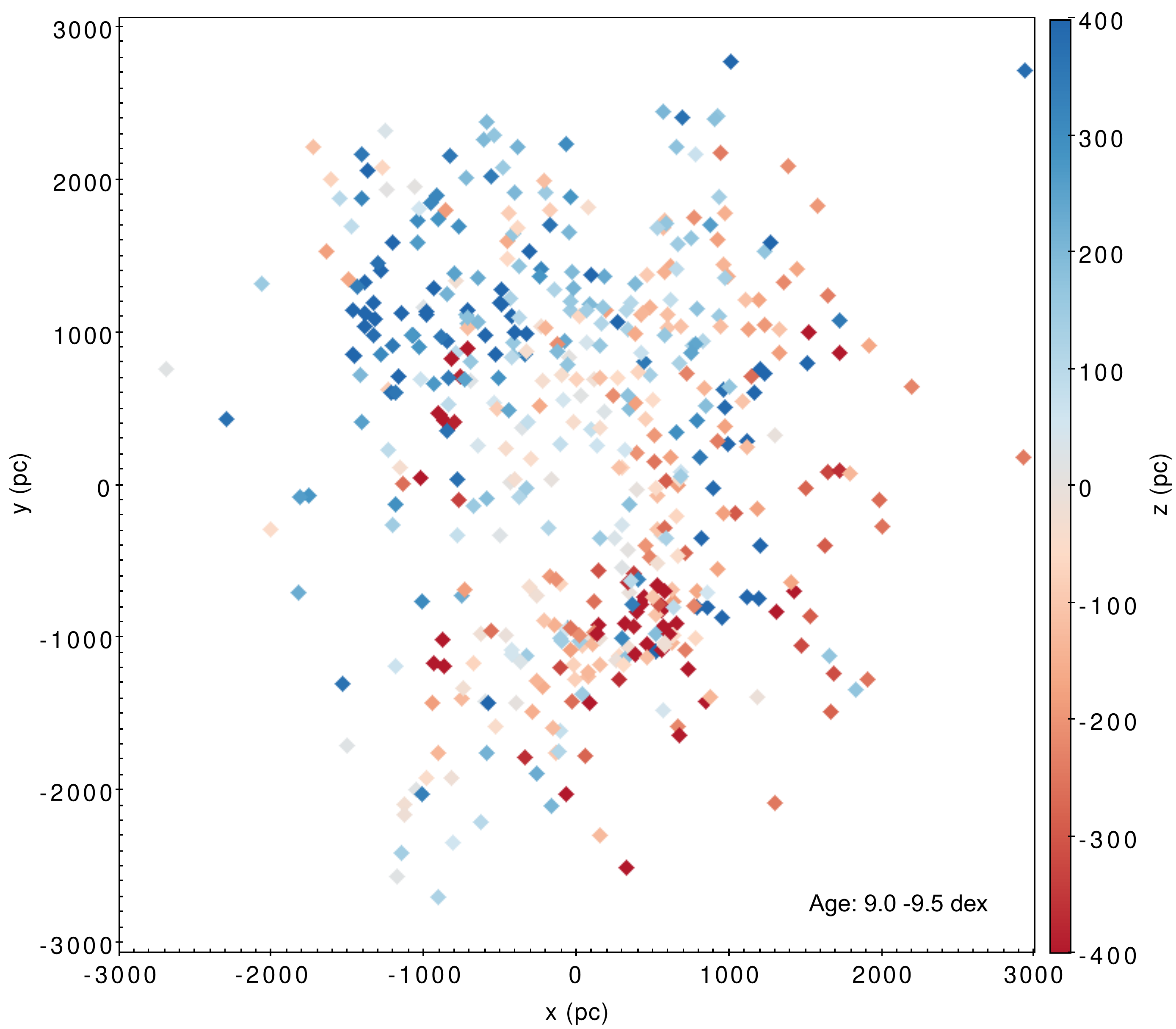}
\caption{Distribution of the height from the Galactic plane as a trace of the ripple of the disk, using the highest confidence sample. Top: populations younger than 7.8 dex, middle: 8.3--8.8 dex, bottom: populations older than 9 dex. Note the difference in the scale height between panels.
\label{fig:warp}}
\end{figure}

\begin{figure*}
\epsscale{1}
		\gridline{
             \fig{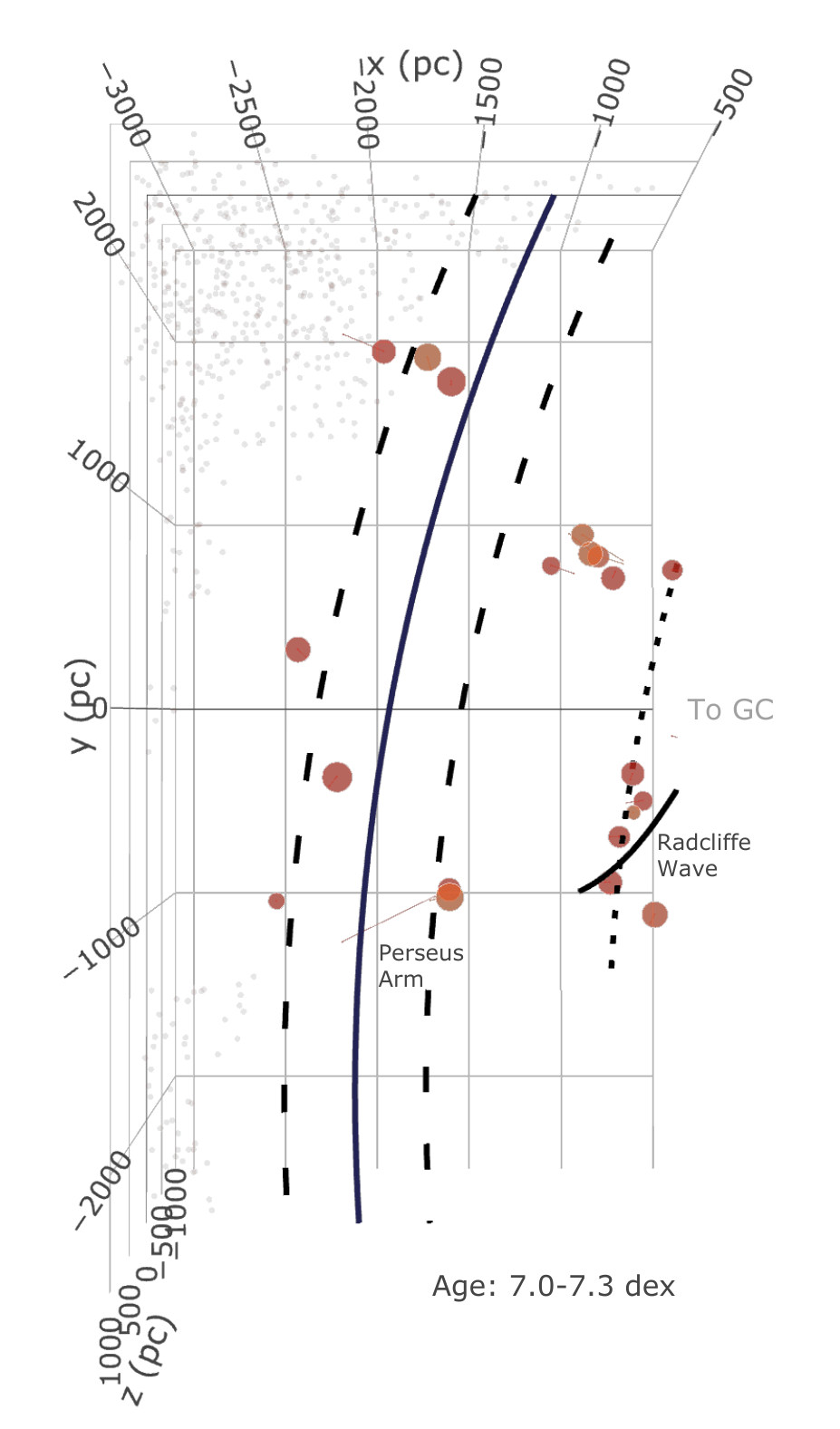}{0.25\textwidth}{}
             \fig{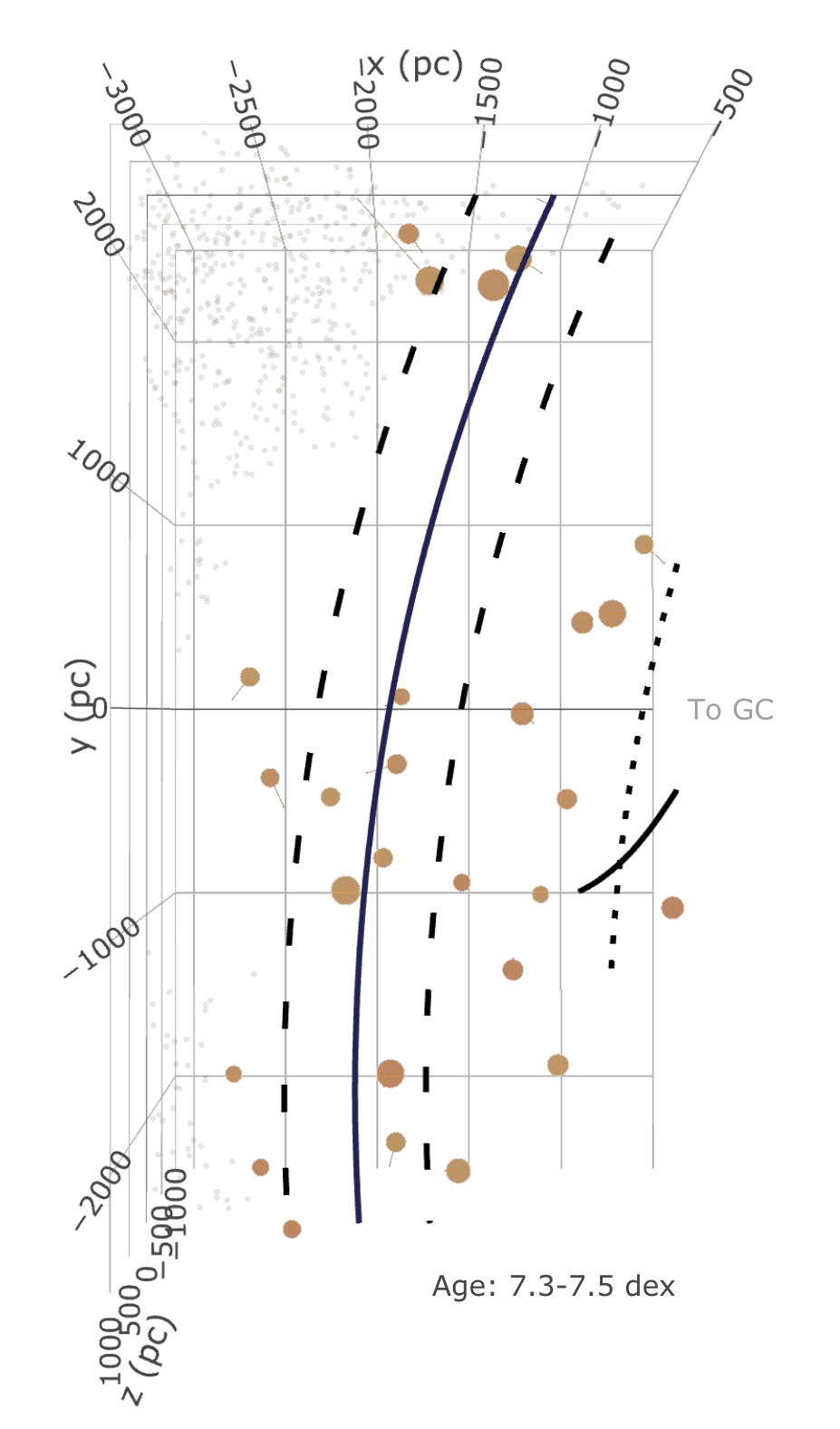}{0.25\textwidth}{}
             \fig{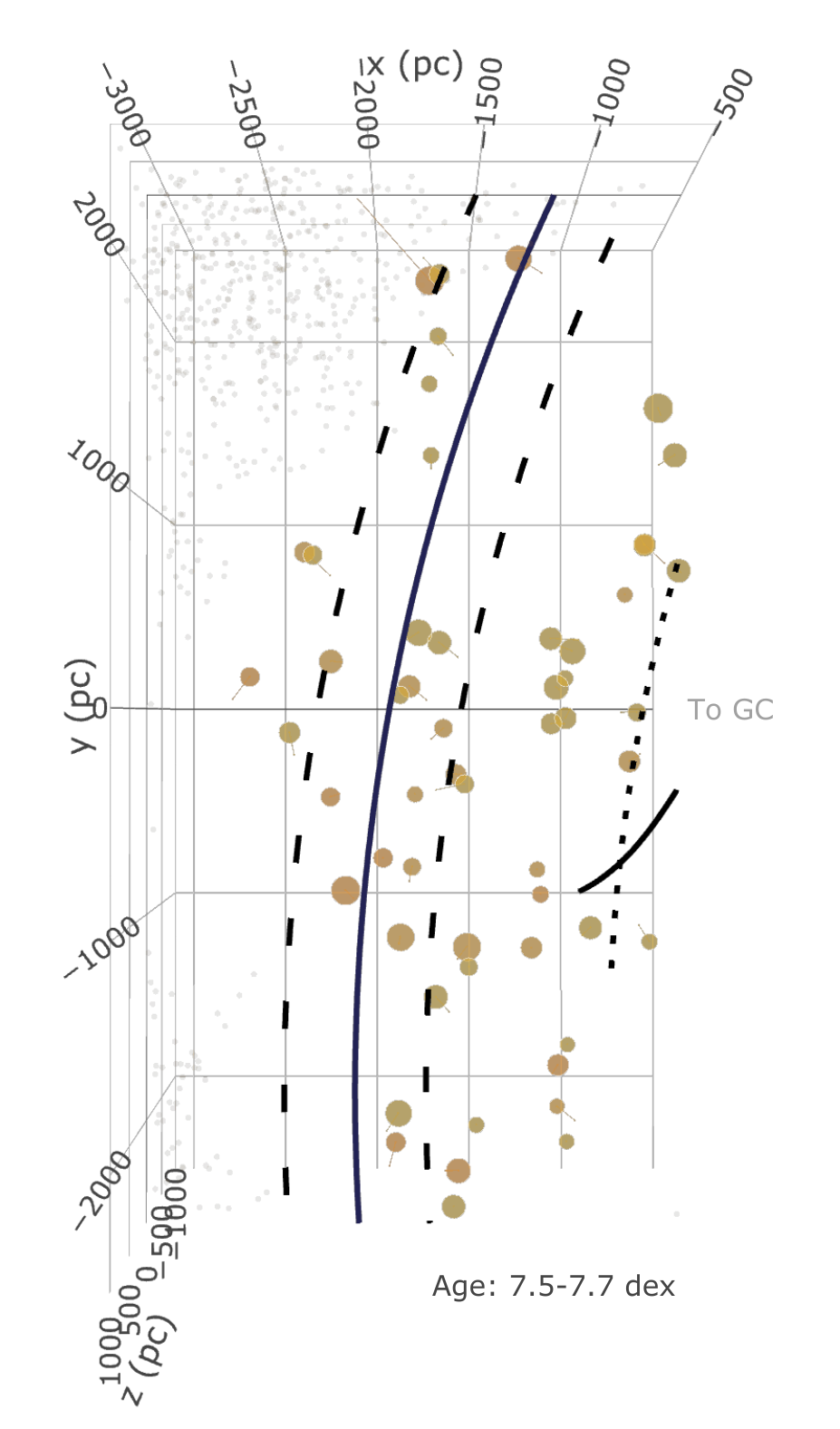}{0.25\textwidth}{}
             \fig{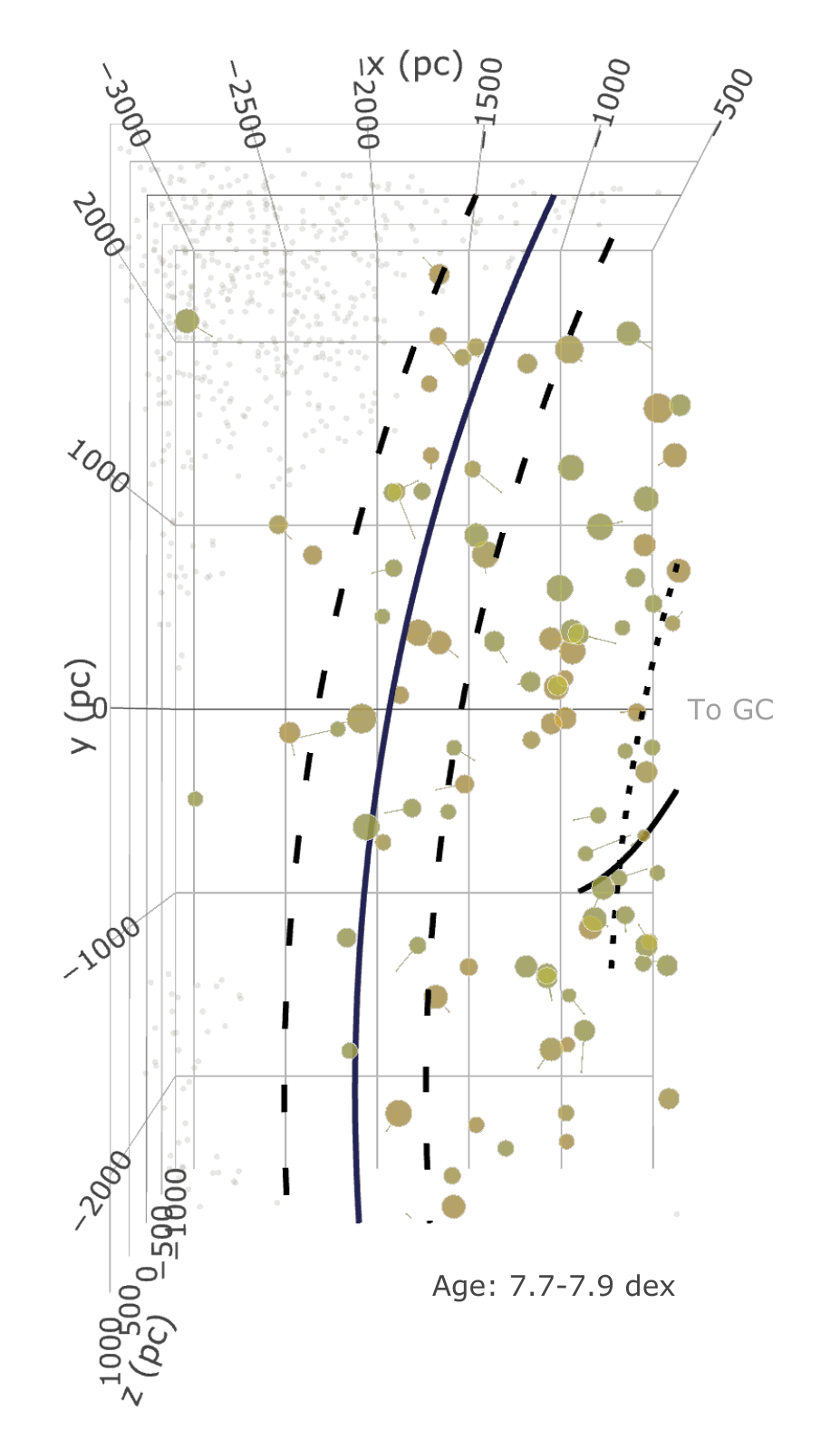}{0.25\textwidth}{}
        }\vspace{-1 cm}
\caption{Slices from Figure \ref{fig:3dstat} (highest confidence sample), but zooming in at the young populations towards the Perseus arm, at the age ranges of 7--7.3 dex, 7.3--7.5 dex, 7.5--7.7 dex, and 7.7--7.9 dex. Black line show the location of the Perseus arm from \citet{reid2019} \label{fig:perseus}}
\end{figure*}

In contrast to \citetalias{kounkel2019a}, with the sample extending up to 3 kpc, it is now possible to reach both the Sagittarius and the Perseus spiral arms. The separation between the arms (in their current form) persists up to the age of 7.8 dex.

\subsubsection{Sagittarius Arm}

Examining the young populations at $900<X<2500$ pc, most likely associated with the Sagittarius arm shows that the arm undergroes a progressive evolution of its position: 60 Myr ago it was more than 500 pc closer to the solar position than at the current epoch (Figure \ref{fig:sagittarius}). The trace of this arm from \citet{reid2019} show a better agreement with the population with an age of 7.8 dex than those that have an age of 7 dex. It is most clearly shown in the bottom panel of Figure \ref{fig:sagittarius}. Examining the difference between the $X$ position of the populations that can be associated with the arm and the $X$ position of the trace (at the corresponding $Y$ position of each particular group), the groups younger than 20 Myr appear to be the most distant, on average, whereas those that are older than 60 Myr are the most nearby. The difference for the different age bins persists even with a more restrictive selection on quality of the identified groups (i.e, using only the groups with small uncertainty in age and those that have statistically significant differences in their HR diagram relative to the field). As the star formation rate over the last 100 Myr remained more or less constant in the region of space currently associated with the Arm (Figure \ref{fig:agebin}), this displacement is unlikely to be caused by the contamination from a separate and unrelated structure, although we cannot rule out a possibility of  the Sagittarius arm supplanting another arm-like structure by forming in a similar place while the latter disappeared. 

The trace of the spiral arms from \citet{reid2019} is global for the Galaxy based on various molecular masers, and there is substantial scatter in the position of those masers on smaller scales. However, the offset between the trace of the arm and the youngest populations of stars exceeds the reported width of 270 pc.

The displacement of the Sagittarius arm is not an artifact due to the variable extinction limit of the \textit{Gaia} catalog. This limit, which we calculate and account for in our analysis, is not substantially different between the Sagittarius and Perseus arms when averaged across the entire arm in the volume of space of this study, with both arms having a comparable $A_V$ distribution. Therefore, extinction would not explain the difference in the evolution of structure as a function of age between these two populations. Similarly, as the brightest, high mass members of 20 and 60 Myr stellar populations are not significantly different, our sensitivity should not be strongly age dependent in the presence of a consistent magnitude+extinction limit.

If an estimation of distances we have underestimated the systematic offset in parallax from Gaia DR2, everything would shift inwards, with the younger populations being located closer to the trace of the arm. The relative separation in distance between them and the older populations would persist, however, pushing the $\sim$60 Myr populations away from the trace, also closer towards the Sun. However, we note that there is no significant difference in position between the trace of the arm and the identified stellar populations towards it at any age bin within past 100 Myr. As both Sagittarius and Perseus arms are close to being equidistant from us, and the extinction along the line of sight near their vicinity is comparable, if there were significant systematics affecting either our ability of identifying groups at these age ranges as a function of distance or estimating distances in general, we would expect these systematics to be observed towards both arms. Nonetheless, it is not clear why there is such a disagreement between the trace from \citet{reid2019} which was derived from masers presumably associated with the young star forming regions and the actual populations younger than $<$20 Myr.

Based on the difference in position of the young populations over the last 100 Myr, we estimate the speed of motion of the spiral arm of $\sim$8--10 \kms, consistent with estimates of the density wave propagation from \citet{dobbs2014}. This true spatial velocity is not to be confused with the pattern speed of 28.2 \kms\ kpc$^{-1}$ from \citet{dias2019}, as the latter is referring to the angular frequency of rotation uniform throughout the disk that is necessary to preserve the spiral arms, assuming they are a standing wave. It does not refer to the velocity, nor does it take into the account temporal evolution of the spiral arms that is commonly seen in the simulations \citep[e.g.,][]{li2019}.

\subsubsection{Local Arm}

Along the Local arm a recent study by \citet{alves2020} have found a long, coherent 2.7 kpc molecular gas structure, the Radcliffe wave that perturbs out of the Galactic plane with the maximum amplitude of 160 pc. The formation of this filament may be responsible for the apparent tilt of local star forming regions relative to the plane that has long been identified as the Gould's belt. Our clustering recovers stellar populations all along this wave. Almost all of the stellar groups that appear associated with it are younger than 12 Myr (7.1 dex), and it is no longer apparent at the ages older than 15 Myr (7.2 dex). Thus, it's formation within the Local arm is a relatively recent phenomenon.

There has been some debate in the literature whether the regions of the Gould's belt/Radcliffe wave that are displaced from the Galactic plane to high Z (ripples) are caused by a specific event, such as a collision with a high velocity HII cloud \citep{comeron1994}, impact with dark matter \citep{bekki2009}, or a series of supernovae eruptions warping the disk locally \citep{poppel2000}, presenting the conditions in the solar neighborhood as somewhat unique in causing star formation in the solar neighborhood to occur at comparatively high altitude relative to what was commonly found elsewhere.
 
The apparent ripples in the Local arm relative to the plane (independently of the Radcliffe wave, as it is neither a standing, nor a traveling wave with respect to time) does appear to persist up to the ages of 50 Myr (7.7 dex), possibly as old as 100 Myr (8 dex), after which the vertical scale height of the disk exceeds the amplitude of the ripples inside the Local arm. This significantly exceeds the lifetime of the populations that are associated with it. Thus either the origin of the ripples precedes the formation of the wave, or had to have occurred multiple times throughout these last 100 Myr. Some ripples can also be observed in the Sagittarius arm as well. (Figure \ref{fig:warp}), which has been previously also observed by \citet{alfaro1992}. Thus, any mechanism that would cause such ripples unlikely to be unique or particularly rare.

A rather curious feature of the Radcliffe wave is that despite its youth, very few strings are associated with it, both within 1 kpc where their census should be largely complete, and outside of it. Rather, most of the structures recovered along it tend to be compact and isolated. In \citetalias{kounkel2019a} we found that the vast majority of stars younger than $\sim$8 dex tend to be a part of extended strings that are oriented preferentially perpendicular to the Local arm. Thus, the Radcliffe wave may represent a different mode of star formation than most other young stellar populations. Alternatively, it is possible that strings will later develop, as molecular gas continues to accrete in the vicinity of these regions. Star formation in a particular region could persist for $\sim$10 Myr \citep[e.g., the Orion Complex,][]{kounkel2018a}, with the molecular gas still infalling to form clouds in one part of it, even while the gas is fully consumed and/or dispersed in a different part of the same region. Given that all of the populations are very young, and, indeed, still associated with molecular gas, their assembly may still be ongoing.

It is also notable that the Radcliffe wave (nor, arguably, any other stellar populations along the Local Arm younger than 30 Myr (7.5 dex) to some extent) does not extend all the way towards the completeness edge imposed by the extinction, which, at these ages allows to peer up to almost 3 kpc in any direction. Rather, the wave truncates outside of $-1000 \lesssim Y \lesssim +1500$ pc. Similarly, although the Sagittarius arm reaches the edge of completeness along the $X$ axis, few young stellar populations are found outside of $-1000 \lesssim Y \lesssim +1000$ pc. Finally, very few groups younger than 25 Myr (7.4 dex) have been recovered along the Perseus arm within the completeness limit, although it does have a greater presence at somewhat older ages (Figures \ref{fig:perseus}, \ref{fig:agebin}).

\subsection{Populations older than 100 Myr}\label{sec:old}
\begin{figure*}
\epsscale{1.1}
 \centering
\plotone{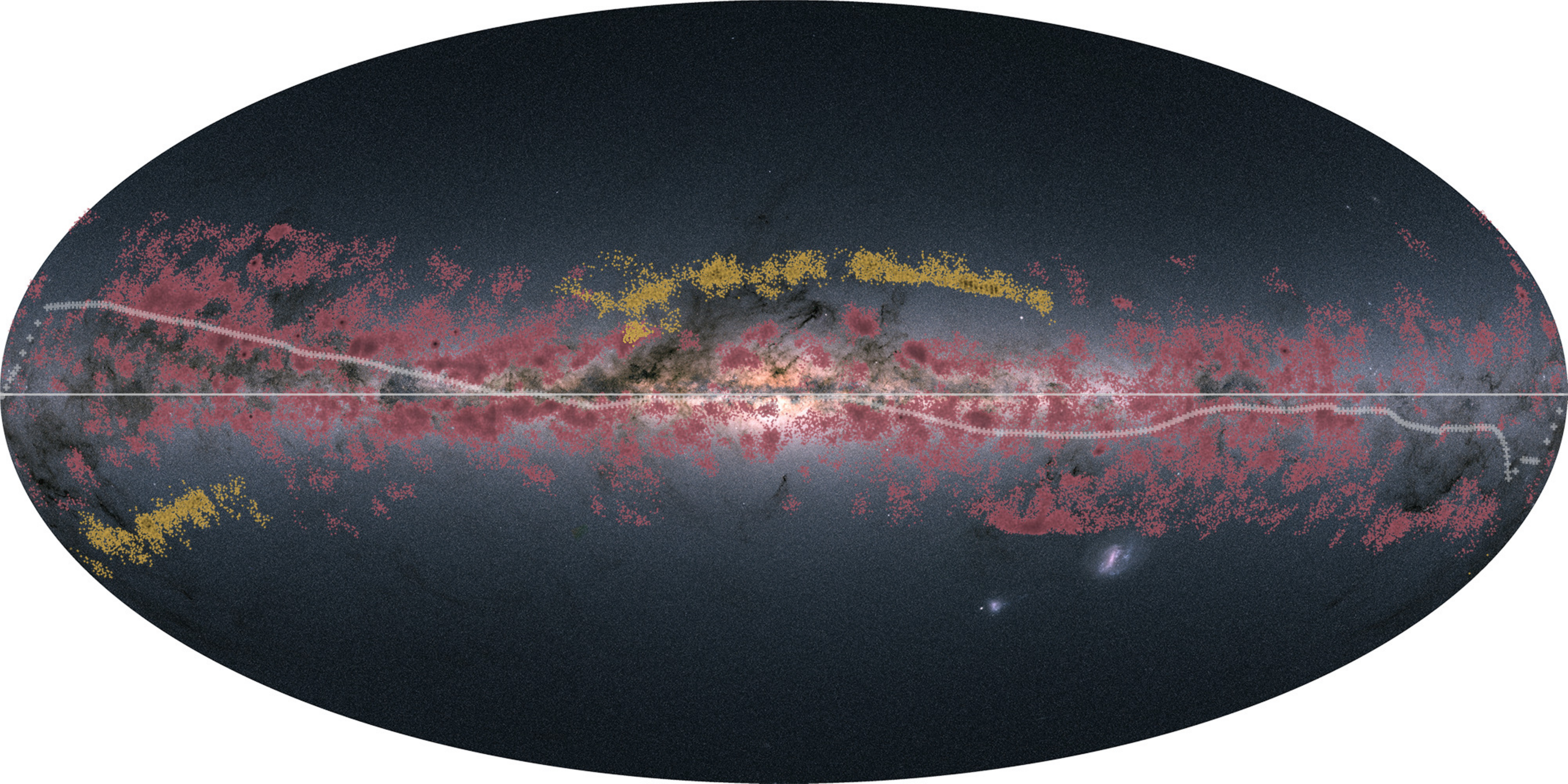}
\caption{Distribution of the clustered populations older than 9 dex located at distances larger than 750 pc, plotted on top of \textit{Gaia} DR2 color map \citep[courtesy of ESA/Gaia/DPAC,][]{gaia-collaboration2018}, to highlight the areas of large extinction Note that extinction can clear out certain lines of sight, and the ``fingers of god'' that are discussed in Section \ref{sec:complete} can be attributed to the extinction patterns. The heliocentric stream is highlighted in yellow. White crosses show the smoothed median $b$ position of the remaining sources as a function of $l$, and the horizontal line shows the location of the Galactic plane. The full sample is plotted; this underlying distribution of stars is also consistent in the highest confidence sample only. 
\label{fig:old}}
\end{figure*}

Unfortunately, the resolution to large scale structure is not as precise as what is offered by the 1 kpc sample. The few strings of stars that have been identified continue showing the similar correlation as in the previous work: these extended populations tend to form perpendicular to the spiral arms, however, their catalog is not complete even at the younger ages, and no new strings were added at ages greater than 8.5 dex. The uncertainty in parallax also becomes significant beyond 1 kpc, making it difficult to conclusively analyze their geometry, compared to the sample from \citetalias{kounkel2019a}. Furthermore, due to a turn-off that occurs at increasingly fainter magnitude in the older populations, much of the Galactic midplane beyond 1 kpc is truncated due to extinction, leaving behind only a few select lines of sight along which the structure could be recovered (Figure \ref{fig:old}, producing the apparent ``fingers'' that could be seen in e.g., Figure \ref{fig:3d}. And although it is possible to recover groupings at larger distances at higher galactic latitudes, they may not be as reliable tracers of the overall structure.

Even considering a single average position of the identified populations, without incorporating the strings, the overall distribution of overdensities is consistent with \citetalias{kounkel2019a}. The Local arm in the current form is no longer apparent in the population older than 7.8 dex. At $\sim$8 dex, the distribution of density of the populations is mostly uniform, lacking any gaps that could be attributable to the spiral arms. However an overdensity in the distribution of identified groups that may be related to an ancient spiral arm at a similar position but with different orientation is apparent at $\sim$8.2--8.8 dex. At these ages, there may be older analogues to the Perseus and Sagittarius arms as well.

These overdensities which may be remnant spiral arms are labeled in Figure \ref{fig:3d}, and, unaided all three are most apparent in Figure \ref{fig:3dbright}, which excludes lower mass populations. Figure \ref{fig:3dstat} particularly shows the orientation of the older counterpart to the Local arm within 1 kpc. This particular feature can also be recovered in analyzing the 3d distribution at this age range in the catalog of known clusters from \citet{cantat-gaudin2018a,cantat-gaudin2020}. 

At ages of 7.9--8.1 dex, there are no apparent overdensities that correspond to spiral arms, with a very sharp transition period.

Similarly, at the ages older than 8.8 dex, two different overdensities are apparent. The relative orientation of these structures results in a substantial deficit of old populations in the direction of the Galactic anticenter (Figures \ref{fig:3d}, \ref{fig:old}), even within the volume of space in which we should be complete at these ages. Although there is some difference in the recovery of groups by HDBSCAN in layers found at different distances along the seam lines, such biases should not affect the populations outside of these seams. Furthermore, the gap separating the two different structures is robust even if the sample is restricted to the highest confidence group (Figure \ref{fig:3dstat}). Although this gap has not been seen prior to \citetalias{kounkel2019a}, such as in the distribution of previously known clusters \citep[e.g.,][]{cantat-gaudin2018a}, few clusters are known in this volume of space at these age bins to be able to fully analyze their density distribution. Furthermore, the lack of old clusters in the solar neighborhood (which in part can be explained by the gap we observe) has long since been known \citep{oort1958}.

Improvement in astrometry from the future data releases from \textit{Gaia}, as well as ground based spectroscopic follow-up of the stars in the identified populations would be beneficial for confirming that these overdensities are indeed related to the ancient spiral arms and not occur for any other reason. We should note, however, that tracing the position of clusters in the galactic simulations over time \citep[such as, e.g.,][]{li2019}, it should be indeed possible to recover diffuse ancient spiral arms that have formed even up to the oldest epochs of the simulations (up to 1 Gyr for this particular one). We defer the full comparison of observations to the simulations to the follow up works.

Some rippling of the galactic disk may be apparent for the populations older than 9 dex (Figures \ref{fig:warp}, \ref{fig:old}). Examining the median $b$ position of the stars at these ages as a function of $l$ shows that their distribution may be tilted relative to the galactic plane by as much as 10$^\circ$ (Figure \ref{fig:old}). It is broadly consistent with \citet{romero-gomez2019} -- although asymmetries in $Z$ distribution in that work are found on much larger scales, there is a similar trend of populations found below the Galactic plane being predominantly located in the $-Y$ directions, and those that are above being located in the $+Y$ direction. No such ripples are apparent in the populations of intermediate age (8--9 dex).

\subsection{Heliocentric Stream}

Although extinction does shape a substantial portion of substructure that is seen in the distribution of the old populations, there is one feature that does stand apart from the bulk of the identified old groups, and it does not clearly trace the edge of the dusty clouds along the line of sight (Figure \ref{fig:old}). This feature resembles a stream of some kind. It is deconvolved into dozens of different groups that do not show a strong relative kinematic coherence, even though it does appear to be continuous in the plane of the sky. It is most apparent towards the Galactic center, elevated at $b\sim25^\circ$ away from it at the maximum separation, possibly continuing towards $b\sim-25^\circ$ near the Galactic anticenter, apparently forming a great circle in the sky. A part of this stream is also present in the 1 kpc sample from \citetalias{kounkel2019a}. In the 3d plot in Figure \ref{fig:3d}, this feature appears to form a plane centered at the Sun, reaching upwards of 600 pc from the plane at the furthest point.

Almost all of the identified groups associated with this heliocentric stream have measured ages of $>8.8$ dex. Such a distribution of ages would not be surprising at these distances at $b\sim25^\circ$, as they are dominated by the thick disk and halo stars \citep{rix2013}. The stream does persist as a distinct feature in the highest confidence sample, but it is difficult to say how statistically significant the coherence in ages of the groups along the stream is. Furthermore, although there has been a number of remnants of tidally stretched globular clusters and dwarf galaxies that exist as extended streams, it is difficult to imagine why should any such stream be centered at the Solar position, making it more likely that it is artificial. On the other hand, it's position has no correlation to \textit{Gaia} scanning law that would explain the reason behind the systematic preference in the coherence of velocities along this specific plane. Further investigation would be necessary to determine the true nature of the heliocentric stream.

\section{Discussion}\label{sec:disc}

In this section, we address several points of concern that could be raised regarding sample that could affect our results. 

\begin{figure*}
\epsscale{1.1}
\plottwo{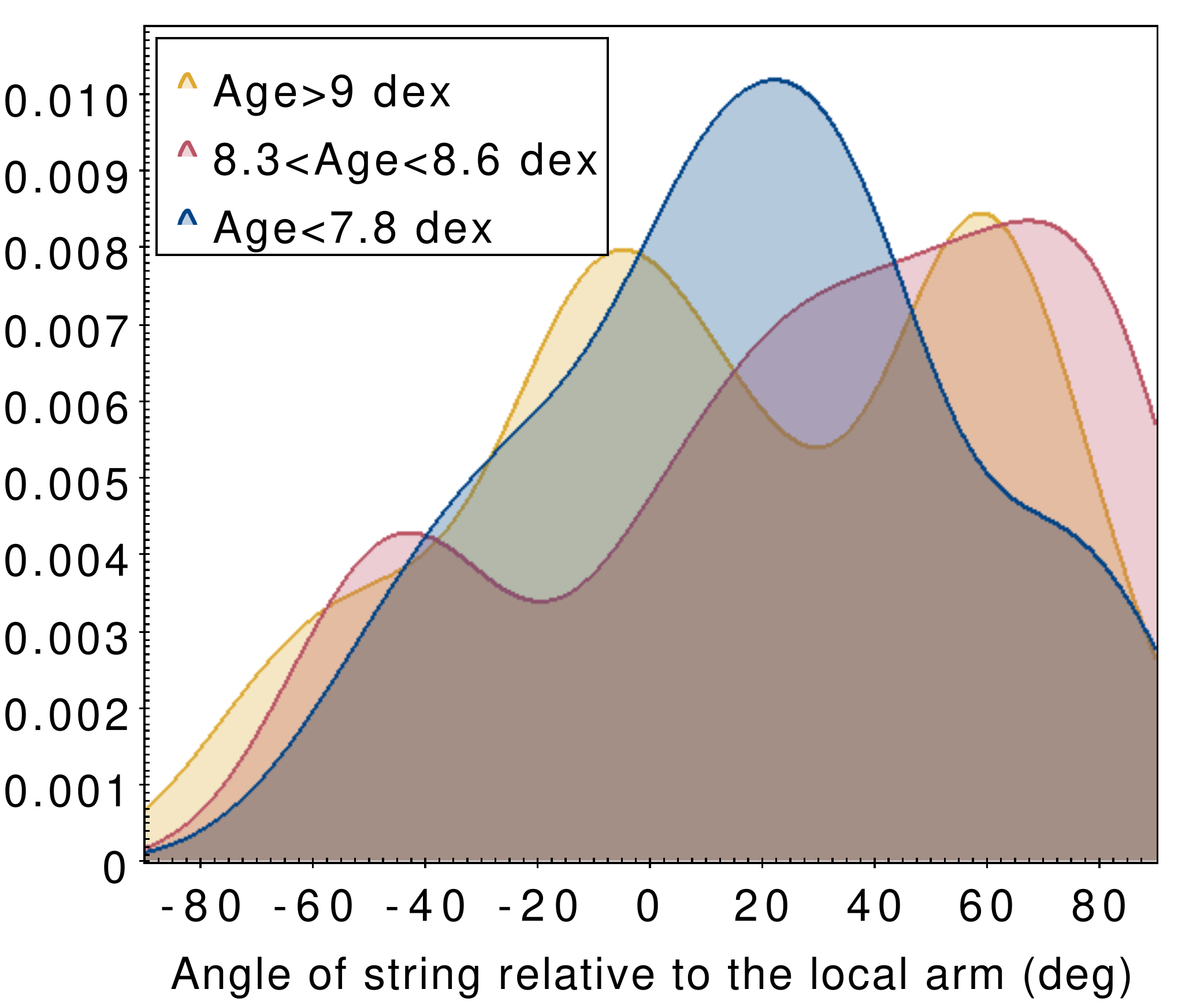}{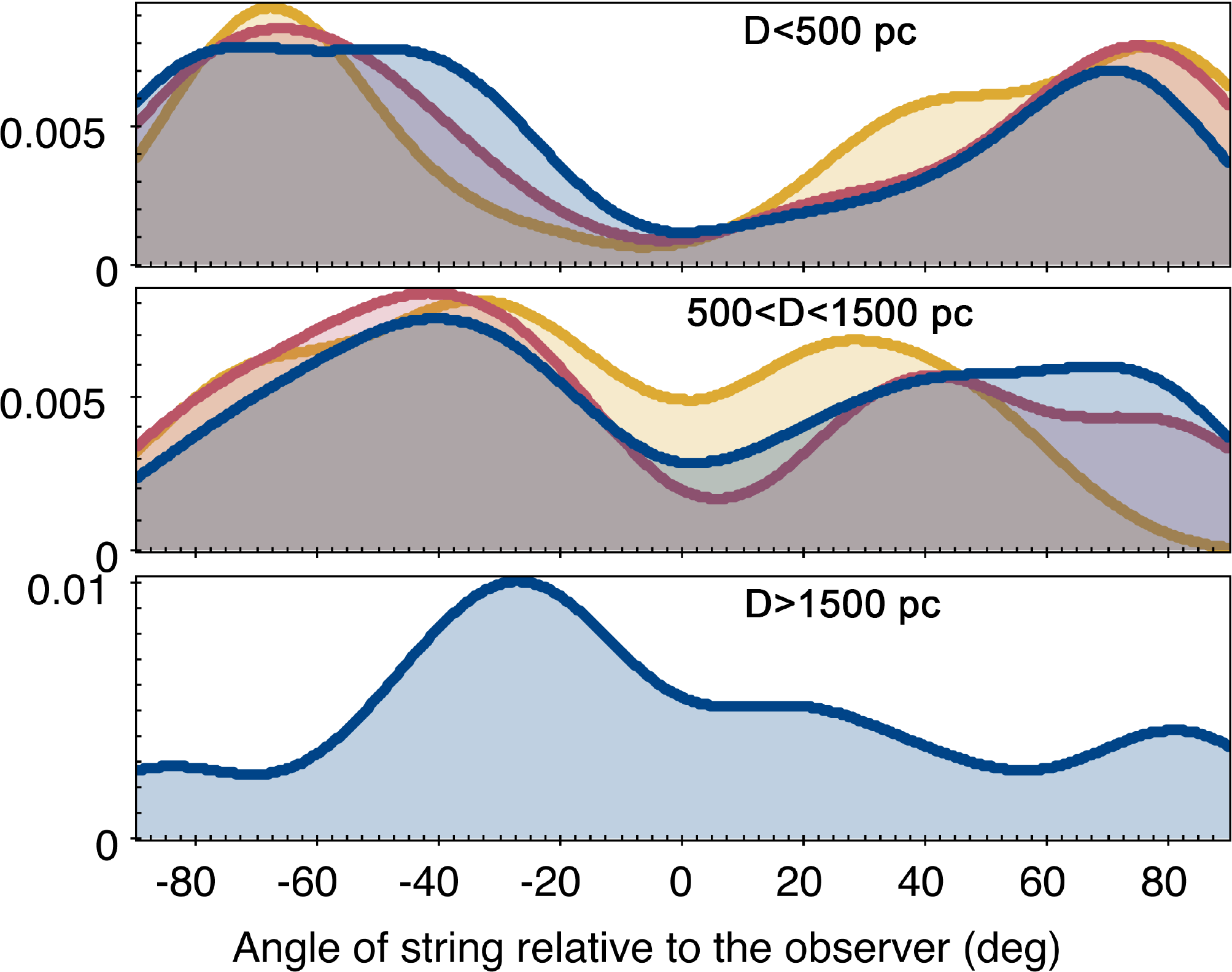}
\caption{Left: The kernel density estimate (KDE) of the angle of strings near Local Arm relative to the spiral arm trace from \citet{reid2019}. The angle of 0$^\circ$ is perpendicular to the arm, $\pm90^\circ$ is along it. We note that the strings in the two older age bins are not necessarily related to the Local Arm in its current form; the comparison is done primarily to highlight their relative orientation to each other and to the strings at other ages. Right: The relative orientation of the identified strings obtained through linear fitting relative to the observer as a function of age and distance. Angle of $\pm$90$^\circ$ refers to the strings located in the plane of the sky, with preferentially tangential orientation, 0$^\circ$ refers to radial orientation. \label{fig:angle}}
\end{figure*}

\subsection{Orientation of Strings}

As in \citetalias{kounkel2019a}, we have found that the young strings prefer an orientation that is close to perpendicular to the Local Arm (Figure \ref{fig:angle}, left), which has implications for their origins. Although there may be some evidence for this behavior towards the Perseus and Sagittarius arms, it is difficult to state conclusively due to their highly incomplete census. There are observational effects that could in principle mimic this behavior, such as ``Fingers of God" pointed radially along the observer's line of sight due to parallax/distance errors. However, as shown in Figure \ref{fig:angle} (right), tangential orientation of strings in the plane of the sky relative is slightly preferred within the closest few 100 pc (as they are on average $\sim$200 pc long, this can be expected). Beyond that, up to 1 kpc, averaging across all ages, there is no preferred orientation of strings relative to the observer, only to the other strings in the similar age bin. Beyond 1 kpc, there is a preference for radial orientation. It is difficult to say if it is due to greater uncertainty in distance or due to them being oriented perpendicular to Sagittarius and Perseus spiral arms. As we are able to recover strings of only populations $\lesssim$100 Myr, this would also be preferred orientation in the latter scenario. In future data release from \textit{Gaia}, with and improvement in astrometric measurements (Gaia EDR3 is expected to improve the parallaxes by 20\%), it would be possible to better disambiguate orientation of the more distant strings from the parallax uncertainty.

\subsection{In Situ String Formation}

Our findings suggest that each generation of stars that was formed inside the region associated with the Sagittarius arm since then has been progressively forming closer towards the Galactic center over the last 100 Myr. It should be noted, however, that there is no strong systematic trend in the velocity vector corrected for the average Galactic rotation in any particular direction, along U, V, or W (Figure \ref{fig:sagittarius}). The average U may be 2--5 \kms\ towards the Galactic anticenter in several age bins. If we assumed that all of the generations of stars have formed where the current epoch of star formation is, then this is the direciton in which they would need to migrate over time. However, the populations in the 40--60 Myr age bin show the average U $\sim$3 \kms\ in the opposite direction, towards the Galactic center, despite being located, on average, between 20--40 and 60--80 Myr populations. Furthermore, the overall velocity distributions are broad. For these reasons, deprojecting the position of the populations backwards in time (with velocities corrected for the Galactic rotation and local standard of rest) would not result in them being more compact, or appear to originate from a more similar position. Thus, this change in position cannot necessarily be attributable to the young stellar populations inside the arm moving or migrating significantly away from their initial position after their formation. As populations form, they remain roughly comoving in the reference frame of the Galaxy. Although as they evolve and the velocity dispersion increases, they do scatter away from their position of birth, the direction is random. Furthermore, many of them lack the speed to travel the necessary distances in the given period of time. It is impossible to reproduce the current distribution with stellar dynamics alone. Rather, the motion is most likely be attributable the density wave itself. The spiral arm density wave then triggers the molecular cloud formation, with clouds continuously forming at a slightly different position with respect to the Galactic center in the volume of space covered by the data. Then from these clouds young populations from which we trace the spiral arms then forms. If there is any motion with respect to Y axis as well, it is difficult to conclusively measure due to the overall shape of the spiral arms.

We note that there is no such similar displacement of either Local or Perseus arms. The corotation radius of stars to be comoving with respect to the spiral arms is expected to be comparable to the Solar radius \citep{dias2019}, thus, a lack of evolution in position of the Local arm may be expected. However, the separation from the Perseus arm to the corotation radius is comparable to the separation from the Sagittarius arms. Thus, the change in position of only one of them but not the other is noteworthy.

\section{Conclusions}\label{sec:concl}

In this paper we present a sample of 987,376 stars that can be clustered in \textit{Gaia} DR2 catalog to reveal 8292 comoving groups. Although individual membership would require further vetting, particularly in the more distant populations, their bulk allows for a detailed look at the temporal evolution of the structure of the Galaxy.

Such analysis is only possible if the ages of all the stellar populations can be derived in a robust and uniform manner. For that purpose, we developed a neural network, Auriga, that predicts population properties based on the input \textit{Gaia} and 2MASS photometry of the members. The advantage of a neural network compared to the traditional isochrone fitting is the speed (the entire sample is processed simultaneously instead of one population at a time, deriving parameters for the entire catalog in this work in only a few minutes). It also doesn't require initial estimates for the parameters of any of the populations, which also significantly automates the process. Finally, outside of constructing the initial training set, deriving ages using neural networks is agnostic to any potential differences between the synthetic isochrones and the real data beyond what is necessary to produce an initial training set. As more and more coherent stellar populations are found in \textit{Gaia} catalog, far exceeding the number of clusters that can be carefully visually inspected, neural networks like Auriga are ideally suited for deriving parameters of these populations.

This work builds on \citetalias{kounkel2019a}, expanding the distance reach from 1 kpc up to the boundary defined by the extinction map in the magnitude-limited sample. This boundary is age dependent: youngest ($<8$ dex) populations are complete up to the distance $>3$ kpc (Figure \ref{fig:3d}). Older populations lack bright high mass stars, and thus it is more difficult to recover them in the areas of high extinction. No old ($>9$ dex) structures are found beyond 1 kpc along the Galactic plane, although they can extend up to 2 kpc at higher galactic latitudes.

Analyzing the highest confidence sample, accounting for the completeness limits, we find that:
\begin{itemize}
  \item The Local arm in the current epoch of star formation has a finite size, not extending to the completeness limits of the survey.
 \item The nearby portion of the Perseus arm has experienced a lull in the star formation activity over the last 25 Myr. Most of the populations along it have an age of 25--60 Myr.
 \item Unlike the Perseus arm, position of which has remained stable over time, the Sagittarius arm has been continuously shifting closer towards the Galactic Center in the last 100 Myr years, having been displaced by more than 500 pc during that time. This corresponds to the velocity of $\sim$8--10 \kms. This effect cannot be accounted by stellar migration, rather, it appears to be an evolution of the position of the arm itself.
 \item A number of youngest populations in the Local Arm are positionally consistent with the Radcliffe Wave - a recently discovered 2.7 kpc gaseous structure, that oscillates from the Galactic plane with an amplitude of 160 pc. Many such populations along it tend to be less extended, more compact than what is typically observed in other young populations.
 \item The ripple in the disk is not limited to just the Radcliffe Wave, but exists among both the young ($<8$ dex), and the old ($>9$ dex) populations.
 \item Similarly to \citetalias{kounkel2019a}, we find evidence of the Local as well as Sagittarius and Perseus Arms becoming distinct $\sim$ 100 Myr ago. At the older age bins there are other overdensities and gaps in the distribution of the identified populations that may be attributed to the older spiral arms.
 \item The relative position of these overdensities resulted in a current apparent lack of populations older than 1 Gyr in the solar neighborhood or towards the galactic anticenter.
 \item There is a peculiar heliocentric stream, elevated at $\sim25^\circ$ above the Galactic plane, extending to $\sim$1--2 kpc in distance. It is unclear what the origin of this stream may be.
 \end{itemize} 

\appendix

\restartappendixnumbering
\section{Auriga Neural Network for structure parameter determination}\label{sec:auriga}

\subsection{Background}

With the discovery of increasingly larger number of star clusters and associations with \textit{Gaia}, it becomes increasingly more important to use isochrone fitting to determine their ages to fully understand the dynamical evolution of the Galaxy. Unfortunately, increased sample makes commonly used techniques quite impractical. Some studies frequently rely on "chi-by-eye" estimates (and thus may be prone to produce erroneous and non-optimal measurements) as few robust fitters are available. The ones that are available may be strongly depended on the initial guesses of the fitted parameters, the corresponding choice of priors. They may struggle to converge, may be prone to overfitting and underestimating the uncertainties, slow to run, and require careful vetting of final results. Finally, the isochrones themselves may not necessarily offer the perfect representation of the actual stellar photometry - e.g., the low mass stars may appear to be inflated compared to the isochrones \citep{jackson2018}. Although these systematic discrepancies can be managed in the fitting process, they could add an external source of error. Neither approach is optimal when dealing with a large number of populations in a manner that could be considered uniform.

To date, the two largest and most commonly used catalogs of cluster parameters are \citet{dias2002} and \citetalias{kharchenko2013}. For a number of clusters, the parameters are copied between two catalogs verbatim. For the ones that were not duplicated, the dispersion in measured ages is 0.57 dex.

More recently, other studies have been conducted to rederive cluster ages in bulk. \citet{liu2019} have developed an autonomous pipeline for that purpose, estimating ages and distances (but not \av) for 2443 clusters, however, it often obtains incorrect ages, forcing many to the lower limit of the isochrones used, $<$10 Myr, even when the clusters themselves appear to be evolved. \citet{sim2019} have also performed isochrone fitting for 665 stellar groups, of which 188 reliably overlap with previously known clusters. Unfortunately they did not provide membership lists used in their work, nor their fitting algorithm. Furthermore, \citet{bossini2019} have used BASE-9 \citep{base9}, deriving ages, distances, and extinctions (and corresponding uncertainties) for 269 clusters, producing a rather clean, albeit somewhat limited, sample. The consistency in these three aforementioned works is within 0.3 dex -- a factor of 2 better than \citet{dias2002} and \citetalias{kharchenko2013}.

In \citetalias{kounkel2019a} we have attempted to use neural networks in order to derive parameters of stellar populations. An advantage of deep learning is that a fully trained neural network is very fast, capable of processing large volumes of data in a very short period of time in a uniform and repeatable manner, memorizing and constructing a function for the complete data model, instead of fitting each individual object separately. However, it requires a large training set that spans a full parameter range across which it would be possible to interpolate against, and this training training set has to be representative of the data in question. In \citetalias{kounkel2019a} we used \citetalias{kharchenko2013} catalog supplemented by synthetic cluster photometry, but that was proven to be insufficient, producing acceptable predictions of age and \av\ in less then half of all stellar populations, still requiring to use a fitter on top of it. That exercise did result in a large sample of stars, in conjunction with the above catalogs, that could now be used as a training sample in another improved iteration of the neural network.

\subsection{Sample \& Training}

To assemble the training sample, we took the catalog from \citetalias{kounkel2019a}. We deredenned it using the $A_V$ estimates from that work, using the transformations of $A_{BP}/A_V=1.068$, $A_{RP}/A_V=0.652$, $A_{G}/A_V=0.859$, $A_{J}/A_V=0.288$, $A_{H}/A_V=0.178$, $A_{K}/A_V=0.117$ \citep{marigo2017}. Furthermore, we converted the apparent magnitudes to absolute using \textit{Gaia} parllaxes.

Due to proximity of all the sources inside it, resulting in a smaller relative flux error, it is more optimal for flux manipulation than other catalogs. We grouped all stars in age bins of 0.1 dex, mixing different populations together. Each bin would then be placed at a random distance, of up to 20 kpc, reddened up to 10 \av, and noise in all the measurements was applied. Sources fainter than the resulting $G>18$ mag were discarded (this propagates to the faint magnitude limit of the synthetic population consistent with that of the unaltered data in the other bands as well ), and 250 of remaining stars were randomly chosen to create a set. This exercise was repeated 1000 times for each age bin, resulting in x random artificially constructed populations. All the data were arranged into 7$\times$250 tensor for each population, containing $G$, $G_{BP}$, $G_{RP}$, $J$, $H$, and $K$ magnitude, as well as parallax of all 250 stars, ordered by $G$ mag, and all the inputs were normalized in such a way that they fall between -0.5 to 0.5 range.

This approach is more optimal than using isochrones to generate synthetic cluster photometry, as it does not introduce systematic offsets in flux from incompatibilities between the model in the data. The differences between the real photometry and synthetic photometry may be slight and can be understood by humans as representing a population of similar age. Nonetheless, due to how the neural networks interpret the data, pure synthetic photometry can add odd artifacts in the training process if the goal is to this model on the real data.

The augmented sample, however, is essential for the data model produced by the neural network to achieve better convergence compared to the sample produced only from real clusters. But, this sample still ends up being too ``clean'', and the reddening laws that have been used may be imprecise for wider bands without additional corrections in temperature \citep{a¡nders2019}. Thus, although this sample is useful to the network to better learn the underlying trends, to teach it what actual data look like, the arteficial sample was supplemented with photometry from real clusters. We crossmatched the catalog from \citetalias{cantat-gaudin2018a} with the cluster parameters from \citet{bossini2019}, \citet{liu2019}, and \citet{kharchenko2013}, and used the resulting set was added to the unmodified catalog from \citetalias{kounkel2019a}. For the parameters, distances from \citet{kharchenko2013} were replaced with the distances from \citetalias{cantat-gaudin2018a}. Any clusters younger than 10 Myr in \citet{liu2019} were excluded, and as this catalog does not include extinction, this parameter was duplicated from \citetalias{kharchenko2013}.

Similarly to the artificially constructed populations, 250 stars from each real cluster were chosen randomly. Although the photometry was mostly unaltered, all the inputs were scattered by the corresponding flux and parallax uncertainties. If a particular cluster has fewer than 250 objects, the same star can be drawn multiple times, with slightly different fluxes from the errors, and each cluster was represented 20 times. 15\% of the real clusters (including all of their realizations) were set aside for validation purposes, and were not included in training.

We note that when the network was trained on just the artificial populations, or just on the real clusters, the performance was worse than in the combination of the two. The real data did not have the necessary bulk, and was missing prominent parts of the parameter space, while the artificial data had minute inconsistencies with the real photometry. By combining the two, it was possible to overcome limitations of either dataset, and achieve a more general solution.

The neural network that we refer to as Auriga was constructed in Pytorch \citep{pytorch}, predicting age, \av, and a log of distance. We experimented with various model architectures. The architecture that yielded the best loss was based on the mnist model\footnote{\url{https://github.com/eladhoffer/convNet.pytorch/blob/master/models/mnist.py}}, it resulted in a 20\% better convergence compared to the model in \citetalias{kounkel2019a}, or compared to any other model we explored. The training used stochastic gradient descent to minimize mean squared error (MSE) loss, stopping when the loss in the real cluster data has converged, and it used Adam optimizer with the learning rate of 1e-4.

Neural networks themselves do not generate the uncertainties in predicted parameters. However, it is possible to use a method described in \citet{olney2020} to estimate them. The inputs that were scattered by the errors are statistically comparable, but the neural network treats each realization of the same dataset as independent, producing slightly different solution for each one. This makes it possible to measure scatter between them.

\subsection{Validation} \label{sec:aurigavalidation}

\begin{deluxetable}{ccccc}
\tabletypesize{\scriptsize}
\tablewidth{0pt}
\tablecaption{Parameters of clusters from \citet{cantat-gaudin2018a}\tablenotemark{a}  \label{tab:clusters}}
\tablehead{
\colhead{Cluster} &\colhead{Theia\tablenotemark{b}} & \colhead{Age}& \colhead{\av}& \colhead{Dist.}\\
\colhead{} &\colhead{ID} & \colhead{(dex)}& \colhead{(mag)}& \colhead{(pc)}
}
\startdata
ASCC\_10 &557& 8.27$\pm$0.11 & 0.80$\pm$0.05 & 664.1$\pm$15.4\\
ASCC\_101 &445& 8.60$\pm$0.06 & 0.21$\pm$0.04 & 394.6$\pm$7.1\\
ASCC\_105 &544& 8.17$\pm$0.10 & 0.56$\pm$0.07 & 563.5$\pm$17.6\\
\enddata
\tablenotetext{}{Only a portion shown here. Full table is available in an electronic form.}
\tablenotetext{a}{Includes clusters from \citet{cantat-gaudin2019a}, \citet{castro-ginard2019}, and \citet{castro-ginard2020}}
\tablenotetext{b}{Refers to the closest matched populations in Table \ref{tab:theia}. Note that multiple clusters can be associated with the same underling population}
\end{deluxetable}

\begin{figure}
\epsscale{1.1}
 \centering
		\gridline{
             \fig{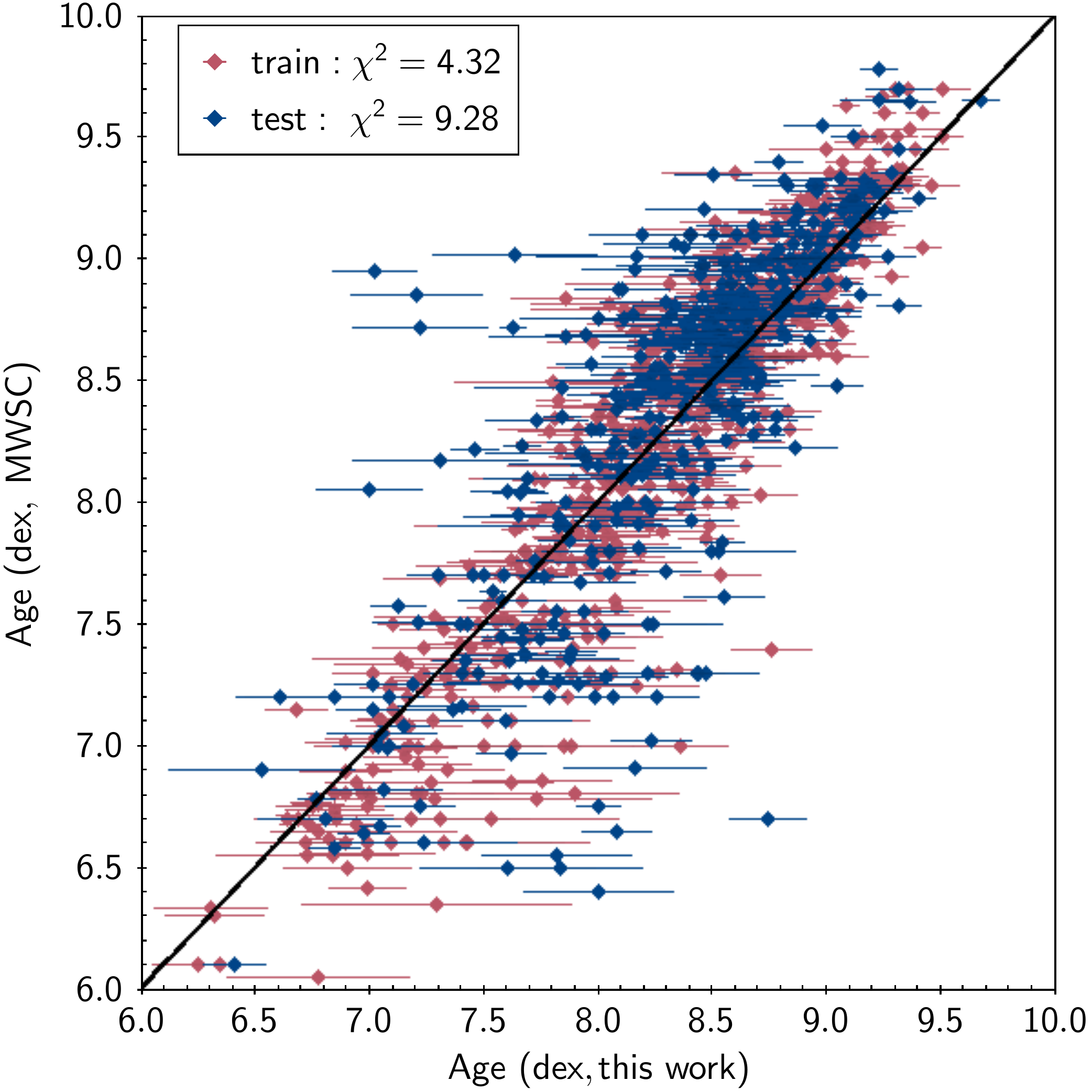}{0.25\textwidth}{}
             \fig{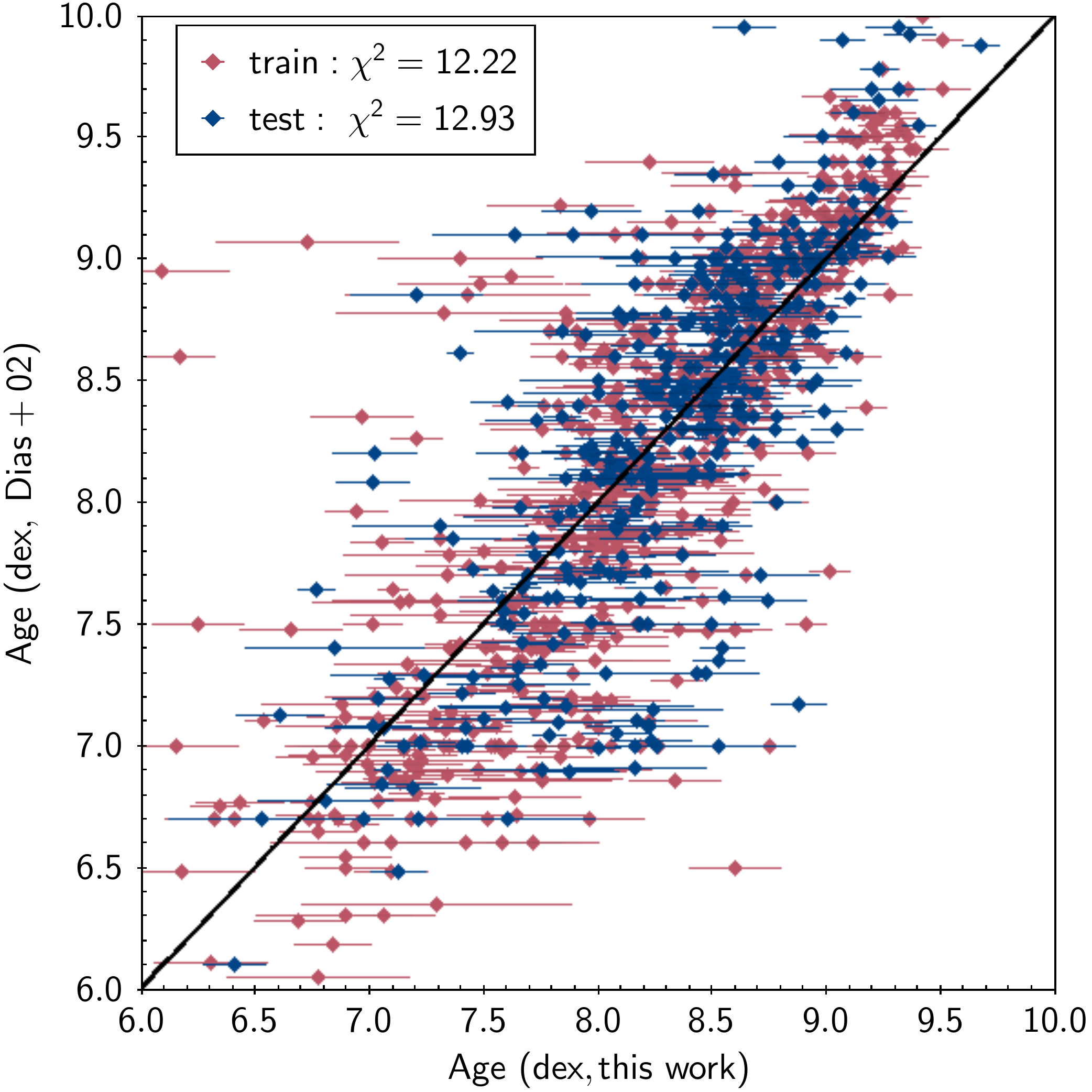}{0.25\textwidth}{}
        }\vspace{-1 cm}\gridline{
		\fig{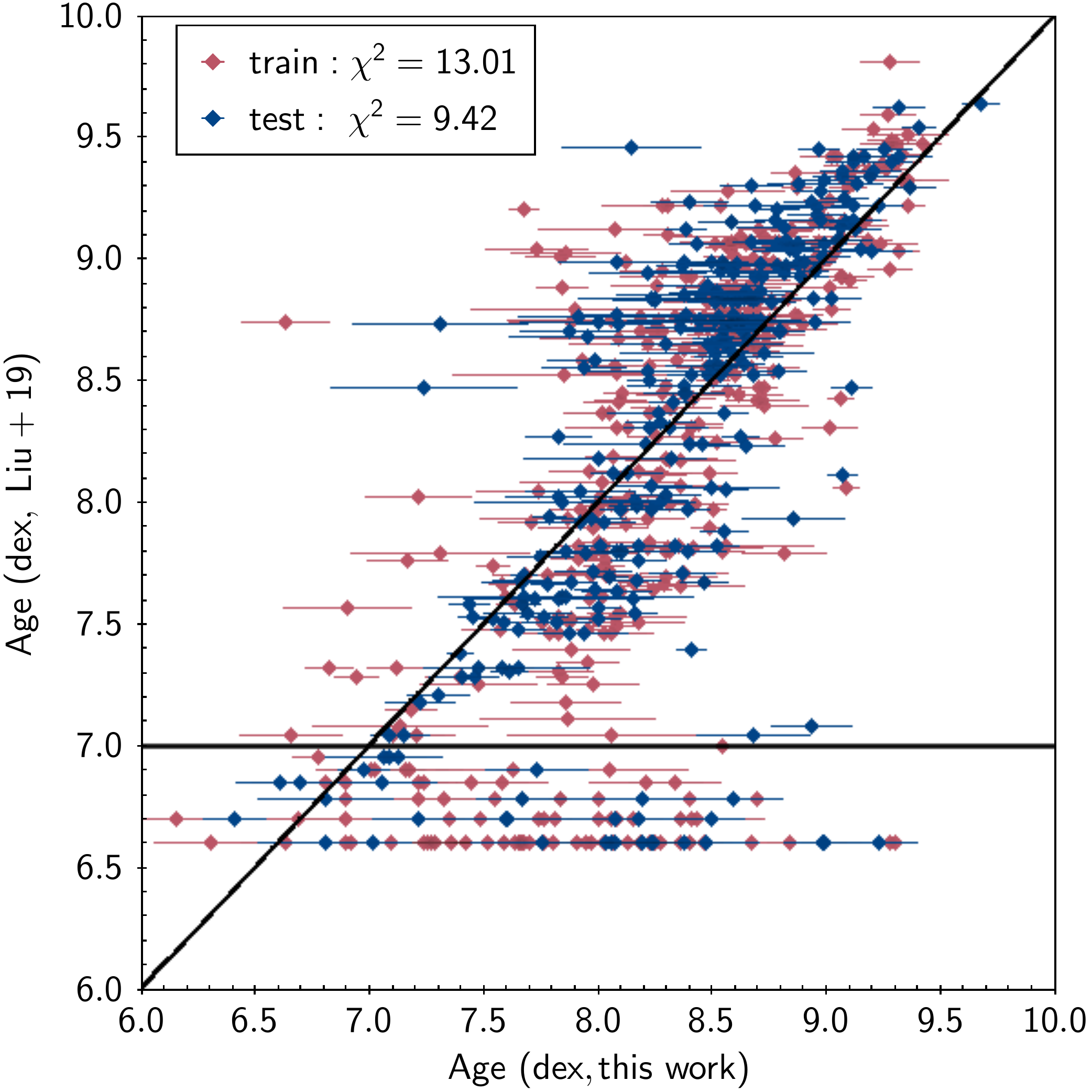}{0.25\textwidth}{}
             \fig{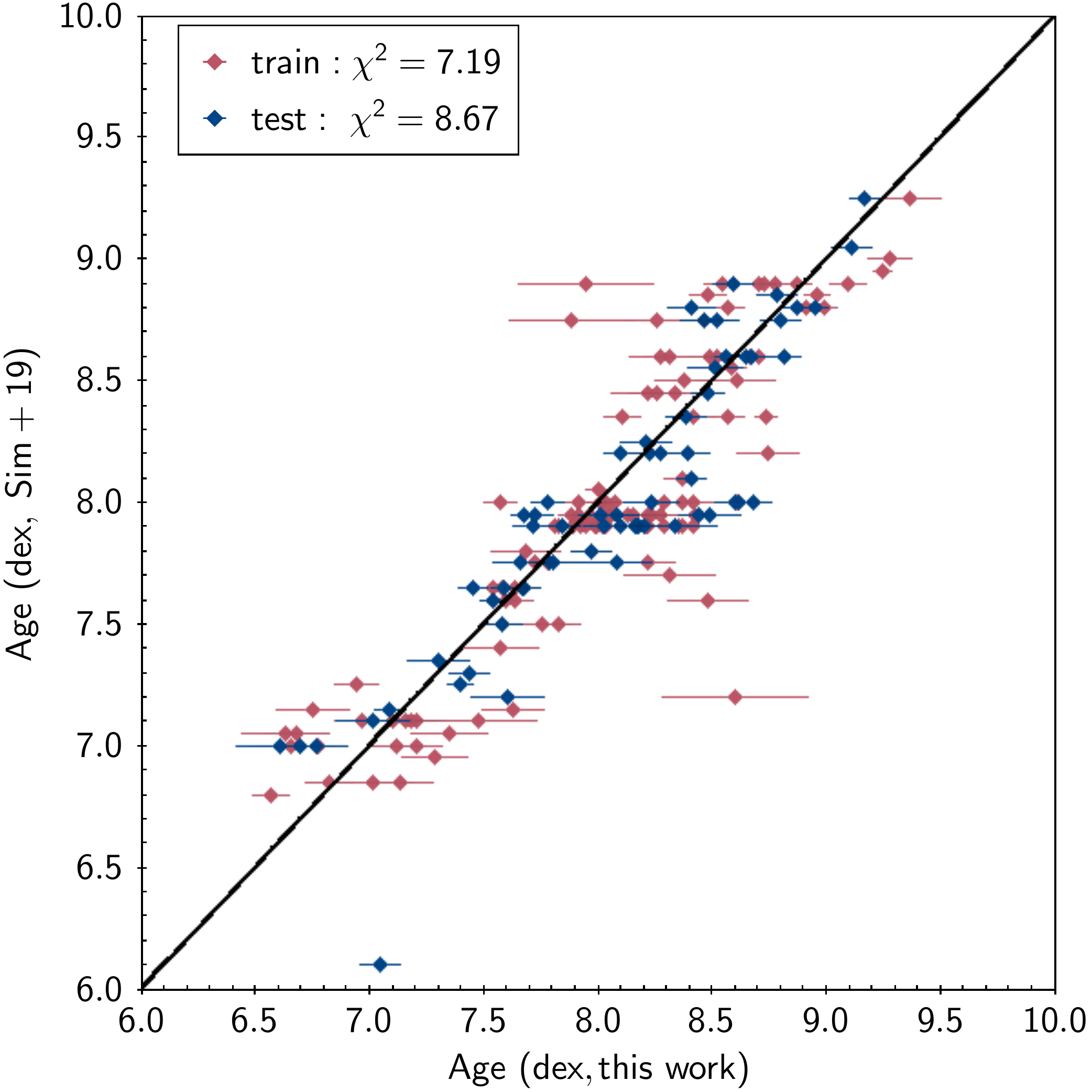}{0.25\textwidth}{}
             }\vspace{-1 cm}\gridline{
             \fig{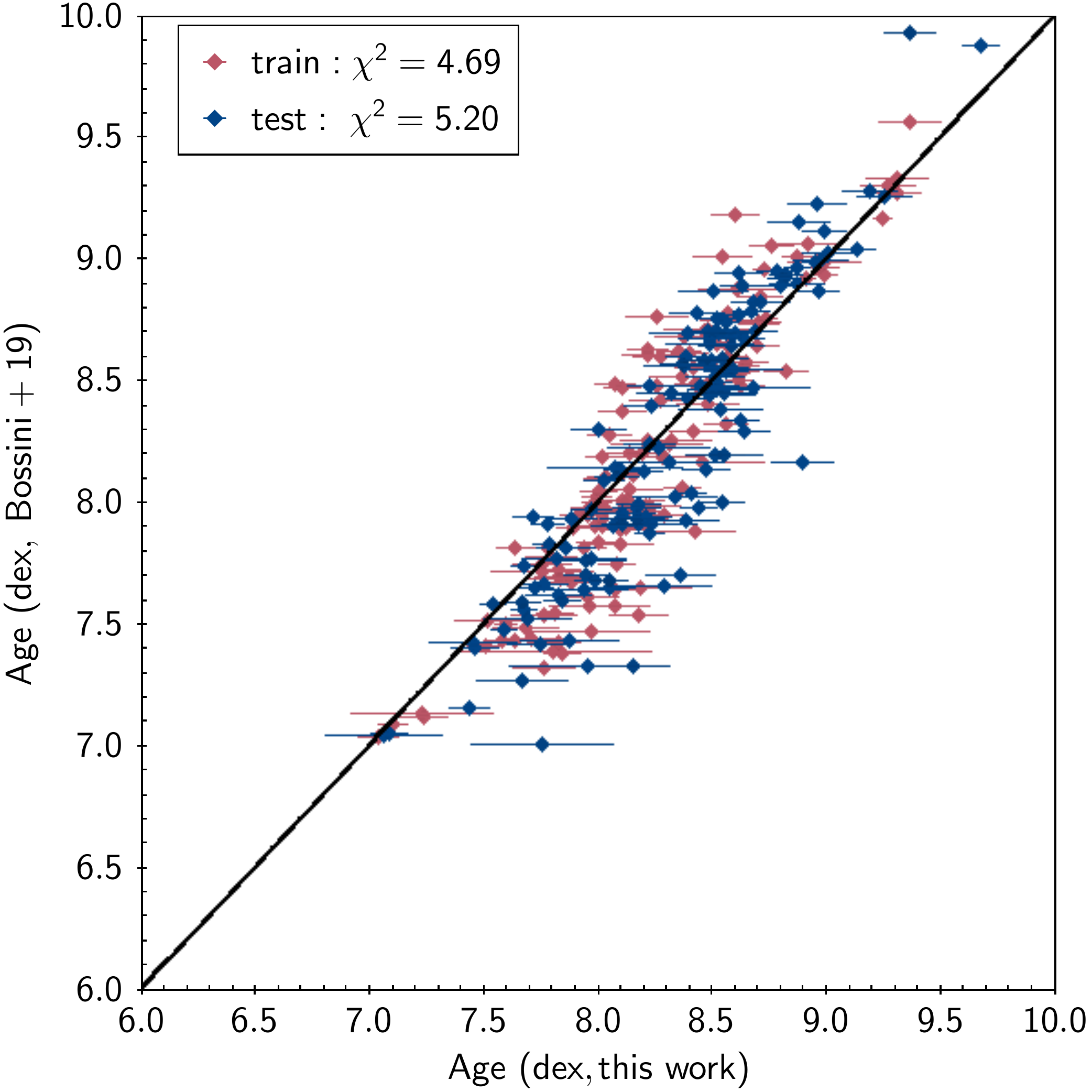}{0.25\textwidth}{}
             \fig{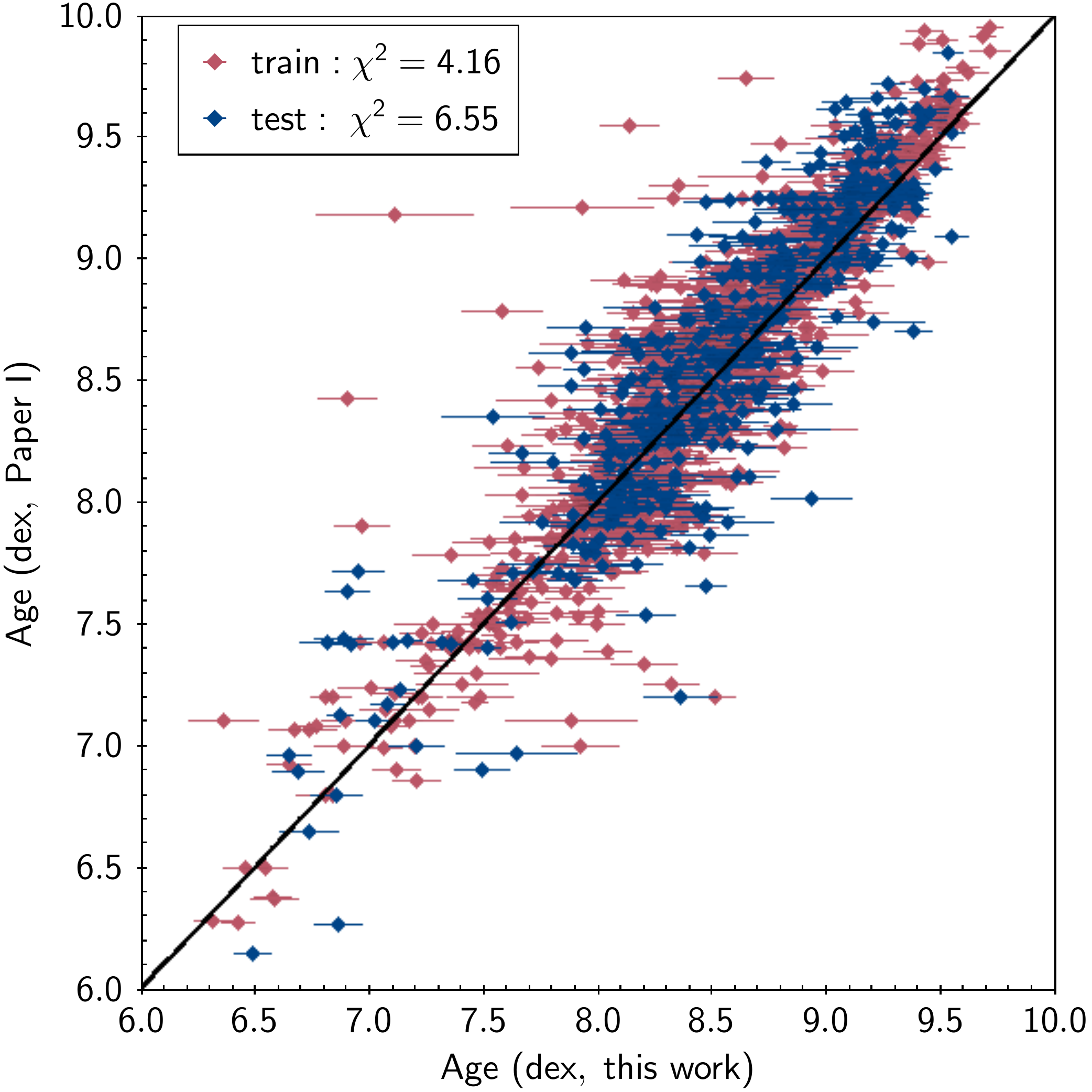}{0.25\textwidth}{}}
        \vspace{-0.75 cm}
\caption{Comparison of ages derived by the Auriga to those from other studies. The train sample is in red, and the test sample is in blue.
\label{fig:agecomp}}
\end{figure}

The trained Auriga model is made available on GitHub\footnote{\url{https://github.com/mkounkel/Auriga}}. The predicted parameters for clusters from \citetalias{cantat-gaudin2018a} are presented in Table \ref{tab:clusters}.

\begin{figure}
\epsscale{1.1}
 \centering
		\gridline{
             \fig{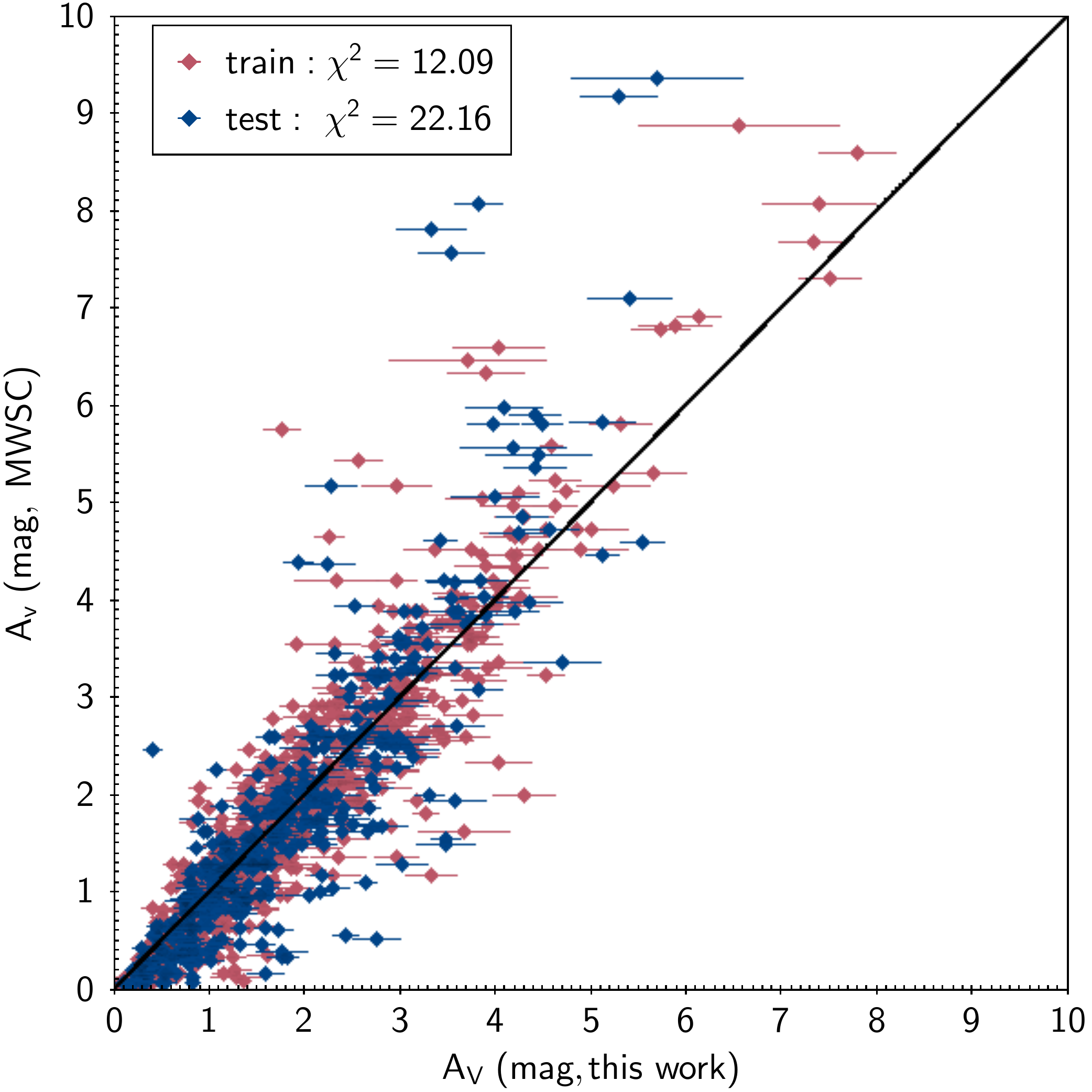}{0.25\textwidth}{}
             \fig{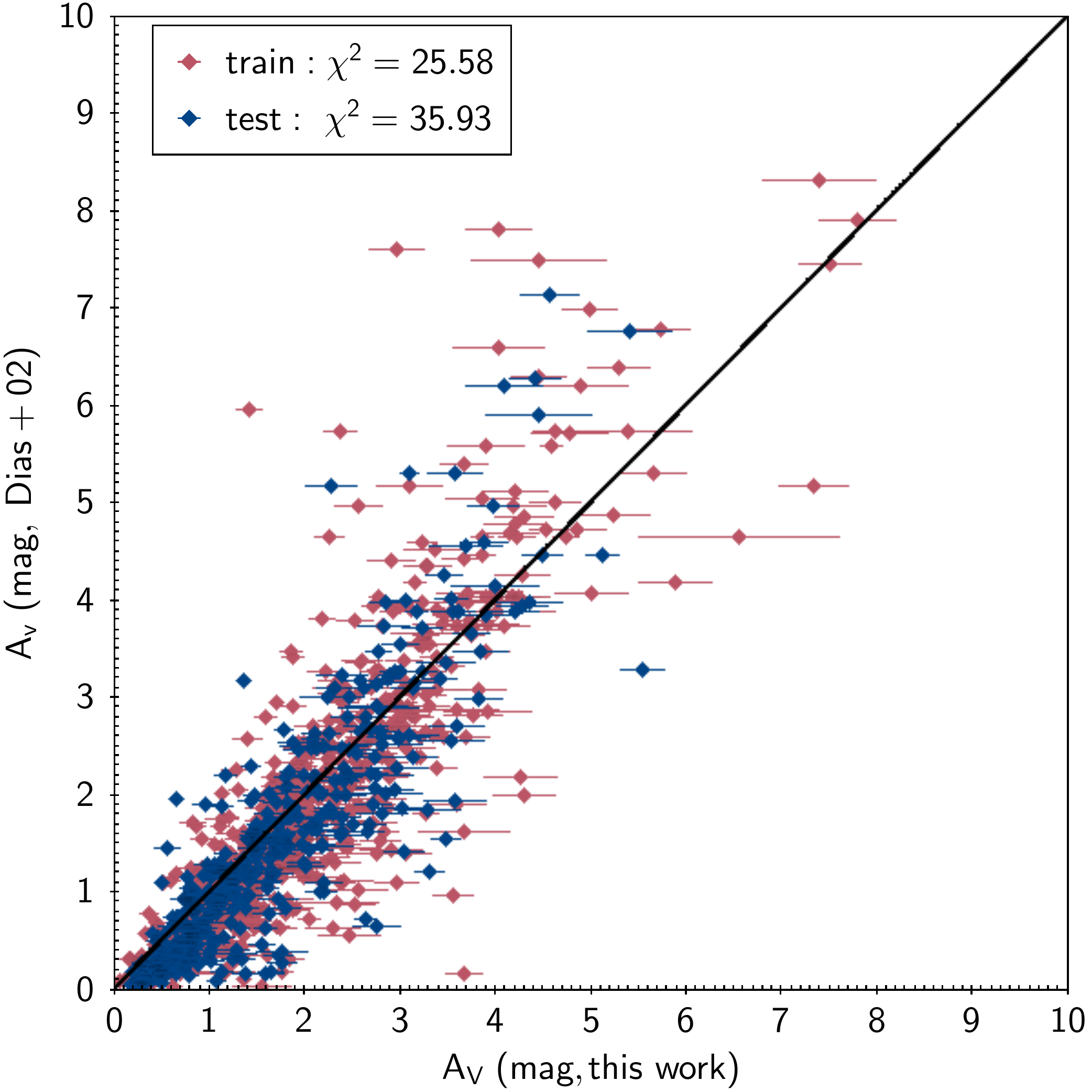}{0.25\textwidth}{}
        }\vspace{-1 cm}\gridline{
		\fig{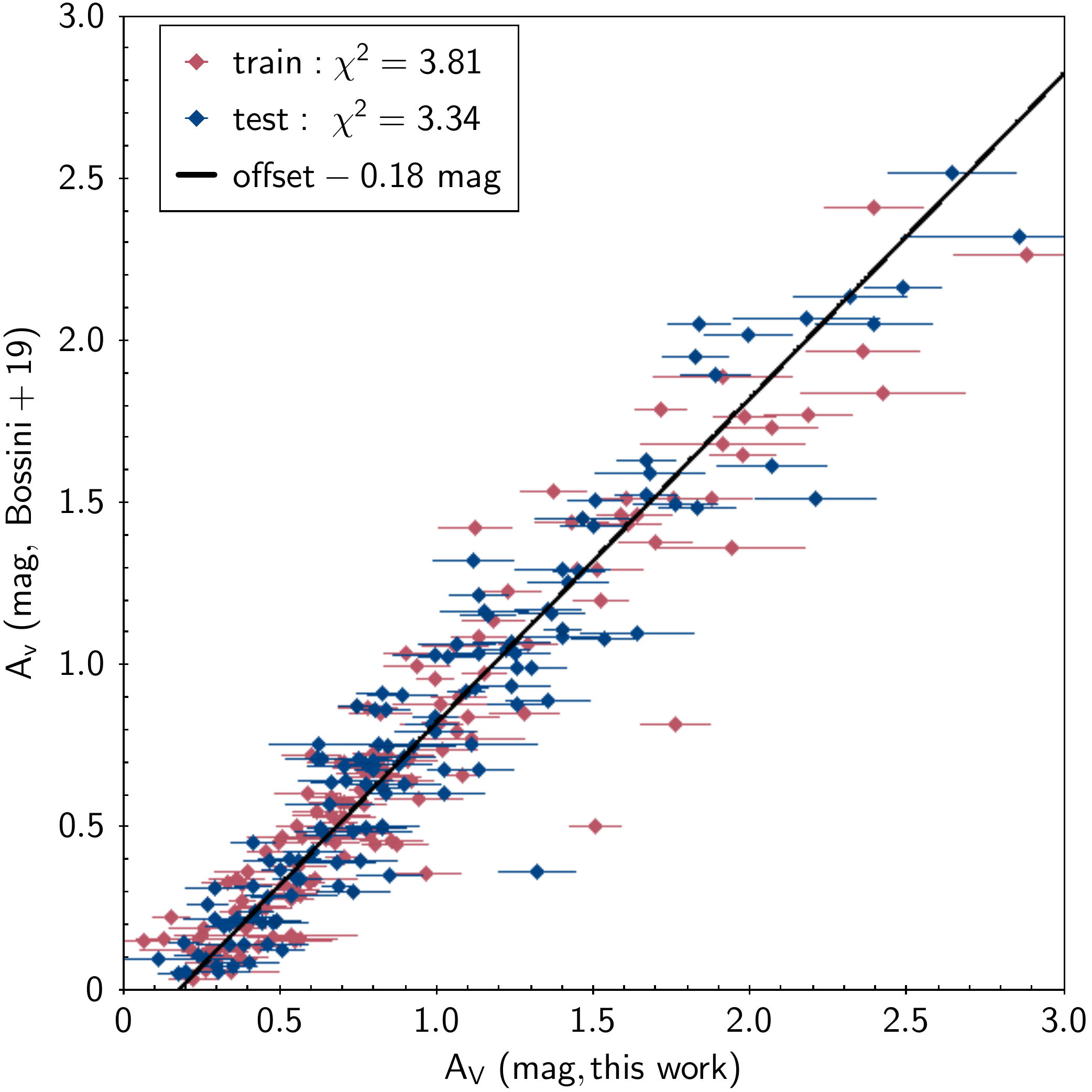}{0.25\textwidth}{}
             \fig{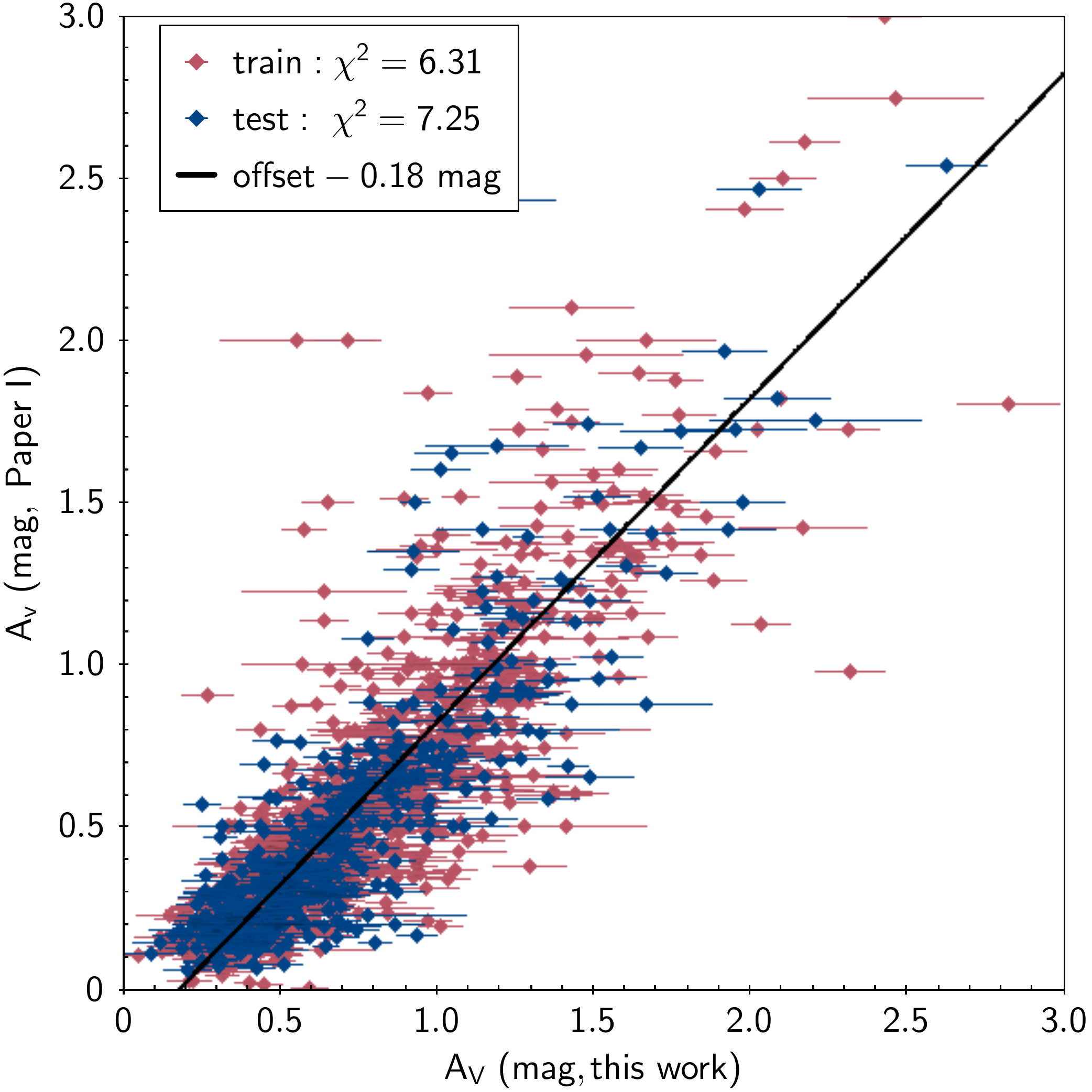}{0.25\textwidth}{}
             }
        \vspace{-0.75 cm}
\caption{Comparison of \av\ derived by Auriga to those from other studies. The train sample is in red, and the test sample is in blue.
\label{fig:avcomp}}
\end{figure}

\begin{figure}
\epsscale{1.1}
 \centering
		\gridline{
             \fig{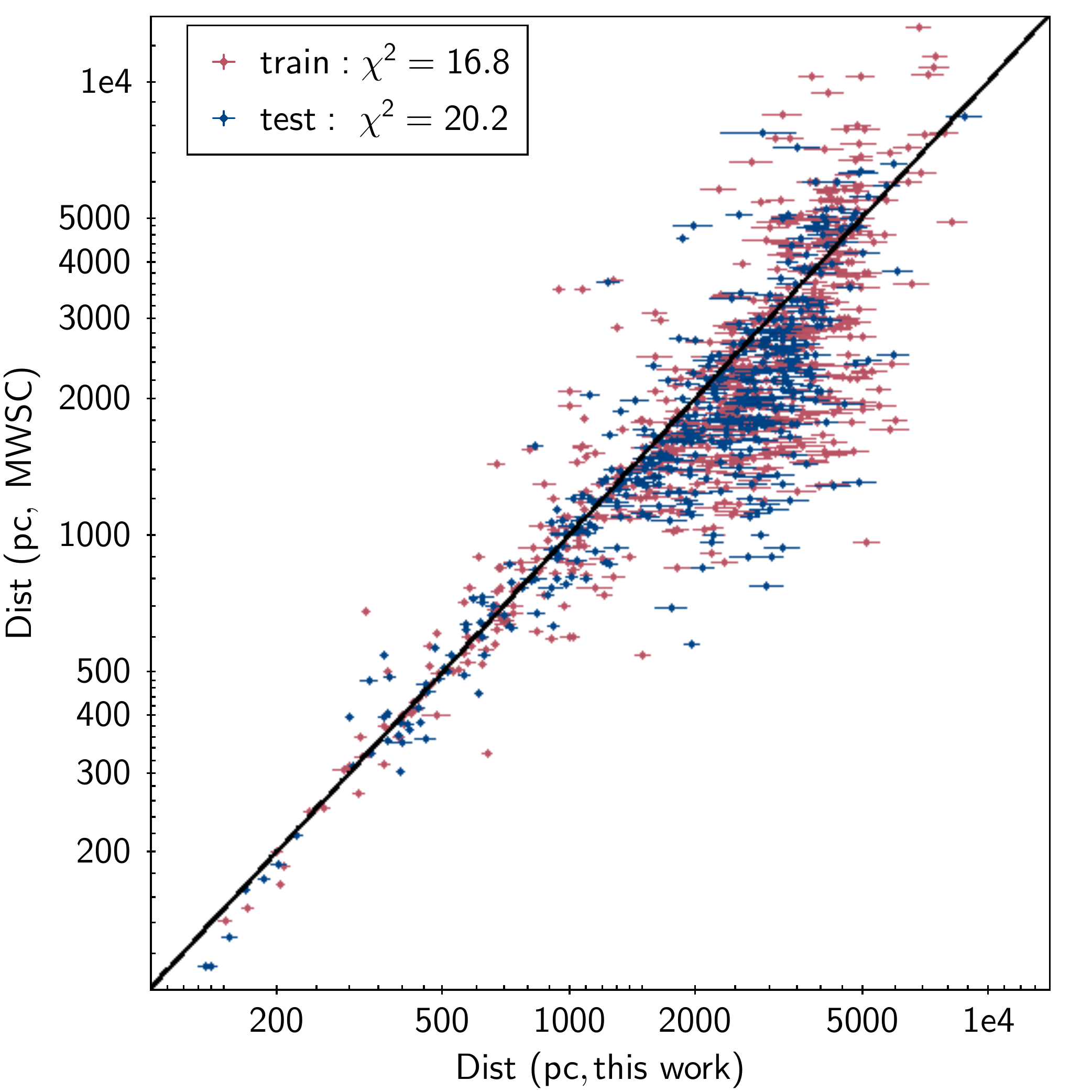}{0.25\textwidth}{}
             \fig{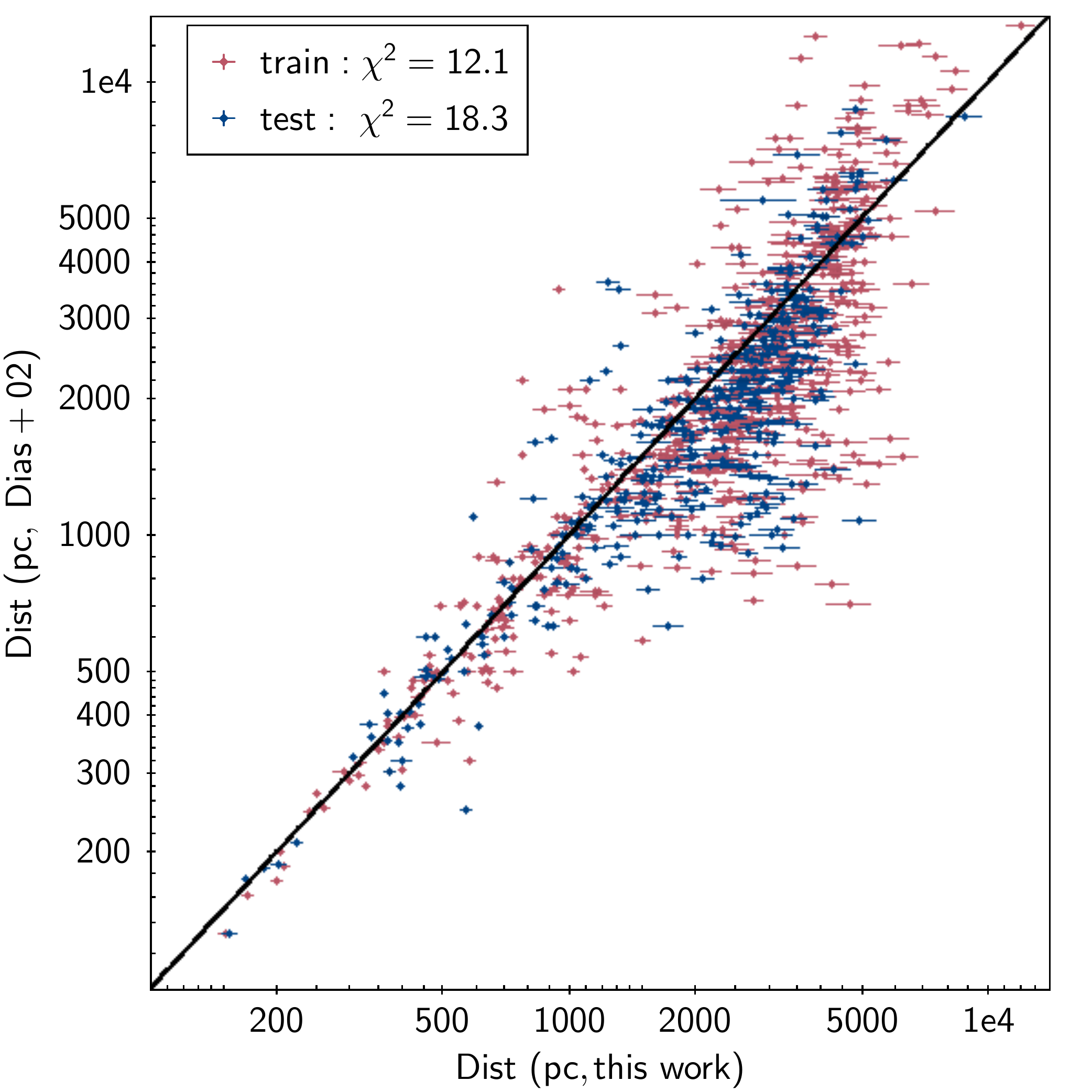}{0.25\textwidth}{}
        }\vspace{-1 cm}\gridline{
		\fig{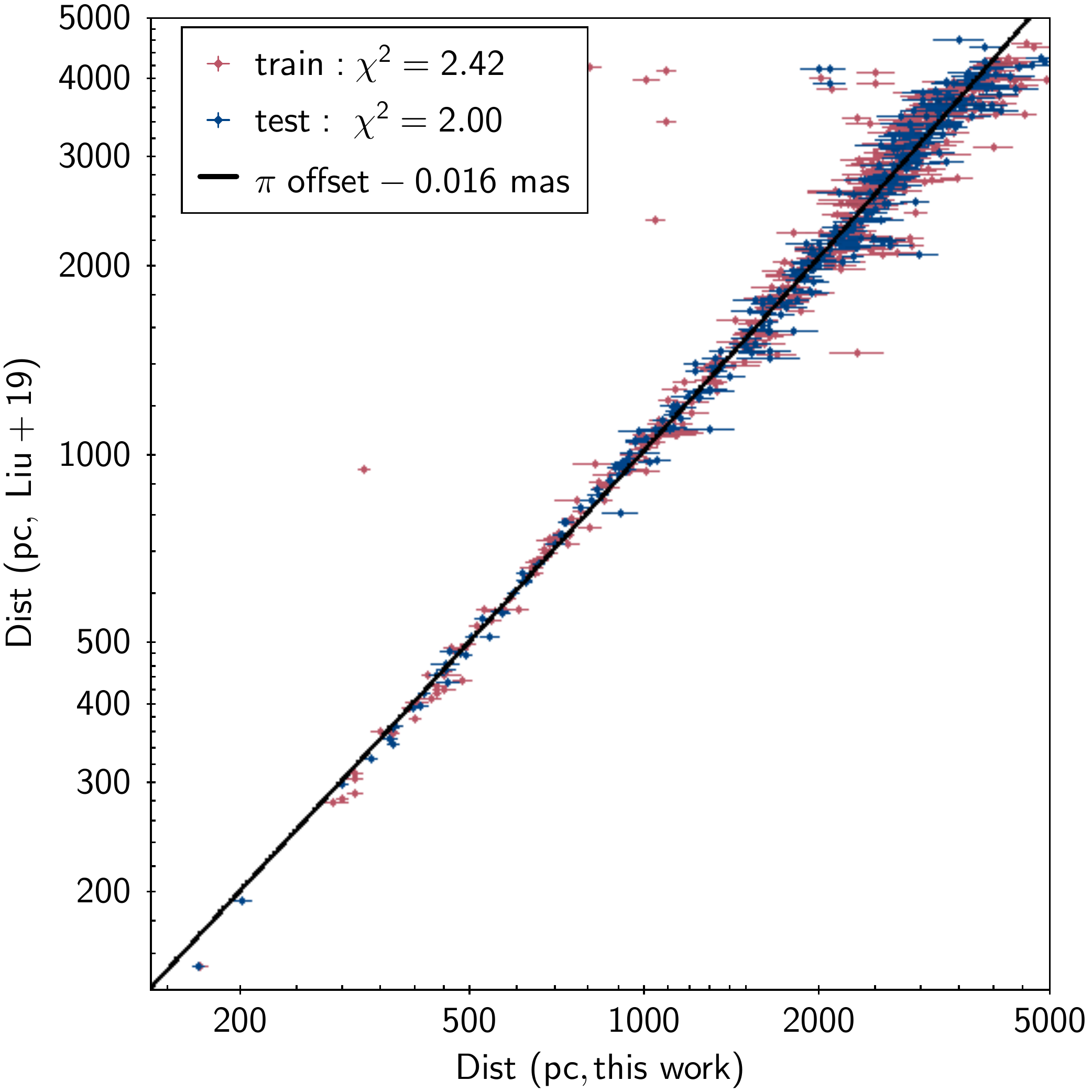}{0.25\textwidth}{}
             \fig{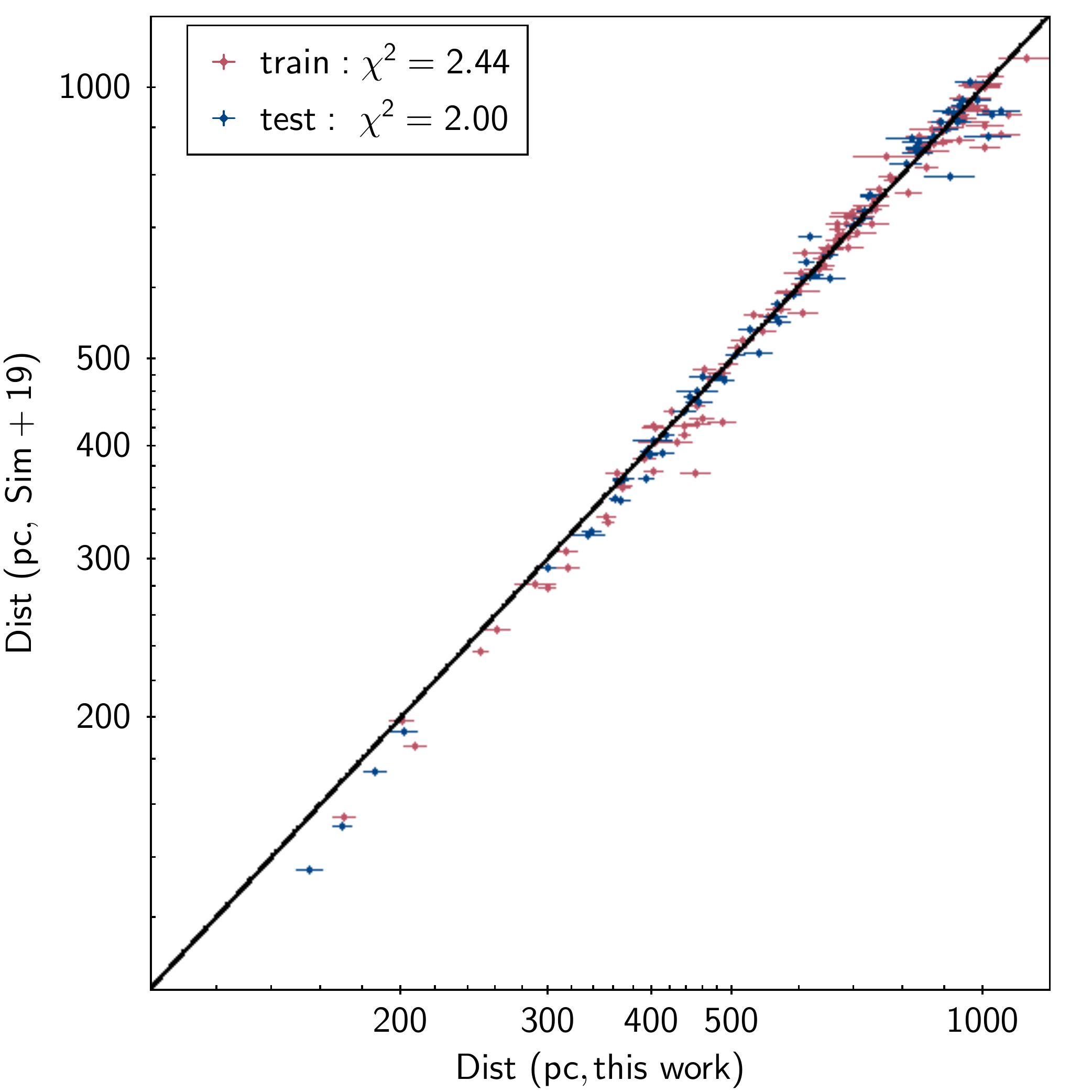}{0.25\textwidth}{}
             }\vspace{-1 cm}\gridline{
             \fig{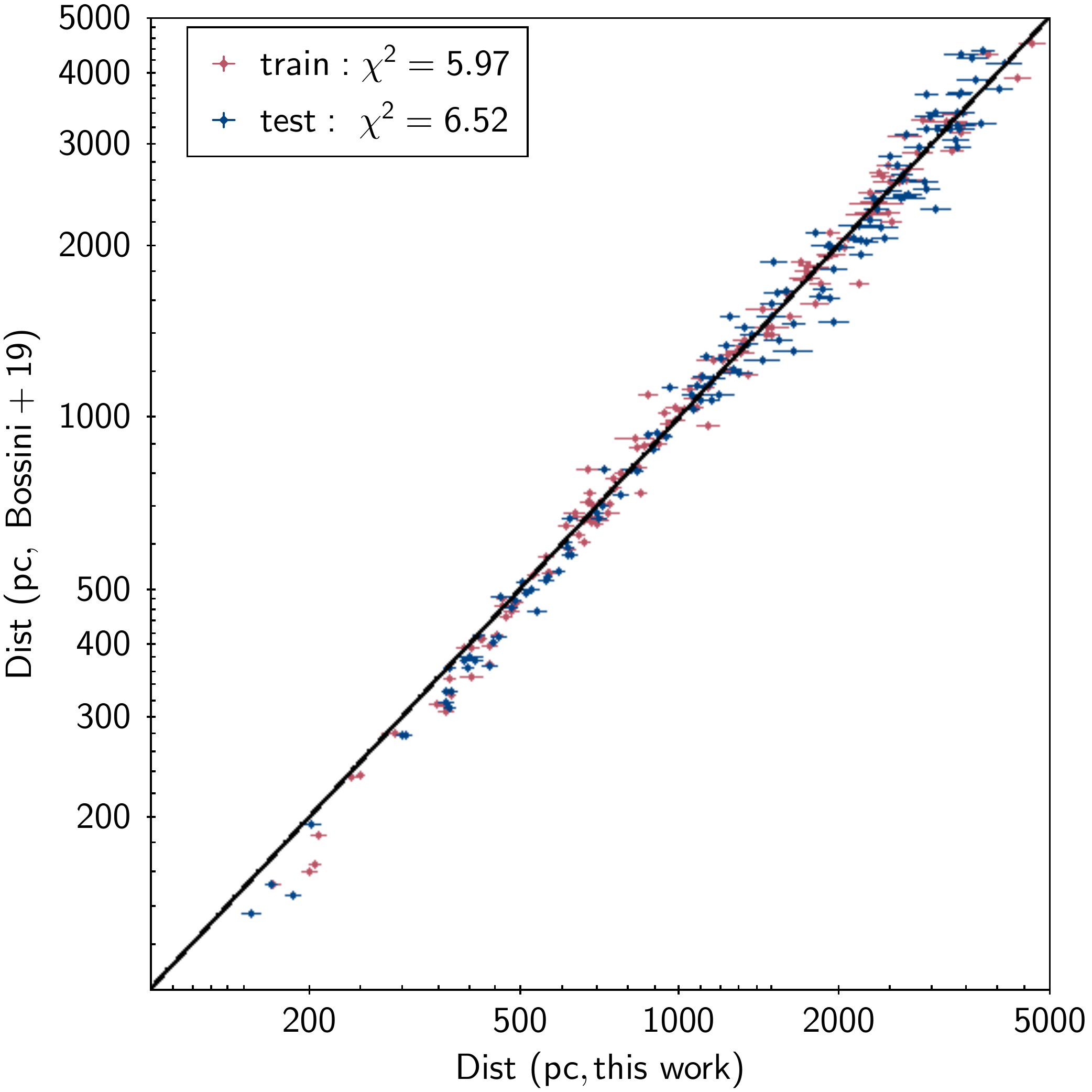}{0.25\textwidth}{}
             \fig{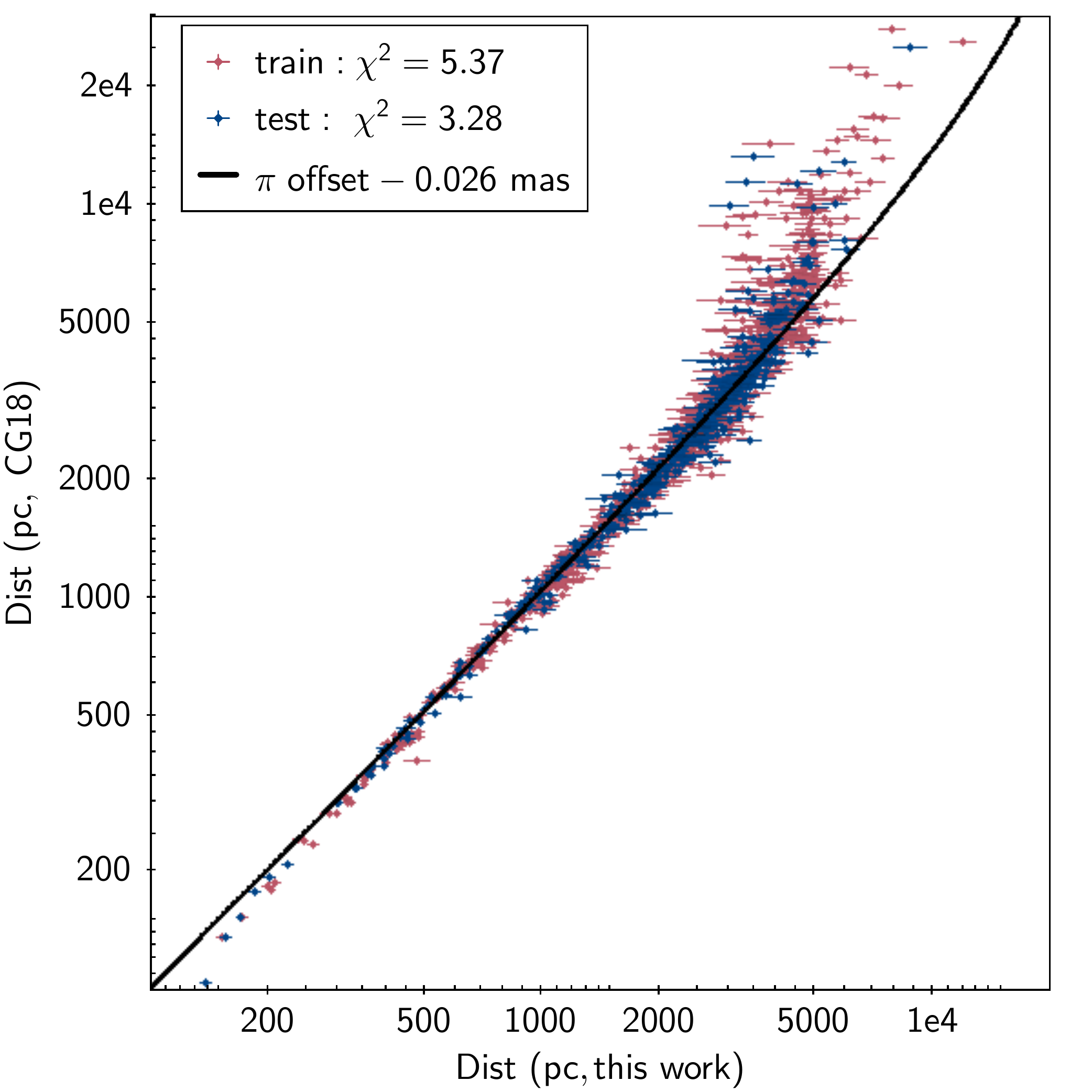}{0.25\textwidth}{}}
        \vspace{-0.75 cm}
\caption{Comparison of distances derived by Auriga to those from other studies.
\label{fig:distcomp}}
\end{figure}

The comparison of these predictions to the fits presented in other papers are shown in Figure \ref{fig:agecomp} for age, Figure \ref{fig:avcomp} for extinction, and Figure \ref{fig:distcomp} for distance. Although, as expected, there is a somewhat better convergence on the train compared to test samples that the network has not seen or learned from, they tend to achieve a comparable performance.

The reported uncertainties could reproduce the difference between the parameters derived in this work to those in other surveys with a factor of 2--3, depending on the underlying precision of the survey and the parameter in question. We note that the corresponding errors (most notably in age) in these surveys are typically not provided -- including them would further account for some of the scatter. Thus, the uncertainties do largely appear to be representative of the underlying error distribution in the data.

There are slight systematic deviations between our work and some datasets. We underestimate the age for old clusters and overestimate the age of the young clusters by $\sim$0.1 dex, with the split between the two occurring at the age of $\sim$120 Myr -- the transition age at which all of the low mass stars would already completely arrive onto the main sequence, and the high mass stars would not yet evolve away from it in bulk. Because the best tracers of age in the young and the old populations are very different, the fitting of their isochrones tend to require different strategy. The slight resulting "S" curve is apparent in most panels in Figure \ref{fig:agecomp}, to a various degree. We note that the ages from \citet{sim2019} show the most consistent linear correlation with our ages, without this discontinuity.

In extinction there is a systematic offset of 0.18 mag in \av\ derived in this work, and that found by \citet{bossini2019} and \citetalias{kounkel2019a}. There is no bulk zero point offset in \av\ relative to \citetalias{kharchenko2013} and \citet{dias2002}, although with a much larger scatter.

Distance to a population can be generally computed in two ways: through averaging the parallax of the individual stars, and through calculating the distance module from their photometry. Auriga considers the weights from both the photometry and the parallax in its interpolation. However, \textit{Gaia} DR2 parallaxes are systematically too small, with the quoted values of this offset as small as -29 $\mu$as, and as big as -82 $\mu$as \citep{lindegren2018,stassun2018,bobylev2019b,xu2019}. Some population studies do take this offset into the account, some do not, but these offsets become increasingly more important the further the cluster is located.

As a result, there is a zero point offset between distances derived here and some of the other works, namely -16 $\mu$as relative to \citet{liu2019} and -26 $\mu$as relative to \citetalias{cantat-gaudin2018a}. There is no offset in comparison to \citet{bossini2019} or \citet{sim2019}. This holds consistently for populations located in between 200 and 5000 pc.  Beyond 5000 pc, the offset for the latter increases to -70 $\mu$as. We note that in cases where in \citetalias{kharchenko2013} training sub-sample the native distances were kept instead of replacing them with the ones from \citetalias{cantat-gaudin2018a}, this tends to result in a more uniform offset of -70 $\mu$as at all distances - although, as that work predates \textit{Gaia}, this results in a much greater uncertainty and scatter in the predicted distance. For the populations located closer than 200 pc, Auriga tends to overestimate their distance by 
$\sim$20 pc - as only a few clusters are found in that space, such an overshoot is probably related to edge effects of the interpolation.

\software{TOPCAT \citep{topcat}, HDBSCAN \citep{hdbscan}, PyTorch \citep{pytorch}, GalPy \citep{galpy}}

\acknowledgments
We thank Tristian Cantat-Gaudin and Luke Bouma for wonderful discussion.
M.K. and K.C. acknowledge support provided by the NSF through grant AST-1449476, and from the Research Corporation via a Time Domain Astrophysics Scialog award (\#24217).
This work has made use of data from the European Space Agency (ESA)
mission {\it Gaia} (\url{https://www.cosmos.esa.int/gaia}), processed by
the {\it Gaia} Data Processing and Analysis Consortium (DPAC,
\url{https://www.cosmos.esa.int/web/gaia/dpac/consortium}). Funding
for the DPAC has been provided by national institutions, in particular
the institutions participating in the {\it Gaia} Multilateral Agreement.

\bibliographystyle{aasjournal.bst}
\bibliography{msv5.bbl}

\begin{thebibliography}{}
\expandafter\ifx\csname natexlab\endcsname\relax\def\natexlab#1{#1}\fi
\providecommand{\url}[1]{\href{#1}{#1}}
\providecommand{\dodoi}[1]{doi:~\href{http://doi.org/#1}{\nolinkurl{#1}}}
\providecommand{\doeprint}[1]{\href{http://ascl.net/#1}{\nolinkurl{http://ascl.net/#1}}}
\providecommand{\doarXiv}[1]{\href{https://arxiv.org/abs/#1}{\nolinkurl{https://arxiv.org/abs/#1}}}

\bibitem[{{Alfaro} {et~al.}(1992){Alfaro}, {Cabrera-Cano}, \&
  {Delgado}}]{alfaro1992}
{Alfaro}, E.~J., {Cabrera-Cano}, J., \& {Delgado}, A.~J. 1992, \apj, 399, 576,
  \dodoi{10.1086/171949}

\bibitem[{Alves {et~al.}(2020)Alves, Zucker, Goodman, Speagle, Meingast,
  Robitaille, Finkbeiner, Schlafly, \& Green}]{alves2020}
Alves, J., Zucker, C., Goodman, A.~A., {et~al.} 2020, Nature,
  \dodoi{10.1038/s41586-019-1874-z}

\bibitem[{{Anders} {et~al.}(2019){Anders}, {Khalatyan}, {Chiappini}, {Queiroz},
  {Santiago}, {Jordi}, {Girardi}, {Brown}, {Matijevi{\v{c}}}, {Monari},
  {Cantat-Gaudin}, {Weiler}, {Khan}, {Miglio}, {Carrillo}, {Romero-G{\'o}mez},
  {Minchev}, {de Jong}, {Antoja}, {Ramos}, {Steinmetz}, \& {Enke}}]{anders2019}
{Anders}, F., {Khalatyan}, A., {Chiappini}, C., {et~al.} 2019, \aap, 628, A94,
  \dodoi{10.1051/0004-6361/201935765}

\bibitem[{{Andrae} {et~al.}(2018){Andrae}, {Fouesneau}, {Creevey}, {Ordenovic},
  {Mary}, {Burlacu}, {Chaoul}, {Jean-Antoine-Piccolo}, {Kordopatis}, {Korn},
  {Lebreton}, {Panem}, {Pichon}, {Th{\'e}venin}, {Walmsley}, \&
  {Bailer-Jones}}]{andrae2018}
{Andrae}, R., {Fouesneau}, M., {Creevey}, O., {et~al.} 2018, \aap, 616, A8,
  \dodoi{10.1051/0004-6361/201732516}

\bibitem[{{Angus} {et~al.}(2019){Angus}, {Morton}, {Foreman-Mackey}, {van
  Saders}, {Curtis}, {Kane}, {Bedell}, {Kiman}, {Hogg}, \&
  {Brewer}}]{angus2019}
{Angus}, R., {Morton}, T.~D., {Foreman-Mackey}, D., {et~al.} 2019, \aj, 158,
  173, \dodoi{10.3847/1538-3881/ab3c53}

\bibitem[{{Angus} {et~al.}(2020){Angus}, {Beane}, {Price-Whelan}, {Newton},
  {Curtis}, {Berger}, {van Saders}, {Kiman}, {Foreman-Mackey}, {Lu},
  {Anderson}, \& {Faherty}}]{angus2020}
{Angus}, R., {Beane}, A., {Price-Whelan}, A.~M., {et~al.} 2020, arXiv e-prints,
  arXiv:2005.09387.
\newblock \doarXiv{2005.09387}

\bibitem[{{Beccari} {et~al.}(2020){Beccari}, {Boffin}, \&
  {Jerabkova}}]{beccari2020}
{Beccari}, G., {Boffin}, H. M.~J., \& {Jerabkova}, T. 2020, \mnras, 491, 2205,
  \dodoi{10.1093/mnras/stz3195}

\bibitem[{{Bekki}(2009)}]{bekki2009}
{Bekki}, K. 2009, \mnras, 398, L36, \dodoi{10.1111/j.1745-3933.2009.00702.x}

\bibitem[{{Bobylev}(2019)}]{bobylev2019b}
{Bobylev}, V.~V. 2019, Astronomy Letters, 45, 10,
  \dodoi{10.1134/S1063773719010018}

\bibitem[{{Bossini} {et~al.}(2019){Bossini}, {Vallenari}, {Bragaglia},
  {Cantat-Gaudin}, {Sordo}, {Balaguer-N{\'u}{\~n}ez}, {Jordi}, {Moitinho},
  {Soubiran}, {Casamiquela}, {Carrera}, \& {Heiter}}]{bossini2019}
{Bossini}, D., {Vallenari}, A., {Bragaglia}, A., {et~al.} 2019, \aap, 623,
  A108, \dodoi{10.1051/0004-6361/201834693}

\bibitem[{{Bovy}(2015)}]{galpy}
{Bovy}, J. 2015, \apjs, 216, 29, \dodoi{10.1088/0067-0049/216/2/29}

\bibitem[{Campello {et~al.}(2013)Campello, Moulavi, \& Sander}]{hdbscan1}
Campello, R. J. G.~B., Moulavi, D., \& Sander, J. 2013, in Advances in
  Knowledge Discovery and Data Mining, ed. J.~Pei, V.~S. Tseng, L.~Cao,
  H.~Motoda, \& G.~Xu (Berlin, Heidelberg: Springer Berlin Heidelberg),
  160--172

\bibitem[{{Cantat-Gaudin} {et~al.}(2018){Cantat-Gaudin}, {Jordi}, {Vallenari},
  {Bragaglia}, {Balaguer-N{\'u}{\~n}ez}, {Soubiran}, {Bossini}, {Moitinho},
  {Castro-Ginard}, {Krone-Martins}, {Casamiquela}, {Sordo}, \&
  {Carrera}}]{cantat-gaudin2018a}
{Cantat-Gaudin}, T., {Jordi}, C., {Vallenari}, A., {et~al.} 2018, \aap, 618,
  A93, \dodoi{10.1051/0004-6361/201833476}

\bibitem[{{Cantat-Gaudin} {et~al.}(2019){Cantat-Gaudin}, {Krone-Martins},
  {Sedaghat}, {Farahi}, {de Souza}, {Skalidis}, {Malz}, {Mac{\^e}do}, {Moews},
  {Jordi}, {Moitinho}, {Castro-Ginard}, {Ishida}, {Heneka}, {Boucaud}, \&
  {Trindade}}]{cantat-gaudin2019a}
{Cantat-Gaudin}, T., {Krone-Martins}, A., {Sedaghat}, N., {et~al.} 2019, \aap,
  624, A126, \dodoi{10.1051/0004-6361/201834453}

\bibitem[{{Cantat-Gaudin} {et~al.}(2020){Cantat-Gaudin}, {Anders},
  {Castro-Ginard}, {Jordi}, {Romero-Gomez}, {Soubiran}, {Casamiquela},
  {Tarricq}, {Moitinho}, {Vallenari}, {Bragaglia}, {Krone-Martins}, \&
  {Kounkel}}]{cantat-gaudin2020}
{Cantat-Gaudin}, T., {Anders}, F., {Castro-Ginard}, A., {et~al.} 2020, arXiv
  e-prints, arXiv:2004.07274.
\newblock \doarXiv{2004.07274}

\bibitem[{{Castro-Ginard} {et~al.}(2019){Castro-Ginard}, {Jordi}, {Luri},
  {Cantat-Gaudin}, \& {Balaguer-N{\'u}{\~n}ez}}]{castro-ginard2019}
{Castro-Ginard}, A., {Jordi}, C., {Luri}, X., {Cantat-Gaudin}, T., \&
  {Balaguer-N{\'u}{\~n}ez}, L. 2019, \aap, 627, A35,
  \dodoi{10.1051/0004-6361/201935531}

\bibitem[{{Castro-Ginard} {et~al.}(2020){Castro-Ginard}, {Jordi}, {Luri},
  {{\'A}lvarez Cid-Fuentes}, {Casamiquela}, {Anders}, {Cantat-Gaudin},
  {Mongui{\'o}}, {Balaguer-N{\'u}{\~n}ez}, {Sol{\`a}}, \&
  {Badia}}]{castro-ginard2020}
{Castro-Ginard}, A., {Jordi}, C., {Luri}, X., {et~al.} 2020, \aap, 635, A45,
  \dodoi{10.1051/0004-6361/201937386}

\bibitem[{{Chen} {et~al.}(2019){Chen}, {Huang}, {Yuan}, {Wang}, {Fan}, {Xiang},
  {Zhang}, {Tian}, \& {Liu}}]{chen2019a}
{Chen}, B.-Q., {Huang}, Y., {Yuan}, H.-B., {et~al.} 2019, \mnras, 483, 4277,
  \dodoi{10.1093/mnras/sty3341}

\bibitem[{{Comeron} \& {Torra}(1994)}]{comeron1994}
{Comeron}, F., \& {Torra}, J. 1994, \aap, 281, 35

\bibitem[{{Dias} {et~al.}(2002){Dias}, {Alessi}, {Moitinho}, \&
  {L{\'e}pine}}]{dias2002}
{Dias}, W.~S., {Alessi}, B.~S., {Moitinho}, A., \& {L{\'e}pine}, J.~R.~D. 2002,
  \aap, 389, 871, \dodoi{10.1051/0004-6361:20020668}

\bibitem[{{Dias} {et~al.}(2019){Dias}, {Monteiro}, {L{\'e}pine}, \&
  {Barros}}]{dias2019}
{Dias}, W.~S., {Monteiro}, H., {L{\'e}pine}, J.~R.~D., \& {Barros}, D.~A. 2019,
  \mnras, 486, 5726, \dodoi{10.1093/mnras/stz1196}

\bibitem[{{Dobbs} \& {Baba}(2014)}]{dobbs2014}
{Dobbs}, C., \& {Baba}, J. 2014, \pasa, 31, e035, \dodoi{10.1017/pasa.2014.31}

\bibitem[{{Ernst} {et~al.}(2011){Ernst}, {Just}, {Berczik}, \&
  {Olczak}}]{ernst2011}
{Ernst}, A., {Just}, A., {Berczik}, P., \& {Olczak}, C. 2011, \aap, 536, A64,
  \dodoi{10.1051/0004-6361/201118021}

\bibitem[{{Gaia Collaboration} {et~al.}(2018){Gaia Collaboration}, {Brown},
  {Vallenari}, {Prusti}, {de Bruijne}, {Babusiaux}, {Bailer-Jones}, {Biermann},
  {Evans}, {Eyer}, \& et~al.}]{gaia-collaboration2018}
{Gaia Collaboration}, {Brown}, A.~G.~A., {Vallenari}, A., {et~al.} 2018, \aap,
  616, A1, \dodoi{10.1051/0004-6361/201833051}

\bibitem[{{Green} {et~al.}(2019){Green}, {Schlafly}, {Zucker}, {Speagle}, \&
  {Finkbeiner}}]{green2019}
{Green}, G.~M., {Schlafly}, E., {Zucker}, C., {Speagle}, J.~S., \&
  {Finkbeiner}, D. 2019, \apj, 887, 93, \dodoi{10.3847/1538-4357/ab5362}

\bibitem[{{Jackson} {et~al.}(2018){Jackson}, {Deliyannis}, \&
  {Jeffries}}]{jackson2018}
{Jackson}, R.~J., {Deliyannis}, C.~P., \& {Jeffries}, R.~D. 2018, \mnras, 476,
  3245, \dodoi{10.1093/mnras/sty374}

\bibitem[{{Jerabkova} {et~al.}(2019){Jerabkova}, {Boffin}, {Beccari}, \&
  {Anderson}}]{jerabkova2019a}
{Jerabkova}, T., {Boffin}, H. M.~J., {Beccari}, G., \& {Anderson}, R.~I. 2019,
  \mnras, 489, 4418, \dodoi{10.1093/mnras/stz2315}

\bibitem[{{Kharchenko} {et~al.}(2013){Kharchenko}, {Piskunov}, {Schilbach},
  {R{\"o}ser}, \& {Scholz}}]{kharchenko2013}
{Kharchenko}, N.~V., {Piskunov}, A.~E., {Schilbach}, E., {R{\"o}ser}, S., \&
  {Scholz}, R.-D. 2013, \aap, 558, A53, \dodoi{10.1051/0004-6361/201322302}

\bibitem[{{Kounkel} \& {Covey}(2019)}]{kounkel2019a}
{Kounkel}, M., \& {Covey}, K. 2019, \aj, 158, 122,
  \dodoi{10.3847/1538-3881/ab339a}

\bibitem[{{Kounkel} {et~al.}(2018){Kounkel}, {Covey}, {Su{\'a}rez},
  {Rom{\'a}n-Z{\'u}{\~n}iga}, {Hernandez}, {Stassun}, {Jaehnig}, {Feigelson},
  {Pe{\~n}a Ram{\'\i}rez}, {Roman-Lopes}, {Da Rio}, {Stringfellow}, {Kim},
  {Borissova}, {Fern{\'a}ndez-Trincado}, {Burgasser},
  {Garc{\'\i}a-Hern{\'a}ndez}, {Zamora}, {Pan}, \& {Nitschelm}}]{kounkel2018a}
{Kounkel}, M., {Covey}, K., {Su{\'a}rez}, G., {et~al.} 2018, \aj, 156, 84,
  \dodoi{10.3847/1538-3881/aad1f1}

\bibitem[{{Li} \& {Gnedin}(2019)}]{li2019}
{Li}, H., \& {Gnedin}, O.~Y. 2019, \mnras, 486, 4030,
  \dodoi{10.1093/mnras/stz1114}

\bibitem[{{Lindegren} {et~al.}(2018){Lindegren}, {Hern{\'a}ndez}, {Bombrun},
  {Klioner}, {Bastian}, {Ramos-Lerate}, {de Torres}, {Steidelm{\"u}ller},
  {Stephenson}, {Hobbs}, {Lammers}, {Biermann}, {Geyer}, {Hilger}, {Michalik},
  {Stampa}, {McMillan}, {Casta{\~n}eda}, {Clotet}, {Comoretto}, {Davidson},
  {Fabricius}, {Gracia}, {Hambly}, {Hutton}, {Mora}, {Portell}, {van Leeuwen},
  {Abbas}, {Abreu}, {Altmann}, {Andrei}, {Anglada}, {Balaguer-N{\'u}{\~n}ez},
  {Barache}, {Becciani}, {Bertone}, {Bianchi}, {Bouquillon}, {Bourda},
  {Br{\"u}semeister}, {Bucciarelli}, {Busonero}, {Buzzi}, {Cancelliere},
  {Carlucci}, {Charlot}, {Cheek}, {Crosta}, {Crowley}, {de Bruijne}, {de
  Felice}, {Drimmel}, {Esquej}, {Fienga}, {Fraile}, {Gai}, {Garralda},
  {Gonz{\'a}lez-Vidal}, {Guerra}, {Hauser}, {Hofmann}, {Holl}, {Jordan},
  {Lattanzi}, {Lenhardt}, {Liao}, {Licata}, {Lister}, {L{\"o}ffler},
  {Marchant}, {Martin-Fleitas}, {Messineo}, {Mignard}, {Morbidelli}, {Poggio},
  {Riva}, {Rowell}, {Salguero}, {Sarasso}, {Sciacca}, {Siddiqui}, {Smart},
  {Spagna}, {Steele}, {Taris}, {Torra}, {van Elteren}, {van Reeven}, \&
  {Vecchiato}}]{lindegren2018}
{Lindegren}, L., {Hern{\'a}ndez}, J., {Bombrun}, A., {et~al.} 2018, \aap, 616,
  A2, \dodoi{10.1051/0004-6361/201832727}

\bibitem[{{Liu} \& {Pang}(2019)}]{liu2019}
{Liu}, L., \& {Pang}, X. 2019, \apjs, 245, 32, \dodoi{10.3847/1538-4365/ab530a}

\bibitem[{{Luri} {et~al.}(2018){Luri}, {Brown}, {Sarro}, {Arenou},
  {Bailer-Jones}, {Castro-Ginard}, {de Bruijne}, {Prusti}, {Babusiaux}, \&
  {Delgado}}]{luri2018}
{Luri}, X., {Brown}, A.~G.~A., {Sarro}, L.~M., {et~al.} 2018, \aap, 616, A9,
  \dodoi{10.1051/0004-6361/201832964}

\bibitem[{{Lutz} \& {Kelker}(1973)}]{lutz1973}
{Lutz}, T.~E., \& {Kelker}, D.~H. 1973, \pasp, 85, 573, \dodoi{10.1086/129506}

\bibitem[{{Malmquist}(1922)}]{malmquist1922}
{Malmquist}, K.~G. 1922, Meddelanden fran Lunds Astronomiska Observatorium
  Serie I, 100, 1

\bibitem[{{Marigo} {et~al.}(2017){Marigo}, {Girardi}, {Bressan}, {Rosenfield},
  {Aringer}, {Chen}, {Dussin}, {Nanni}, {Pastorelli}, {Rodrigues}, {Trabucchi},
  {Bladh}, {Dalcanton}, {Groenewegen}, {Montalb{\'a}n}, \& {Wood}}]{marigo2017}
{Marigo}, P., {Girardi}, L., {Bressan}, A., {et~al.} 2017, \apj, 835, 77,
  \dodoi{10.3847/1538-4357/835/1/77}

\bibitem[{McInnes {et~al.}(2017)McInnes, Healy, \& Astels}]{hdbscan}
McInnes, L., Healy, J., \& Astels, S. 2017, The Journal of Open Source
  Software, 2, \dodoi{10.21105/joss.00205}

\bibitem[{{Meingast} {et~al.}(2019){Meingast}, {Alves}, \&
  {F{\"u}rnkranz}}]{meingast2019}
{Meingast}, S., {Alves}, J., \& {F{\"u}rnkranz}, V. 2019, \aap, 622, L13,
  \dodoi{10.1051/0004-6361/201834950}

\bibitem[{{Olney} {et~al.}(2020){Olney}, {Kounkel}, {Schillinger}, {Scoggins},
  {Yin}, {Howard}, {Covey}, {Hutchinson}, \& {Stassun}}]{olney2020}
{Olney}, R., {Kounkel}, M., {Schillinger}, C., {et~al.} 2020, \aj, 159, 182,
  \dodoi{10.3847/1538-3881/ab7a97}

\bibitem[{{Oort}(1958)}]{oort1958}
{Oort}, J.~H. 1958, Ricerche Astronomiche, 5, 507

\bibitem[{Paszke {et~al.}(2017)Paszke, Gross, Chintala, Chanan, Yang, DeVito,
  Lin, Desmaison, Antiga, \& Lerer}]{pytorch}
Paszke, A., Gross, S., Chintala, S., {et~al.} 2017, in NIPS-W

\bibitem[{{P{\"o}ppel} \& {Marronetti}(2000)}]{poppel2000}
{P{\"o}ppel}, W.~G.~L., \& {Marronetti}, P. 2000, \aap, 358, 299

\bibitem[{{Reid} {et~al.}(2019){Reid}, {Menten}, {Brunthaler}, {Zheng}, {Dame},
  {Xu}, {Li}, {Sakai}, {Wu}, {Immer}, {Zhang}, {Sanna}, {Moscadelli}, {Rygl},
  {Bartkiewicz}, {Hu}, {Quiroga-Nu{\~n}ez}, \& {van Langevelde}}]{reid2019}
{Reid}, M.~J., {Menten}, K.~M., {Brunthaler}, A., {et~al.} 2019, \apj, 885,
  131, \dodoi{10.3847/1538-4357/ab4a11}

\bibitem[{{Rix} \& {Bovy}(2013)}]{rix2013}
{Rix}, H.-W., \& {Bovy}, J. 2013, \aapr, 21, 61,
  \dodoi{10.1007/s00159-013-0061-8}

\bibitem[{{Robinson} {et~al.}(2016){Robinson}, {von Hippel}, {Stein},
  {Stenning}, {Wagner-Kaiser}, {Si}, \& {van Dyk}}]{base9}
{Robinson}, E., {von Hippel}, T., {Stein}, N., {et~al.} 2016, {BASE-9: Bayesian
  Analysis for Stellar Evolution with nine variables}, Astrophysics Source Code
  Library.
\newblock \doeprint{1608.007}

\bibitem[{{Romero-G{\'o}mez} {et~al.}(2019){Romero-G{\'o}mez}, {Mateu},
  {Aguilar}, {Figueras}, \& {Castro-Ginard}}]{romero-gomez2019}
{Romero-G{\'o}mez}, M., {Mateu}, C., {Aguilar}, L., {Figueras}, F., \&
  {Castro-Ginard}, A. 2019, \aap, 627, A150,
  \dodoi{10.1051/0004-6361/201834908}

\bibitem[{{R{\"o}ser} \& {Schilbach}(2019)}]{roser2019}
{R{\"o}ser}, S., \& {Schilbach}, E. 2019, \aap, 627, A4,
  \dodoi{10.1051/0004-6361/201935502}

\bibitem[{{R{\"o}ser} {et~al.}(2019){R{\"o}ser}, {Schilbach}, \&
  {Goldman}}]{roser2019a}
{R{\"o}ser}, S., {Schilbach}, E., \& {Goldman}, B. 2019, \aap, 621, L2,
  \dodoi{10.1051/0004-6361/201834608}

\bibitem[{{Schinnerer} {et~al.}(2017){Schinnerer}, {Meidt}, {Colombo}, {Chand
  ar}, {Dobbs}, {Garc{\'\i}a-Burillo}, {Hughes}, {Leroy}, {Pety}, {Querejeta},
  {Kramer}, \& {Schuster}}]{schinnerer2017}
{Schinnerer}, E., {Meidt}, S.~E., {Colombo}, D., {et~al.} 2017, \apj, 836, 62,
  \dodoi{10.3847/1538-4357/836/1/62}

\bibitem[{{Sim} {et~al.}(2019){Sim}, {Lee}, {Ann}, \& {Kim}}]{sim2019}
{Sim}, G., {Lee}, S.~H., {Ann}, H.~B., \& {Kim}, S. 2019, Journal of Korean
  Astronomical Society, 52, 145, \dodoi{10.5303/JKAS.2019.52.5.145}

\bibitem[{{Stassun} \& {Torres}(2018)}]{stassun2018}
{Stassun}, K.~G., \& {Torres}, G. 2018, \apj, 862, 61,
  \dodoi{10.3847/1538-4357/aacafc}

\bibitem[{{Taylor}(2005)}]{topcat}
{Taylor}, M.~B. 2005, in Astronomical Society of the Pacific Conference Series,
  Vol. 347, Astronomical Data Analysis Software and Systems XIV, ed.
  P.~{Shopbell}, M.~{Britton}, \& R.~{Ebert}, 29

\bibitem[{{Xu} {et~al.}(2019){Xu}, {Zhang}, {Reid}, {Zheng}, \&
  {Wang}}]{xu2019}
{Xu}, S., {Zhang}, B., {Reid}, M.~J., {Zheng}, X., \& {Wang}, G. 2019, \apj,
  875, 114, \dodoi{10.3847/1538-4357/ab0e83}

\end{thebibliography}

\end{document}